\documentclass[twocolumn]{aastex631}
\usepackage{tikz}
\usepackage{graphicx}
\usepackage{tabularx}
\usepackage{xcolor,colortbl}
\usetikzlibrary{decorations.markings}
\usetikzlibrary{calc,patterns,angles,quotes}
\usepackage{tkz-euclide}
\usepackage{amsmath}
\usepackage{breqn}
\usepackage{datetime2}

\makeatletter
\newcommand{\HI}{\ensuremath{\textrm{\ion{H}{1}}}}

\newcommand{\kHz}{\ensuremath{\textrm{ kHz}}}
\newcommand{\MHz}{\ensuremath{\textrm{ MHz}}}
\newcommand{\GHz}{\ensuremath{\textrm{ GHz}}}
\newcommand{\Myr}{\ensuremath{\textrm{ Myr}}}
\newcommand{\mas}{\ensuremath{\textrm{ mas}}}
\newcommand{\yr}{\ensuremath{\textrm{ yr}}}
\newcommand{\K}{\ensuremath{\textrm{ K}}}
\newcommand{\m}{\ensuremath{\textrm{ m}}}
\newcommand{\cm}{\ensuremath{\textrm{ cm}}}
\newcommand{\pc}{\ensuremath{\textrm{ pc}}}

\newcommand{\EM}{\ensuremath{\textrm{EM}}}
\newcommand{\RM}{\ensuremath{\textrm{RM}}}

\newcommand{\Jybeam}{\ensuremath{\textrm{ Jy beam}^{-1}}}
\newcommand{\cucm}{\ensuremath{\textrm{ cm}^{-3}}}
\newcommand{\radmsq}{\ensuremath{\textrm{ rad m}^{-2}}}
\newcommand{\kms}{\ensuremath{\textrm{ km s}^{-1}}}
\renewcommand{\Re}{\ensuremath{\textrm{ Re}}}
\renewcommand{\Im}{\ensuremath{\textrm{ Im}}}
\newcommand{\uG}{\ensuremath{\, \mu \textrm{G}}}

\makeatother
\received{May 24, 2024}

\submitjournal{\apj}

\shortauthors{Nasser Mohammed et al.}
\shorttitle{The `tadpole': G137+7 with CHIME}
\begin{document}

\title{Faraday tomography with CHIME: the `tadpole' feature G137+7}

\correspondingauthor{Nasser Mohammed, Anna Ordog}
\email{nmohamme@student.ubc.ca, anna.ordog@ubc.ca}

\newcommand{\ASIAA}{Institute of Astronomy and Astrophysics, Academia Sinica, Astronomy-Mathematics Building, No. 1, Sec. 4, Roosevelt Road, Taipei 10617, Taiwan}
\newcommand{\ASU}{Department of Physics, Arizona State University, Tempe, AZ 85287, USA}
\newcommand{\CCAPS}{Department of Astronomy and Cornell Center for Astrophysics and Planetary Science, Cornell University, Ithaca NY 14853, USA}
\newcommand{\CIFAR}{Canadian Institute for Advanced Research, MaRS Centre, West Tower, 661 University Avenue, Suite 505}
\newcommand{\CIT}{Cahill Center for Astronomy and Astrophysics, MC 249-17 California Institute of Technology, Pasadena CA 91125, USA}
\newcommand{\CITA}{Canadian Institute for Theoretical Astrophysics, 60 St.~George Street, Toronto, ON M5S 3H8, Canada}
\newcommand{\CMU}{Department of Physics, Carnegie Mellon University, 5000 Forbes Avenue, Pittsburgh, 15213, PA, USA}
\newcommand{\CSIRO}{CSIRO Space \& Astronomy, Parkes Observatory, P.O. Box 276, Parkes NSW 2870, Australia}
\newcommand{\CSIROBentley}{ATNF, CSIRO Space \& Astronomy, Bentley, WA, Australia}
\newcommand{\DAA}{David A.~Dunlap Department of Astronomy \& Astrophysics, University of Toronto, 50 St.~George Street, Toronto, ON M5S 3H4, Canada}
\newcommand{\DI}{Dunlap Institute for Astronomy \& Astrophysics, University of Toronto, 50 St.~George Street, Toronto, ON M5S 3H4, Canada}
\newcommand{\DRAO}{Dominion Radio Astrophysical Observatory, Herzberg Research Centre for Astronomy and Astrophysics, National Research Council Canada, PO Box 248, Penticton, BC V2A 6J9, Canada}
\newcommand{\FRQNT}{FRQNT Postdoctoral Fellow}
\newcommand{\ICRAR}{International Centre for Radio Astronomy Research (ICRAR), Curtin University, Bentley WA 6102 Australia}
\newcommand{\INAF}{INAF-Istituto di Radioastronomia, Via Gobetti 101, 40129 Bologna, Italy}
\newcommand{\INAFOAA}{INAF – Osservatorio Astrofisico di Arcetri, Largo E. Fermi 5, 50125 Firenze, Italy}
\newcommand{\LPENS}{Laboratoire de Physique de l’Ecole Normale Supérieure, ENS, Université PSL, CNRS, Sorbonne Université, Université de Paris, F-75005 Paris, France}
\newcommand{\MAN}{Department of Physics and Astronomy, University of Manchester, Manchester, UK}
\newcommand{\MITK}{MIT Kavli Institute for Astrophysics and Space Research, Massachusetts Institute of Technology, 77 Massachusetts Ave, Cambridge, MA 02139, USA}
\newcommand{\MITP}{Department of Physics, Massachusetts Institute of Technology, 77 Massachusetts Ave, Cambridge, MA 02139, USA}
\newcommand{\MKI}{MIT Kavli Institute for Astrophysics and Space Research, Massachusetts Institute of Technology, 77 Massachusetts Ave, Cambridge, MA 02139, USA}
\newcommand{\MPIfR}{Max-Planck-Institut für Radioastronomie, Auf dem Hügel 69, 53121 Bonn, Germany}
\newcommand{\MSI}{McGill Space Institute, McGill University, 3550 rue University, Montr\'eal, QC H3A 2A7, Canada}
\newcommand{\MSIF}{McGill Space Institute Fellow}
\newcommand{\MU}{Department of Physics, McGill University, 3600 rue University, Montr\'eal, QC H3A 2T8, Canada}
\newcommand{\NCRA}{National Centre for Radio Astrophysics, Post Bag 3, Ganeshkhind, Pune, 411007, India}
\newcommand{\NHFP}{NHFP Einstein Fellow}
\newcommand{\NRAO}{National Radio Astronomy Observatory, 520 Edgemont Rd, Charlottesville, VA 22903, USA}
\newcommand{\PI}{Perimeter Institute for Theoretical Physics, 31 Caroline Street N, Waterloo, ON N25 2YL, Canada}
\newcommand{\RU}{Department of Astrophysics/IMAPP, Radboud University, PO Box 9010, 6500 GL Nijmegen, The Netherlands}
\newcommand{\SFU}{Department of Statistics & Actuarial Science, Simon Fraser University, 8888 University Dr, Burnaby, BC V5A 1S6, Canada}
\newcommand{\Skaha}{Skaha Remote Sensing Ltd., 3165 Juniper Drive, Naramata, BC V0H 1N0, Canada}
\newcommand{\SR}{Sidrat Research, 124 Merton Street, Suite 507, Toronto, ON M4S 2Z2, Canada}
\newcommand{\Stanford}{Banting and KIPAC Fellowships: Kavli Institute for Particle Astrophysics \& Cosmology (KIPAC), Stanford University, Stanford, CA 94305, USA}
\newcommand{\TAPIR}{TAPIR, Walter Burke Institute for Theoretical Physics, Mail Code 350-17, Caltech, Pasadena, CA 91125, USA}
\newcommand{\TIFR}{Department of Astronomy and Astrophysics, Tata Institute of Fundamental Research, Mumbai, 400005, India}
\newcommand{\TMU}{Department of Physics, Toronto Metropolitan University, 350 Victoria St, Toronto, ON M5B 2K3, Canada}
\newcommand{\TSI}{Trottier Space Institute, McGill University, 3550 rue University, Montr\'eal, QC H3A 2A7, Canada}
\newcommand{\UBC}{Department of Physics and Astronomy, University of British Columbia, 6224 Agricultural Road, Vancouver, BC V6T 1Z1 Canada}
\newcommand{\UBCO}{Department of Computer Science, Math, Physics, \& Statistics, University of British Columbia, Okanagan Campus, Kelowna, BC V1V 1V7, Canada}
\newcommand{\UCalgary}{Department of Physics and Astronomy, University of Calgary, 2500 University Drive NW, Calgary, Alberta, T2N 1N4, Canada}
\newcommand{\UCHILE}{Department of Electrical Engineering, Universidad de Chile, Av. Tupper 2007, Santiago 8370451, Chile}
\newcommand{\UM}{Department of Physics and Astronomy, University of Manitoba, Winnipeg, MB R3T 2N2, Canada}
\newcommand{\UTas}{School of Natural Sciences, University of Tasmania, Hobart, Tas 7000 Australia}
\newcommand{\UTPHYS}{Department of Physics, University of Toronto, 60 St.~George Street, Toronto, ON M5S 1A7, Canada}
\newcommand{\UVA}{Anton Pannekoek Institute for Astronomy, University of Amsterdam, Science Park 904, 1098 XH Amsterdam, The Netherlands}
\newcommand{\UW}{Department of Physics and Astronomy, University of Waterloo, Waterloo, ON N2L 3G1, Canada}
\newcommand{\UWM}{Department of Physics, University of Wisconsin-Madison, 1150 University Ave, Madison, WI 53706, USA}
\newcommand{\WVUGWAC}{Center for Gravitational Waves and Cosmology, West Virginia University, Chestnut Ridge Research Building, Morgantown, WV 26505, USA}
\newcommand{\WVULCSEE}{Lane Department of Computer Science and Electrical Engineering, 1220 Evansdale Drive, PO Box 6109  Morgantown, WV 26506, USA}
\newcommand{\WVUPA}{Department of Physics and Astronomy, West Virginia University, P.O. Box 6315, Morgantown, WV 26506, USA}
\newcommand{\YU}{Department of Physics, Yale University, New Haven, CT 06520, USA}
\newcommand{\OC}{Department of Physics and Astronomy, Okanagan College, Kelowna, BC V1Y 4X8, Canada}

\author[0009-0008-1224-0382]{Nasser Mohammed}
\affiliation{\UBCO }
\affiliation{\DRAO }
\author[0000-0002-2465-8937]{Anna Ordog}
\affiliation{\UBCO }
\affiliation{\DRAO }
\author[0000-0001-5181-6673]{Rebecca A. Booth}
\affiliation{\UCalgary }
\author[0000-0003-0932-3140]{Andrea Bracco}
\affiliation{\INAFOAA }
\affiliation{\LPENS }
\author[0000-0003-4781-5701]{Jo-Anne C. Brown}
\affiliation{\UCalgary }
\author[0000-0002-3973-8403]{Ettore Carretti}
\affiliation{\INAF }
\author[0000-0002-6300-7459]{John M. Dickey}
\affiliation{\UTas }
\author[0000-0002-0190-2271]{Simon Foreman}
\affiliation{\ASU }
\author[0000-0002-1760-0868]{Mark Halpern}
\affiliation{\UBC }
\author[0000-0002-5288-312X]{Marijke Haverkorn}
\affiliation{\RU }
\author[0000-0001-7301-5666]{Alex S. Hill}
\affiliation{\UBCO }
\affiliation{\DRAO }
\author[0000-0002-4241-8320]{Gary Hinshaw}
\affiliation{\UBC }
\author[0000-0002-3354-3859]{Joseph W Kania}
\affiliation{\WVUPA }
\affiliation{\WVUGWAC }
\author[0000-0001-5953-0100]{Roland Kothes}
\affiliation{\DRAO }
\author[0000-0003-1455-2546]{T.L. Landecker}
\affiliation{\DRAO }
\author[0000-0001-8064-6116]{Joshua MacEachern}
\affiliation{\UBC }
\author[0000-0002-4279-6946]{Kiyoshi W. Masui}
\affiliation{\MITK }
\affiliation{\MITP }
\author[0009-0007-6503-1501]{Aimee Menard}
\affiliation{\UBCO }
\affiliation{\DRAO }
\author[0000-0003-2469-1611]{Ryan R. Ransom}
\affiliation{\OC }
\affiliation{\DRAO }
\author[0000-0002-5313-6409]{Wolfgang Reich}
\affiliation{\MPIfR }
\author[0009-0006-8615-8352]{Patricia Reich}
\affiliation{\MPIfR }
\author[0000-0002-4543-4588]{J. Richard Shaw}
\affiliation{\UBC }
\author[0000-0003-2631-6217]{Seth R. Siegel}
\affiliation{\PI }
\affiliation{\MU }
\affiliation{\TSI }
\author[0000-0001-8749-1436]{Mehrnoosh Tahani}
\affiliation{\Stanford }
\author[0000-0001-9472-041X]{Alec J. M. Thomson}
\affiliation{\CSIROBentley }
\author[0000-0002-9516-3245]{Tristan Pinsonneault-Marotte}
\affiliation{\UBC }
\author[0000-0002-1491-3738]{Haochen Wang}
\affiliation{\MITK }
\affiliation{\MITP }
\author[0000-0001-7722-8458]{Jennifer L. West}
\affiliation{\DRAO }
\author[0009-0006-9479-7509]{Maik Wolleben}
\affiliation{\Skaha }
\author[0000-0001-7314-9496]{Dallas Wulf}
\affiliation{\MU }
\affiliation{\TSI }
\collaboration{29}{CHIME and GMIMS Collaborations}

\begin{abstract} 
A direct consequence of Faraday rotation is that the polarized radio sky does not resemble the total intensity sky at long wavelengths. We analyze G137+7, which is undetectable in total intensity but appears as a depolarization feature. We use the first polarization maps from the Canadian Hydrogen Intensity Mapping Experiment. Our $400-729 \MHz$ bandwidth and angular resolution, $17'$ to $30'$, allow us to use Faraday synthesis to analyze the polarization structure. In polarized intensity and polarization angle maps, we find a ``tail'' extending $10\arcdeg$ from the ``head'' and designate the combined object the ``tadpole''. Similar polarization angles, distinct from the background, indicate that the head and tail are physically associated. The head appears as a depolarized ring in single channels, but wideband observations show that it is a Faraday rotation feature. Our investigations of \ion{H}{1} and H$\alpha$ find no connections to the tadpole. The tail suggests motion of either the gas or an ionizing star through the ISM; the B2(e) star HD~20336 is a candidate. While the head features a coherent, $\sim-8 \radmsq$ Faraday depth, Faraday synthesis also identifies multiple components in both the head and tail. We verify the locations of the components in the spectra using QU fitting. Our results show that $\sim$octave-bandwidth Faraday rotation observations at $\sim 600 \MHz$ are sensitive to low-density ionized or partially-ionized gas which is undetectable in other tracers.
\end{abstract}

\section{Introduction} \label{sec:intro}

Investigations of the Galactic interstellar medium (ISM) have revealed the pervasive presence of magnetic fields and ionized gas \citep{Ferriere:2001}. Observations of radio polarization can probe various scales and phases of the ISM, revealing crucial information about the interplay of magnetic fields with other energy sources. Deriving the three-dimensional configuration of the magnetic field from polarization data can be challenging; nevertheless, recent observations have enabled progress towards a much clearer picture of the evolution of the ISM and the formation of clouds and stars \citep[e.g.][]{2022A&A...660L...7T,2022A&A...660A..97T}.

At wavelengths $\sim 1-100 \cm$, polarized radiation largely arises from synchrotron emission generated by cosmic ray electrons as they spiral around magnetic fields. Polarized radiation beyond the Earth's atmosphere was first detected by \citet{westerhout_brouw_muller_tinbergen_1962} and \citet{wielebinski_shakeshaft_pauliny-toth} and then extensively mapped by \citet{brouw_spoelstra_1976}. Recent surveys of this polarized radiation have Nyquist-sampled wide areas of the sky in different frequency ranges \citep{wolleben:2019,wolleben:2021,Carretti:2019}.

Linearly polarized electromagnetic waves undergo Faraday rotation as they propagate through a magneto-ionic medium. The resulting change in polarization angle informs us of electron density and magnetic field strength and direction. Exploiting this effect, polarized emission from extragalactic sources propagating through the Galaxy has been used to measure the two-dimensional distribution of magnetic fields in the Milky Way and nearby galaxies averaged along the line of sight \citep{Brown:Taylor:2003,Taylor:2009,2012ApJ...755...21M,ordog:2017,Tahani:2018,vaneck:2021,hutschenreuter:2022,thomson:2023}.  The Faraday rotation of the emission of the Galaxy itself is an especially powerful probe of the diffuse ISM because the emitting cosmic rays and the Faraday rotating thermal gas are mixed, where these emission regions illuminate different Faraday rotating regions along the line of sight \citep[e.g.][]{Gaensler:Dickey:2001,2012ApJ...759...25M,vaneck:haverkorn:2017}.

Faraday rotation probes the convolution of ionized gas density and the line-of-sight magnetic field; therefore, it is sensitive to two distinct components of the ISM. The majority of the ionized gas in the Milky Way ISM is found in the warm ionized medium (WIM), traced by H$\alpha$ emission with an ionization fraction $\gtrsim 90\%$ based on observations of the [\ion{O}{1}]$\lambda 6300$ line \citep{Hausen:Reynolds:2002}. However, particularly at low frequencies $\lesssim 1 \GHz$, Faraday rotation is sensitive to very small columns of free electrons and might trace a warm partially-ionized medium (WPIM) \citep{heiles:haverkorn:2012} or the low (few percent; \citealt{Wolfire:2003,Jenkins:2013}) ionization in the warm neutral medium (WNM) \citep{Foster:Kothes:2013,vaneck:haverkorn:2017,bracco:ntormousi:2022}. In fact, low-frequency Faraday rotation observations may not be sensitive to the traditional WIM \citep{Haffner:2009} at all because the electron density there is high enough to cause depolarization even with weak magnetic fields \citep{vaneck:haverkorn:2017}.

\citet{burn_1966} established the formalism for extracting three-dimensional information on the diffuse magneto-ionized ISM with mixed synchrotron emission and Faraday rotation. However, the technique requires data over a wide range of wavelength-squared (${\lambda}^2$), which became feasible only recently with the advent of radio telescopes equipped with wideband receivers and modern digital signal processors \citep{brentjens_de_bruyn_2005,Heald:2009}. Useful $\lambda^2$ coverage typically implies low frequencies and wide bandwidths (ideally octave or more). The use of Faraday synthesis,\footnote{We mostly adopt the terminology described in Table~1 of \citet{Sun:Rudnick:2015}: ``Faraday depth'', ``Faraday spectrum'', ``Faraday synthesis'' \citep{brentjens_de_bruyn_2005}, ``rotation measure spread function'', ``Faraday clean'' \citep{Heald:2009}, and ``3D Faraday synthesis'' \citep{2012A&A...540A..80B}.} a form of Faraday tomography \citep{2023PASJ...75S..50T}, enables studies of the large-scale structure of the magnetic field \citep{dickey:2019,dickey:2022,erceg:jelic:2022}, and also of individual objects and small regions \citep[e.g.][]{schnitzeler:2007,vaneck:haverkorn:2017,vaneck:haverkorn:2019,thomson:2019,thomson:2021}. Direct modelling of the spectra of Stokes parameters $Q$ and $U$ (``QU fitting'') has proven able to detect multiple Faraday depth components in Faraday complex spectra more reliably than Faraday synthesis, but with the drawbacks of needing considerably longer computational times for each LOS and requiring us to assume a Faraday rotation model \citep{Farnsworth:Rudnick:2011,O'Sullivan:Brown:2012,Ideguchi:Takahashi:2014,Sun:Rudnick:2015}. In practice, one often uses Faraday synthesis to inform the selection of models for QU fitting.

Here we study a region, G137+7, first discovered by \citet{Berkhuijsen_Brouw_etal_1964} in a polarization survey of the northern sky at $610 \MHz$ as a ring of low polarization, $2\arcdeg$ in diameter, centered on $\ell = +137\arcdeg, b = +7\arcdeg$. It lies in the so-called Fan region, an area otherwise bright in polarized intensity and uniform in polarization angle at $408 \MHz$ \citep{verschuur_1968}. \citet{verschuur_1969} associated it with the star HD 20336 (distance $246 \pm 20 \pc$; \citealt{gaia_2022}), suggesting that the B2(e) star had tunnelled its way through a cloud of neutral hydrogen, disrupting it and ionizing a portion of the hydrogen gas. \citet{haverkorn_2003} detected this polarized circular object at 350 \MHz, but considered HD 20336 an unlikely progenitor for the feature on account of the star's high proper motion of 18 mas/year being too large to maintain a circular Str\"{o}mgren sphere. Instead, they suggest that the structure would be elongated in the direction opposite to the motion of the star. \citet{iacobelli_haverkorn_katgert_2012} used $150 \MHz$ Faraday synthesis to place the object at a distance of $\sim{100}$\,pc, possibly in the wall of the Local Bubble. The lack of a detectable \ion{H}{2} region led to the suggestion that the source of ionization might be an unidentified  white dwarf. In this paper, we present observations which reveal a tail-like prominence extending approximately $10\arcdeg$ from a prominent circular region of Faraday rotation, coincident with the structure studied by earlier authors. This head-tail structure has led us to name G137+7 the \emph{tadpole}.

The outline of this paper is as follows. In Section~\ref{ssec:faraday_rot}, we review Faraday rotation and Faraday synthesis. In Section~\ref{sec:data_and_calibration}, we describe the Canadian Hydrogen Intensity Mapping Experiment (CHIME) data (Section~\ref{ssec:chime_lbn}), the Dominion Radio Astrophysical Observatory (DRAO) Synthesis Telescope (ST) data (Section~\ref{ssec:st}), and published data sets to which we compare the CHIME maps (Section~\ref{ssec:ancillary}). We present the observed features of the tadpole in Section~\ref{sec:features} and discuss its origin in Section~\ref{sec:origins}. We summarize the paper in Section~\ref{sec:summary}. In Appendix~\ref{sec:A1}, we present simulations of the impact of marginally-resolved Faraday complexity on Faraday synthesis observations. In Appendix~\ref{sec:A2}, we present our QU fitting results.

\section{Faraday rotation and Faraday synthesis} \label{ssec:faraday_rot}

If a polarized photon is emitted with an intrinsic polarization angle $\chi_0$, the polarization angle we measure at wavelength $\lambda$ after this signal has been Faraday rotated by the Galactic ISM is
\begin{equation}
    \chi = \chi_0 + \phi \cdot \lambda^2.
    \label{eq:chi}
\end{equation}
The Faraday depth, $\phi$, is defined as \citep{ferriere:west:2021}
\begin{equation}
    \frac{\phi}{(\text{rad m}^{-2})} = 0.81 \int_{\text{emission}}^{\text{observer}} \frac{n_e}{\text{cm}^{-3}} \frac{\boldsymbol{B}}{\mu\text{G}} \cdot\frac{d\boldsymbol{l}}{\text{pc}}
    \label{eq:RM}
\end{equation}
where $n_e$ is the electron density of the ISM, and we are projecting the magnetic field, $\boldsymbol{B}$, along our line of sight (LOS), $d\boldsymbol{l}$. There is a Fourier transform-like relationship between the observable complex polarization\footnote{We use tilde to indicate complex values.} $\tilde{P}(\lambda^2) = Q + iU$ and the Faraday spectrum $\tilde{F}(\phi)$, which is recoverable through a process known as Faraday synthesis \citep{burn_1966, brentjens_de_bruyn_2005}. The polarized intensity at each Faraday depth is $p(\phi)I \equiv \sqrt{Q(\phi)^2 + U(\phi)^2}$, where $p(\phi)$ is the fractional polarization and $I$ is total intensity. The Faraday spectrum is the spectrum of polarized intensity as a function of Faraday depth for each LOS. A weighting function $W(\lambda^2)$ introduces a rotation measure spread function (RMSF), $\tilde{R}(\phi)$, the Fourier transform of $W(\lambda^2)$. Gaps in $W(\lambda^2)$ result in sidelobes in $\tilde{R}(\phi)$.
Utilizing polarization data over a wide range of wavelengths allows the creation of a (complex) Faraday depth cube of $\tilde{F}(\ell, b, \phi)$ for Galactic coordinates $(\ell, b)$.

The Faraday depth resolution, measured as the FWHM of the main lobe of $R(\phi)=|\tilde{R}(\phi)|$, can be approximated as \citep{Schnitzeler:2009}
\begin{equation} \label{eq:res}
\delta \phi \approx \frac{3.8}{\lambda_{\mathrm{max}}^2 - \lambda_{\mathrm{min}}^2}
\end{equation}
for reference wavelength $\lambda_0$ (the wavelength to which the polarization angles are derotated; \citealt{brentjens_de_bruyn_2005}, Equations 25-26) set to the average $\lambda^2$ in the band (as we do in this paper). The largest-scale feature in Faraday depth space that is not depolarized is \citep{brentjens_de_bruyn_2005}
\begin{equation} \label{eq:Wbrentjens}
\phi_{\mathrm{max-scale}} \approx \frac{\pi}{\lambda_{\mathrm{min}}^2}.
\end{equation}
\citet{Rudnick:Cotton:2023} argue that setting $\lambda_0 = 0$ recovers additional information and modifies Equations~\ref{eq:res} and \ref{eq:Wbrentjens}, but we use traditional Faraday synthesis in this work.

\section{Data \& Instruments} \label{sec:data_and_calibration}

\subsection{CHIME and GMIMS surveys}

This work is a part of the Global Magneto-Ionic Medium Survey (GMIMS), a collaboration using telescopes in the Northern and Southern hemispheres to obtain all-sky diffuse polarization maps with frequency coverage and resolution designed to, in combination, be sensitive to large Faraday depth scales ($\phi_{\mathrm{max-scale}} \sim 100 \radmsq$) with fine sensitivity to small Faraday depth structures ($\delta \phi \sim 10 \radmsq$). This requires frequency coverage from $\sim 300 - 1800 \MHz$ with thousands of frequency channels. GMIMS also requires sensitivity to all spatial scales above the resolution limit, which prefers the use of single-antenna telescopes. GMIMS low-band south \citep[$300-480 \MHz$ using the Murriyang Telescope at the Parkes Observatory;][]{wolleben:2019} and high-band north \citep[$1280-17 50 \MHz$ using the John A.\ Galt Telescope at DRAO;][]{wolleben:2021} data products are public. This work uses pre-release data from the low-band north survey using CHIME at DRAO. The CHIME instrument is described in detail by \citet{chime_overview_2022}. The CHIME data pipeline, including RFI excision, complex gain calibration, averaging over redundant baselines, and stacking over sidereal days, is described in \citet{chime-21-emission}. The CHIME/GMIMS low-band north polarization data set will be described in detail elsewhere (CHIME \& GMIMS Collaborations in prep, hereafter ``CHIME/GMIMS survey paper''). We present a brief overview of the data and processing steps here.

\subsection{CHIME/GMIMS Low Band North} \label{ssec:chime_lbn}

We use all-sky diffuse polarization data from CHIME for polarization observations of the tadpole feature and as the basis for Faraday synthesis of the region. CHIME consists of four stationary parallel cylindrical reflectors, measuring 1024 frequency channels across its range of $400 - 800 \MHz$. Oriented north-south, each $100 \times 20 \m^2$ cylinder has 256 dual-polarization linear feeds ($X$ and $Y$) spaced every $30 \cm$ along the central $78 \m$ of the focal line. CHIME observes the entire meridian at any given moment with baselines from $0.3 - 78  \m$, mapping the northern sky every sidereal day. This gives the telescope the ability to collect data at a range of angular scales, resulting in an effective angular resolution of approximately $30'$ at $400 \MHz$ \citep{masui:shaw:2019}. We exclude autocorrelations of feeds with themselves. As an interferometer, CHIME is missing the very largest scales corresponding to baselines $< 0.3 \m$ or angular scales $\gtrsim 50\arcdeg$. A future survey with the DRAO 15~m Telescope (Ordog et al. in prep) will provide the largest scales for the final GMIMS-low-band-north data product.

We generate full sky maps known as ``ringmaps'' by performing a one-dimensional Fourier transform on $\sim21$~s samples of the visibilities in the North-South direction along the meridian. As the sky passes through the primary beam, we sample the full range of Right Ascension (RA) values with these one-dimensional images each sidereal day, which we combine to produce the full map. We employ stacked ringmaps, using nighttime-only visibilities from 102 nights of data collected from January to November, 2019. The CHIME beam profile, which is declination-dependent and somewhat different in the $XX$ and $YY$ polarizations due to the cylindrical and stationary design of the telescope, is one of our largest uncertainties; for this reason, in this paper we report CHIME data in \Jybeam\ rather than brightness temperature units and confine ourselves to a relatively small declination range at low zenith angle ($\le 21\arcdeg$) where the beam is nearly constant. 

The ringmaps we use do not have beam deconvolution applied. There are small artifacts in the image resulting from this which we describe in Section~\ref{ssec:artifacts}, however, their presence is not detrimental to studying structures on the scale of several degrees, such as the tadpole. In this analysis, we use the $400 - 729 \MHz$ subset of the full CHIME band, as the highest frequencies are contaminated by aliasing, which makes the maps unreliable in the region of interest. 

\subsubsection{Polarization angle calibration}\label{ssec:polcal}

To calibrate CHIME polarization angles $\chi \equiv 0.5 \tan^{-1}(U/Q)$ (calculated using the full $\pm \pi$ range accounting for the signs of $U$ and $Q$), we rely only on CHIME co- and cross-polar data products and two key assumptions: that Stokes $V = 0$ (averaged over RA at every declination), and that the gain difference $\Delta G$ between the $X$ and $Y$ polarizations is small. Because of the declination-dependent beam properties of CHIME, we calibrate each declination in the raw maps using data within a $1\arcdeg$ strip, centered on that declination, and covering the full RA range. We focus on the narrow declination range around G137+7, $+60\arcdeg \lesssim \delta \lesssim 70\arcdeg$. This covers zenith angles $11\arcdeg$ to $21\arcdeg$, a region in which the CHIME beam model is best \citep{chime_overview_2022,chime-21-emission}. 

We represent the co-polar power from the orthogonal feeds as $X_C'X_C'$ (east-west feeds) and $Y_C'Y_C'$ (north-south feeds) and the (complex) cross-polar term as $X_C'Y_C'$. We use the stacked $X_C'X_C'$, $Y_C'Y_C'$, and $X_C'Y_C'$ ringmaps. Here $X_C$ and $Y_C$ refer to the CHIME coordinate system \citep{chime_overview_2022}, in which the naming of the variables $X$ and $Y$ is interchanged from the IAU convention \citep{iau:1973}. (The spherical CHIME cosmology $X_C$ and $Y_C$ coordinate system used here is also different from the Cartesian CHIME/FRB coordinate system; \citealt{2021ApJS..257...59C}.) The prime notation indicates detected values after passing through the full CHIME system, so the observed Stokes vector is
\begin{equation}
\mathcal{S}' \equiv \begin{pmatrix} I' \\ Q' \\ U' \\ V' \end{pmatrix} = \begin{pmatrix} 0.5(Y_C'Y_C' + X_C'X_C') \\ 0.5(Y_C'Y_C' - X_C'X_C') \\ \Re(Y_C'X_C') \\ -\Im(Y_C' X_C') \end{pmatrix}
 = M \mathcal{S}
\end{equation}
for M\"uller matrix $M$ and true sky Stokes vector $\mathcal{S}$. This follows the IAU convention with linear position angle increasing counter-clockwise when looking at the source \citep{iau:1973}. 

Using linear feeds to measure Stokes $Q$ requires the careful subtraction of two large numbers (the autocorrelations), whereas Stokes $U$ involves measuring the small cross-correlation. For calibration of $Q$ and $U$, we perform data-based estimates of the cross terms in $M$ following \citet{heiles_2001}. First, we assume that the difference in gains $G_X$ and $G_Y$ is small enough to approximate $\Delta G = (G_Y - G_X) / (G_Y+G_X) \ll 1$. We calculate the average $\Delta G$ at each declination and frequency from a linear regression of $Y_C'Y_C'$ as a function of $X_C'X_C'$, using data within a $1\arcdeg$ declination bin (overlapping bins centered on each declination) and covering the full 24-hour RA range. Over the declination range of the tadpole region ($45\arcdeg$ to $80\arcdeg$) and the frequencies used in this analysis, we find the mean and median of $\Delta G$ to be $-0.43$ and $-0.38$ respectively, with $73\%$ of declinations and unflagged frequencies having $|\Delta G| <0.5$. Although this does not strictly satisfy $\Delta G \ll 1$, it is sufficiently small so the defined $\Delta G$ term dominates over second-order corrections. The gain difference is dominated by differences in the beam solid angle between the two polarizations due to the asymmetric design of the CHIME telescope. We then apply this declination- and frequency-dependent $\Delta G$ to correct for leakage between Stokes $I$ and $Q$,
\begin{equation}
Q = \frac{Q' - I' \Delta G/2}{1-(\Delta G/2)^2}.
\end{equation}

Stokes $U$ and $V$ are measured from the cross-correlation products. We assume that $\langle V \rangle = 0$ from the sky in diffuse emission because synchrotron emission in low-density astrophysical environments does not produce circular polarization. Leakage between $V$ and $U$ arises from phase offsets. We measure a mean phase shift $\langle \psi \rangle(\delta, \nu)$ at each declination and frequency assuming that $\langle V \rangle = 0$ and calculate
\begin{equation}
U = U' \cos \langle \psi \rangle + V' \sin \langle \psi \rangle.
\end{equation}
The $\langle V \rangle=0$ assumption leads to high-quality fits even in fast radio burst (FRB) observations, where the assumption has less clear physical justification than in the diffuse polarized emission we investigate \citep{McKinven:2023}. We find that the phase shift is linear in frequency, consistent with a cable delay $\tau = \langle \psi\rangle / 2 \pi \nu \sim 1 \textrm{ ns}$ for the diffuse emission, as \citet[their Appendix~A]{McKinven:2021} found in CHIME/FRB data.

\begin{figure}
    \plotone{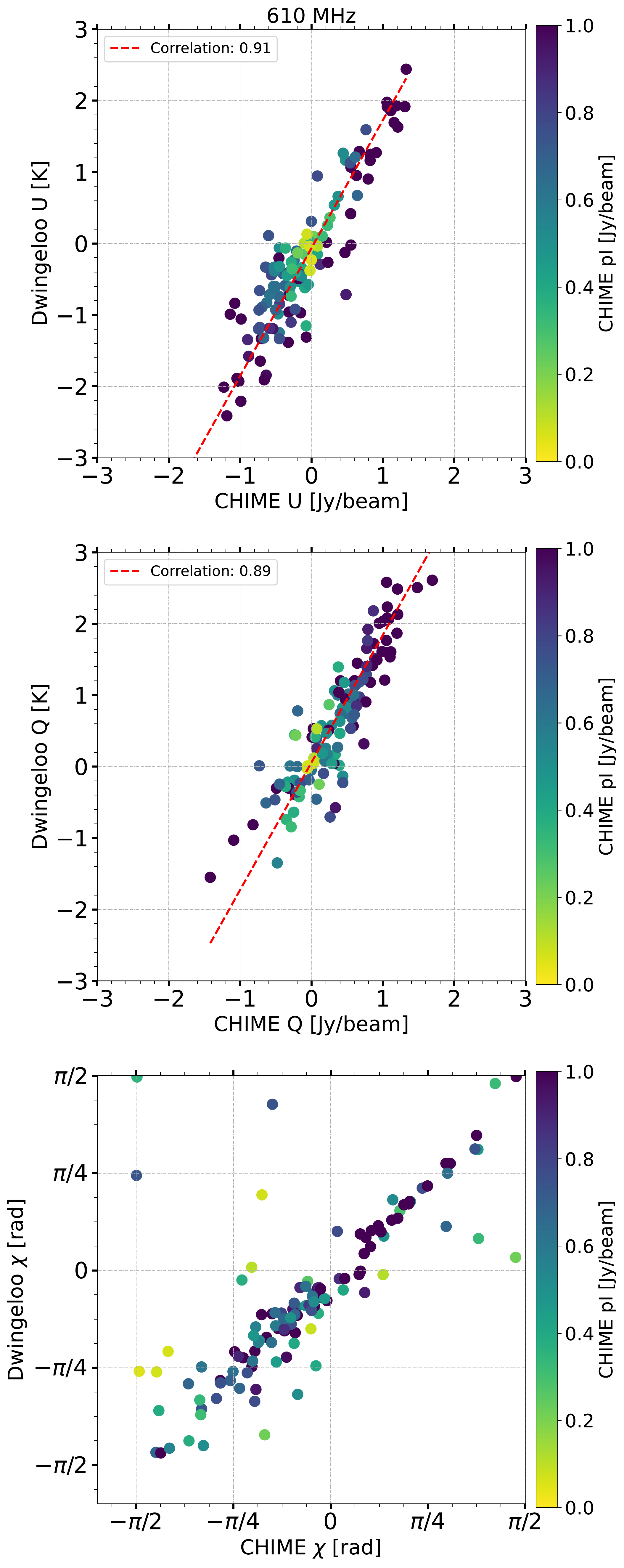}
    \caption{Comparisons between the Dwingeloo \citep{brouw_spoelstra_1976} absolutely calibrated $Q$ and $U$ observations in the Fan region ($120^{\circ} \lesssim \ell \lesssim 180^{\circ}$, $-2^{\circ} \lesssim b \lesssim 30^{\circ}$)  at $610 \MHz$ and CHIME $614 \MHz$. All data points show a Dwingeloo pointing and the median of corresponding CHIME data within the Dwingeloo beam. The color bar is the CHIME polarized intensity for these points. For Stokes $Q$ and $U$, the Dwingeloo data are shown in brightness temperature units and CHIME data in \Jybeam.}
    \label{fig:fig1}
\end{figure}

In Figure \ref{fig:fig1}, we compare the calibrated data to the Dwingeloo telescope survey at $610 \MHz$ in the Fan region \citep{brouw_spoelstra_1976}. There is a strong correlation between Dwingeloo $U$ and CHIME $U$ and Dwingeloo $Q$ and CHIME $Q$ in those directions for which there is Dwingeloo data, with correlation coefficient $R$ values of 0.91 for $U-U$ and 0.89 for $Q-Q$ comparisons. This is a significant improvement from the uncalibrated correlation coefficients of 0.76 and 0.59 respectively. We find a remaining leakage of up to 20\% in Stokes $Q$ based on unresolved point source measurements. Using the mean orthogonal distance between each point and the fitted line, we find that noise from CHIME and Dwingeloo data describe $\approx 70\%$ of the scatter in Figure \ref{fig:fig1}. The polarization angle correlation, also shown in Figure~\ref{fig:fig1}, is also improved through calibration, and most outliers are points with low polarized intensity (yellow dots), where the uncertainty in derived $\chi$ is high.

\begin{figure*}[tb]
\plotone{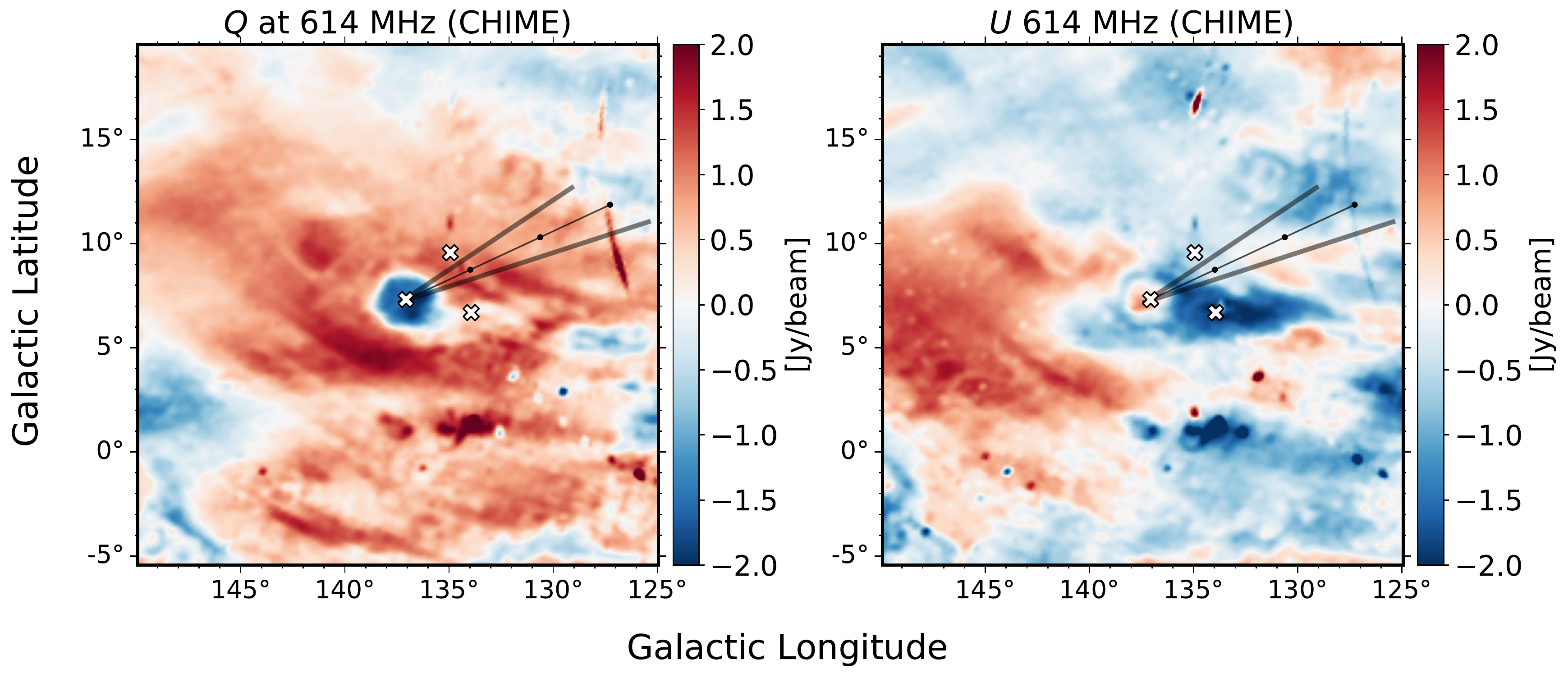}
\caption{Images of the tadpole region in Stokes $Q$ and $U$ at $614 \MHz$ in Galactic coordinates. The `$\times$' markers indicate the position of B2(e) star HD 20336 (the $\times$ near the center of the circular tadpole head) as well as the selected spectra shown in Figure \ref{fig:fig7}.  The thin black line represents the Local Standard of Rest (LSR)-corrected proper motion of HD 20336, projected backwards in time over $3 \Myr$, with each dot representing $1 \Myr$. The translucent lines represent the error cone, which is dominated by the uncertainty in the LSR correction.}
\label{fig:fig2}
\end{figure*}

We show the resulting CHIME $Q$ and $U$ maps, with the $\chi=0$ reference axis rotated to the north Galactic pole, in Figure~\ref{fig:fig2}. While Stokes $I$ to $Q$ leakage does exist in our data, the tadpole structure cannot simply be the result of leakage. Although there is total intensity emission over the entire Fan Region, including the tadpole, this emission is featureless on small scales and thus cannot produce spurious polarization matching the tadpole in morphology. Furthermore, the tadpole cannot be the product of Stokes $I$ emission originating at large angular distances (such as the Galactic plane) and seen in far sidelobes. While the far sidelobes have poor polarization properties, their polarization averages to low values over sizable areas. Moreover, with linear feeds, leakage from $I$ is primarily into $Q$, not $U$ (in the native equatorial coordinates of CHIME), but the tadpole is already evident in Stokes $U$ in equatorial coordinates (not shown).

The slopes of $U-U$ and $Q-Q$ inform the calculation of the beam solid angle in converting \Jybeam\ to brightness temperature units. From these slopes, we deduce that 1 \Jybeam\ corresponds to $1.79 \K$. The effective area of CHIME is therefore 4900 square metres at 610 MHz, confirming that, in our application, CHIME acts like a large single-antenna telescope. This paper focuses on Faraday rotation effects, meaning that the primary data product is polarization angle and its variation with frequency. Absolute calibration of the amplitude scale is thus of minor significance for the present work, and we report intensities in \Jybeam\ units, leaving a more careful conversion to temperature units for the CHIME/GMIMS survey paper. 

\begin{figure}[tb]
\plotone{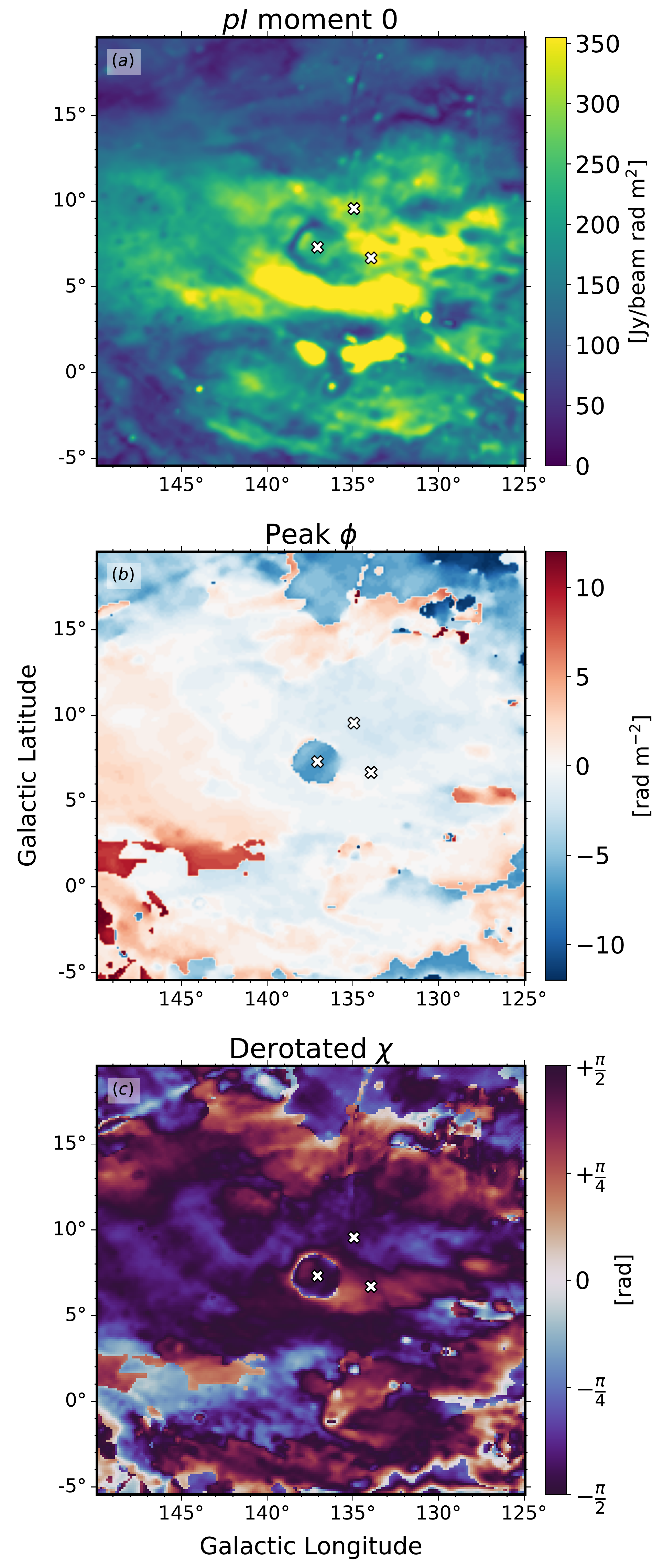}
\caption{Maps of products of Faraday synthesis generated from the $400 - 729 \MHz$ channels of CHIME. $(a)$ Moment zero integrated polarized intensity. $(b)$ Faraday depth of highest peaks of the Faraday depth spectra. $(c)$ the derotated polarization angle (i.e. angle at point of emission). The `$\times$' markers are as in Figure \ref{fig:fig2}.}
\label{fig:fig3}
\end{figure}    

The CHIME observations were made in 2019, at solar minimum. Only night-time data were used. The observations were not corrected for Faraday rotation in the ionosphere, but we expect the effect is smaller than $1 \radmsq$, with an rms value of $0.3 \radmsq$. The result is a small decrease in polarized intensity since we averaged together observations from different nights.

\subsubsection{Faraday synthesis on CHIME data}\label{ssec:RMsynths}

\begin{figure}
\plotone{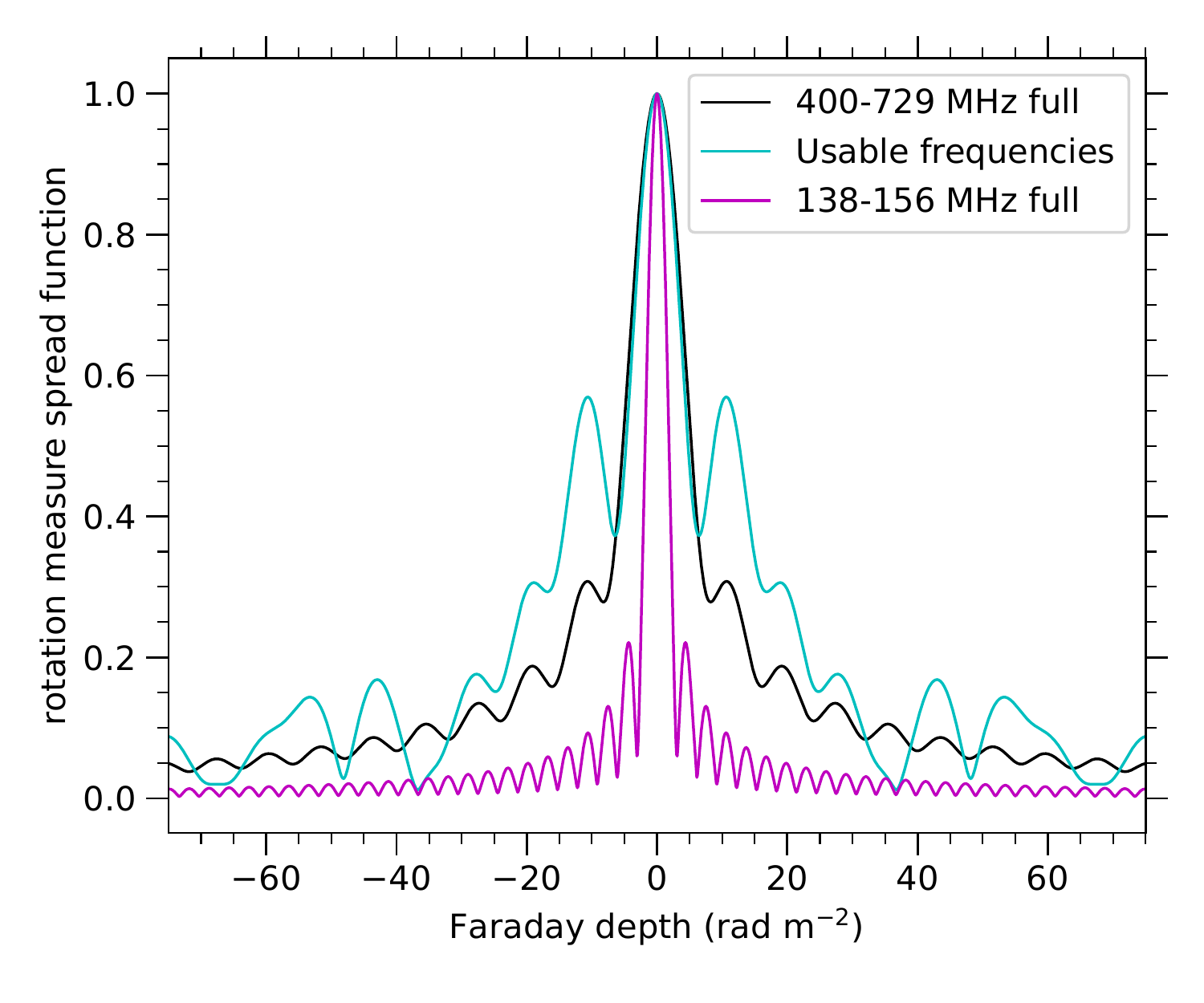}
\plotone{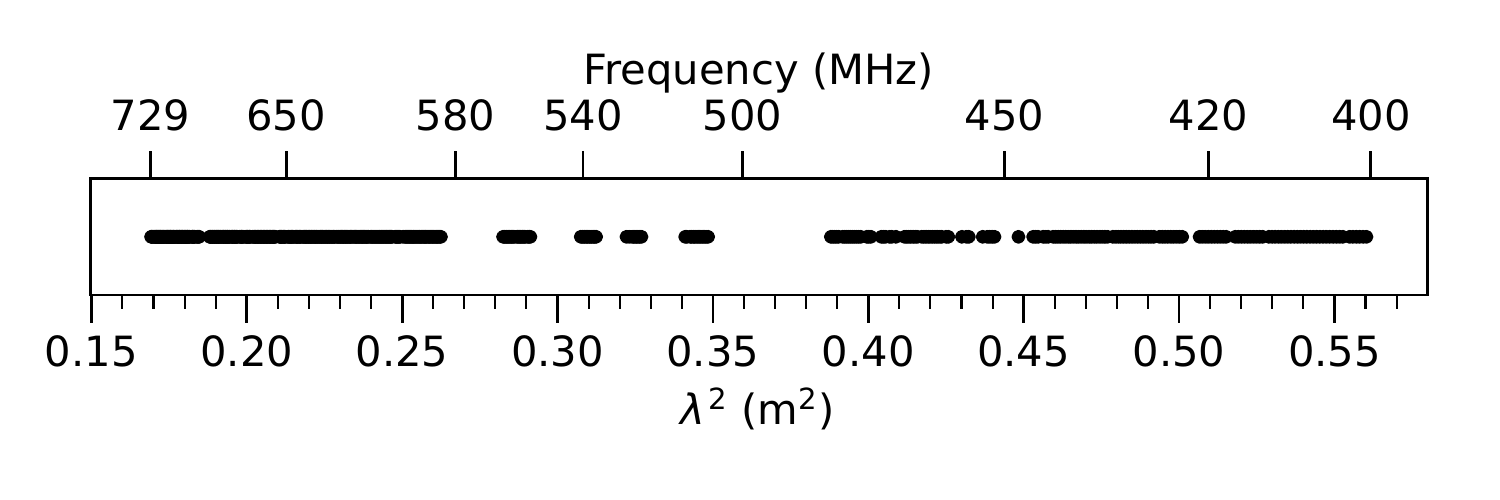}
\caption{{\em Top:} Magnitude of the RMSF for the full CHIME frequency range (black lines), the usable CHIME frequency channels (cyan line), and the full WSRT $150 \MHz$ frequency range (magenta line). See Table~\ref{tab:studies}. {\em Bottom:} Usable CHIME frequencies, $W(\lambda^2)$ (bottom axis) or $W(\nu)$ (top axis). The WSRT RMSF shown here assumes full frequency coverage from $138-156 \MHz$ because we have no record of the missing frequencies.} \label{fig:fig4}
\end{figure}

We apply Faraday synthesis to the Stokes $Q$ and $U$ cubes from the calibrated CHIME polarization data in Galactic coordinates using \texttt{RMTools_3D} in the \texttt{RM-Tools} software package \citep{purcell:2020}. We show images of the main products of Faraday synthesis in Figure~\ref{fig:fig3}, and the magnitude of the RMSF along with frequency and $\lambda^2$ coverage in Figure~\ref{fig:fig4}. For each pixel in the maps, we calculate the dirty Faraday spectrum out to $\pm 200 \radmsq$ sampled every $0.5 \radmsq$. This extent in $\phi$ is sufficient to include the major structures without being dominated by sidelobes. We do not convolve the CHIME data to a common resolution prior to Faraday synthesis as this may be unreliable due to uncertainties in the synthesized beam shape. With the resolution only varying between $\sim 30'$ (at 400 MHz) and $\sim 17'$ (at 729 MHz), structures on the scale of the tadpole (several degrees) are largely unaffected.

The $400-729 \MHz$ frequency coverage yields $\delta \phi = 9.7 \radmsq$ and $\phi_{\mathrm{max-scale}} = 19 \radmsq$. With $\delta \phi < \phi_{\mathrm{max-scale}}$, we expect CHIME data to resolve somewhat extended Faraday depth structures. For all lines of sight in the tadpole region (Section~\ref{ssec:FD_of_tadpole}), we apply the \texttt{RM-Tools} implementation of the RM CLEAN algorithm \citep{Heald:2009} to reduce the sidelobes in the spectra, using a CLEAN threshold of $0.2 \Jybeam$ based on an estimate of the noise in the dirty spectra. We present CLEANed spectra in Section~\ref{ssec:FD_of_tadpole} below.

Using the \texttt{rmtools_peakfitcube} algorithm in \texttt{RM-Tools}, we obtain the peak Faraday depth and its associated error for every spectrum along all lines of sight. The resulting map is shown in Figure~\ref{fig:fig3}$b$. We use peak Faraday depths rather than a first moment \citep{dickey:2019} to focus on the Faraday depth of the brightest feature in each LOS rather than a weighted mean Faraday depth in Faraday complex regions. 

We show the integrated polarized intensity across the Faraday depth spectra as a zero moment map in Figure~\ref{fig:fig3}$a$. A polarization angle map derotated to $\chi_0$ by the peak Faraday depth at each pixel is shown in Figure~\ref{fig:fig3}$c$.

\subsection{DRAO Synthesis Telescope Observations} \label{ssec:st}

We use new and previously published polarized continuum and $\HI$ observations from the DRAO ST \citep{Landecker:2000}. The ST continuum observations combine four $7.5 \MHz$-wide frequency channels within a $35 \MHz$ bandpass centered on $1420.4 \MHz$ to measure Stokes $Q$ and $U$ using dual circular polarization feeds. The ST $\HI$ observations are produced by a 256-channel spectrometer with a velocity range of $211 \kms$ and $1.32 \kms$ spectral resolution. Existing observations come from the Canadian Galactic plane survey \citep[CGPS;][]{Taylor:2003,Landecker:2010}, covering Galactic latitudes $-3\arcdeg < b < +5\arcdeg$. We use a variety of ST fields observed and calibrated in the same manner as CGPS fields for a number of projects, some previously published \citep{W3_west_landecker,2014ApJ...784L..26K}, and some unpublished, covering $+5\arcdeg < b < +12\arcdeg$.

In all ST polarized continuum observations, single-channel $1410 \MHz$ polarization observations from the Effelsberg Medium-Latitude Survey \citep[EMLS;][Reich et al. in prep.]{Uyaniker:1998,Uyaniker:1999,2004mim..proc...45R} provide single-antenna information. The EMLS data are based on Effelsberg 100~m Telescope observations from latitude $|b| < 20\arcdeg$, observed mostly in blocks $7\arcdeg \times 7\arcdeg$ in size but with many exceptions to avoid strong source complexes at the map boundaries. Emission exceeding the map size is missing and was restored by combining with the northern-sky Galt Telescope survey \citep{Wolleben:2006}. We then combine this single-antenna data with the ST $1420 \MHz$ data as follows. First, we mosaicked the ST fields together following \citet{Taylor:2003}. Then we convolve the ST data to $10.4'$, slightly larger than the $9.35'$ EMLS resolution. We then subtract the convolved ST data (Stokes $Q$ and $U$ separately) from the EMLS data; the residual is the large-scale structure missed in the ST data  \citep{1990A&AS...85..633R}. Finally, we add the residual to the original ST data to produce $Q$ and $U$ images including scales from $1'$ to the full sky. This technique is somewhat different from that used for the CGPS, where ST and EMLS data were combined in Fourier space \citep{Landecker:2010}, but produces very similar results. We refer to this combined data set of interferometric observations from the ST with single-antenna data from the Effelsberg and Galt telescopes as `ST+EMLS', and show the resulting polarized intensity and polarization angles maps for the tadpole region in Figure~\ref{fig:fig5}$ij$.

In our ST atomic hydrogen observations, we incorporate the HI4PI Survey of $\HI$ 21 cm brightness temperature, with an angular resolution of $16.2'$, velocity coverage of $\le 600 \kms$, and spectral resolution of $1.49 \kms$ \citep{ben_bekhti_et_al._2016}, to provide the majority of the single-antenna information. The fields we incorporated that were previously processed for the CGPS used the DRAO 26-m data to provide short-spacing information on account of HI4PI being unavailable at the time. We process a wider field of view for the $\HI$ observations than for the polarized continuum observations --- covering Galactic latitudes $-4\arcdeg < b < +12\arcdeg$ and longitudes $126\arcdeg < l < 149\arcdeg$ --- to highlight typical fluctuations seen across the diffuse neutral medium (see \S \ref{ssec:H_I}). For each ST field in this region we determine continuum emission by averaging the $\HI$ channels void of 21 cm emission, then subtracting this to isolate only $\HI$ emission. We calibrate, merge short-spacing and ST data, and mosaic the $\HI$ fields by the same proceedure used for the CGPS  \citet{Taylor:2003}. This results in the $\HI$ brightness temperature map, which we refer to as `ST+HI4PI', shown in Figure \ref{fig:fig10} at a velocity channel about $11.4 \kms$.

\subsection{Ancillary Data Sources}\label{ssec:ancillary}

\begin{table*}[tb]
\centering
\caption{Observing parameters}
\label{tab:studies}
\begin{tabular}{lcccc}
\hline
\hline
                        & \parbox[c]{1.5in}{\center\citet{bernardi_2009}, \citet{iacobelli_haverkorn_katgert_2012}} & \parbox[c]{1in}{\center\citet{haverkorn_2003}}	& 
                        this paper 		& \parbox[c]{1.5in}{\center this paper,  \\ \citet{Landecker:2010}} \\
                        \hline
Instrument              & WSRT                      			& WSRT	&	
CHIME  			& ST+EMLS \\
observed field        	& $\sim 12\arcdeg \times 12\arcdeg$ 	& $\sim 7\arcdeg \times 7\arcdeg$	& all-sky  			& \parbox[c]{1.5in}{\center Galactic plane plus $20\arcdeg \times 7\arcdeg$ extension} \\
frequencies			& $138-156 \MHz$							& $341-375 \MHz$	& $400-729 \MHz$   	& 1.4 GHz        \\
channels	& $2048 \times 9.8 \kHz$					& $5 \times 5 \MHz$	& $844 \times 390 \kHz$	& $1 \times 35 \MHz$ \\
$\lambda^2$ coverage	& $3.7-4.7 \m^2$	& $0.64-0.77 \m^2$	& $0.17-0.56 \m^2$	& $0.04 \m^2$ \\
$\delta \phi$ (eq~\ref{eq:res})	&	$3.8 \radmsq$	& $29 \radmsq$ & $9.7 \radmsq$ & n/a \\
$\phi_{\mathrm{max-scale}}$ (eq~\ref{eq:Wbrentjens}) & $0.8 \radmsq$ & n/a			& $19 \radmsq$ & n/a \\
angular resolution   			& $2' \times 2.2'$     					& $5.0' \times 5.5' $ 	& $\approx 17'$ to $30'$  		& $\sim 1'$ \\
baselines	& $36-2760 \m$	&  $36-2760 \m$	& 	$0.3-80 \m$	& $0-614 \m$ \\
\hline       
\end{tabular}
\tablecomments{\citet{haverkorn_2003} fit $\textrm{RM} = d\chi/d(\lambda^2)$ without Faraday synthesis and thus have no ability to separate multiple Faraday depth components.}
\end{table*}   

\begin{figure*}
\plotone{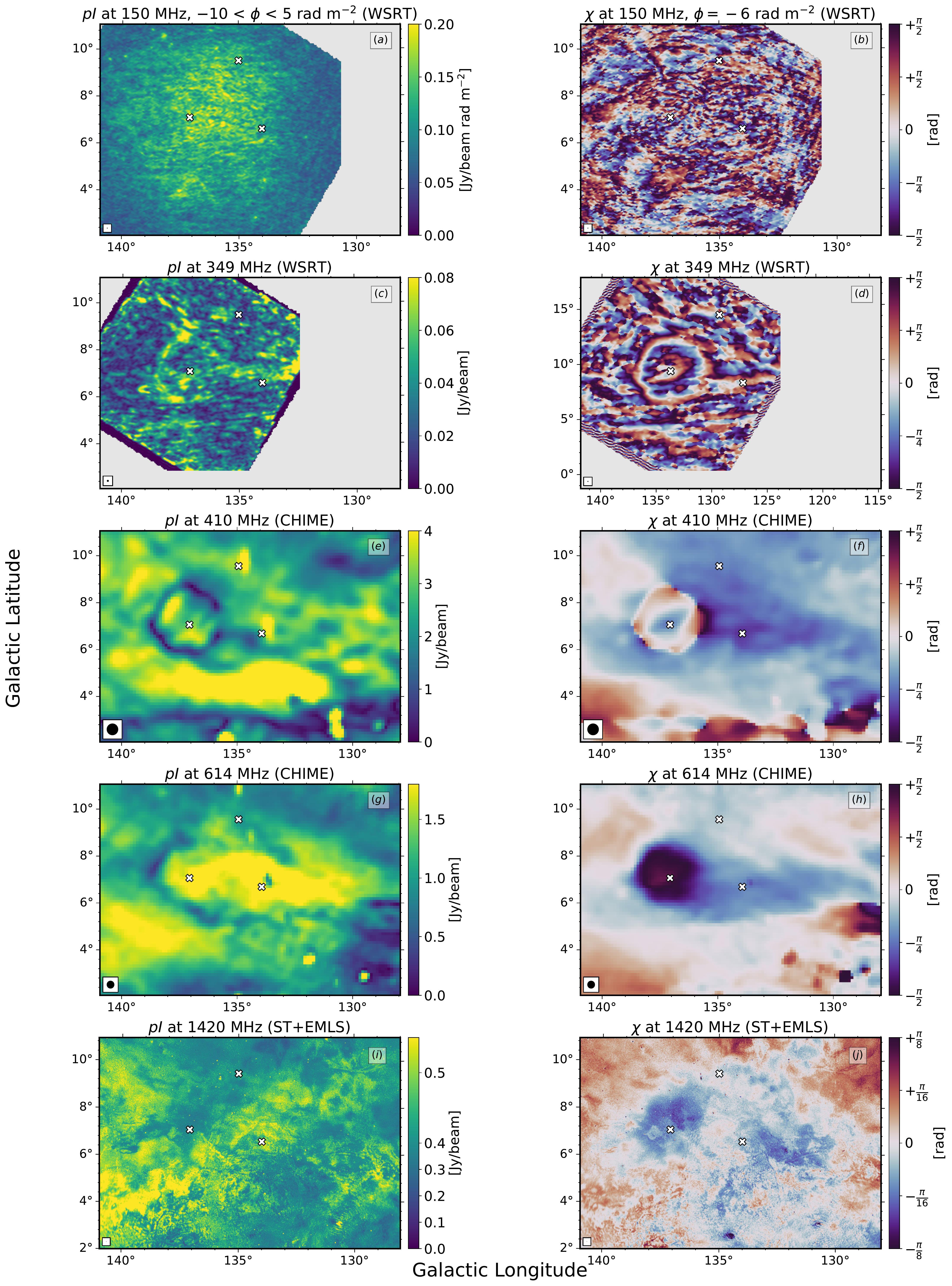}
\caption{The tadpole depicted in polarized intensity (left) and polarization angle (right). From top to bottom: $(a, b)$ \citet{bernardi_2009} Faraday synthesis data at frequencies $\sim 150 \MHz$ with the Westerbork Synthesis Radio Telescope at $\phi = -2 \radmsq$ (a frequency data cube is unavailable), $(c, d)$ \citet{haverkorn_2003} single-frequency polarized intensity and polarization angle at $349 \MHz$, CHIME data at $(e, f)$ $410 \MHz$ and $(g, h)$ $614 \MHz$, and $(i, j)$ $1.4 \GHz$ DRAO ST+EMLS data. `$\times$' markers are as in Figure \ref{fig:fig2}. Beams are shown within each image. Beams for the WSRT and ST+EMLS data are too small to be easily visible (see Table \ref{tab:studies}).} 
\label{fig:fig5}
\end{figure*}

We use data from the Westerbork Synthesis Radio Telescope (WSRT) at $150 \MHz$ \citep[provided by M.\ Iacobelli, 2023, private communication]{bernardi_2009,iacobelli_haverkorn_katgert_2012} and $350 \MHz$ \citep{haverkorn_2003} to supplement discussions of observed depolarization. We summarize the observing parameters for all four polarization data sets used in this paper in Table~\ref{tab:studies}. The five channels in the $350 \MHz$ data are not sufficient for Faraday synthesis, measuring only the rotation measure $\textrm{RM} = d\chi / d \lambda^2$. For the $150 \MHz$ data, only post-Faraday synthesis data are available, so we cannot show single-frequency images or perform QU fitting on these data. We show the WSRT polarized intensity and polarization angle maps in Figure~\ref{fig:fig5}$a$--$d$.

 We use the Wisconsin H-Alpha Mapper (WHAM) survey of H$\alpha$ emission to study ionized hydrogen in the region \citep{wham_nss,Haffner2010}. WHAM has an angular resolution of 1\textsuperscript{o} and provides a kinematically resolved map of H$\alpha$ emission within $\approx 100\kms$ of the local standard of rest with $12 \kms$ spectral resolution.

\section{Features of the Tadpole}\label{sec:features}

\subsection{Morphology in single-frequency images}

We show images of the tadpole region in $Q$ and $U$ at $614 \MHz$ in Figure~\ref{fig:fig2}, image products derived from Faraday synthesis with CHIME in Figure~\ref{fig:fig3}, and $pI$ and $\chi$ from the polarization data sets described above, covering $150-1420 \MHz$, in Figure~\ref{fig:fig5}. The tadpole is immediately apparent in the single-channel CHIME $Q$ and $U$ images in Figure~\ref{fig:fig2} with a circular feature we call the ``head'' near $(\ell, b) = (137\arcdeg, +7\arcdeg)$ and a ``tail'' extending to the right as far as $(127\arcdeg, +6\arcdeg)$, most clearly in $U$. The structure as a whole strongly resembles the larval stage of amphibians, leading us to nickname it the \emph{tadpole}. We use the name `G137+7' to refer to the circular region first identified by \citet{verschuur_1968}, and the name `tadpole' to describe the entire feature, including the tail. The head (G137+7) is the feature which has been studied since \citet{verschuur_1968}; it is visible in all channels in Figure~\ref{fig:fig5}.

At $150 \MHz$ (Fig.~\ref{fig:fig5}$ab$), the head is evident as a large, diffuse structure in $pI$. There is a circular pattern to the polarization angles, suggesting rapid wrapping through $\pi$ radians as one moves outward radially from the center of the head.

At $350 \MHz$ (Fig.~\ref{fig:fig5}$cd$), the head appears as a ring in $pI$ (Fig.~\ref{fig:fig5}$c$). \citet{haverkorn_2003} measured $\RM = -8 \radmsq$ in the center of the head. At this frequency, the head is also a ring in $\chi$, with approximately two full rotations through $\pi$ radians in the radial direction. This is consistent with the RM changing from $-8 \radmsq$ in the center of the head to $0 \radmsq$ outside the head, which would correspond to an angle change $\Delta \chi = \phi \lambda^2 = 6$ radians. \citet{haverkorn_2003} point out an elongated structure of high $pI$ extending northwest which they say does not necessarily have the same origin as the ring; this is evident in both $pI$ (Fig.~\ref{fig:fig5}$c$) and $\chi$ (Fig.~\ref{fig:fig5}$d$). This appears to be the tail. 

At $410 \MHz$ in the CHIME $pI$ data, the clearest signature of the head is a narrow ring of {\em low} polarized intensity which extends in a nearly-complete circle. At $614 \MHz$, there is a similar ring of low polarized intensity, but only in a semicircle on the left (east) side of the head. This feature is one beam wide, a clear signature of beam depolarization, with the polarization angle changing within the beam such that there is destructive interference, reducing the polarized intensity \citep{sokoloff:1998,Gaensler:Dickey:2001}. The same feature is evident in $pI$ from the Faraday synthesis products in Figure~\ref{fig:fig3}, but it does not stand out as much: by using information at a wide range of $\lambda^2$, the Faraday synthesis product is less sensitive to beam depolarization than single-frequency images. Furthermore, if the depolarized ring arises from beam depolarization, we would not expect to see it in the ST+EMLS data due to the signficantly smaller beam ($1'$ compared to $30'$). This is in fact the case in Figure~\ref{fig:fig5}$i$: none of the head, the depolarized ring, or the tail is evident in $pI$.

In polarization angle, the $410 \MHz$ CHIME data show a clear wrap through $\pi$ radians moving radially from the center of the head to outside the ring (Fig.~\ref{fig:fig5}$f$). The tail of the tadpole stands out clearly, especially as a polarization angle feature in Figure~\ref{fig:fig3}$c$ and Figure~\ref{fig:fig5}$fh$. The tadpole, both head and tail, is also visible at $1420 \MHz$ in the ST+EMLS polarization angle image (Fig.~\ref{fig:fig5}$j$), despite not being evident in $pI$. The large-scale structure at $1420 \MHz$ is similar to that observed in CHIME $614 \MHz$ polarization angle (Fig.~\ref{fig:fig5}$h$), where the $\chi$ values agree in sign with that of ST+EMLS. The values of $|\chi|$ are smaller at $1420 \MHz$ than at $614 \MHz$, which is consistent with the expected reduction in Faraday rotation at higher frequencies. We note some smaller structures on sub-tadpole scales in ST+EMLS data not present in CHIME which may arise from probing larger physical depths at a higher frequency and with a much smaller beam, combining to yield a more-distant polarization horizon \citep{Uyaniker:Landecker:2003}. In contrast, although the WSRT data have an angular resolution on the order of magnitude of the ST+EMLS data, the larger $\lambda^2$ means we expect the polarization horizon to be closer, possibly probing physical depths more similar to CHIME than to the ST+EMLS. 

\subsection{Faraday depths} \label{ssec:FD_of_tadpole}

\begin{figure*}[tb]
    \plotone{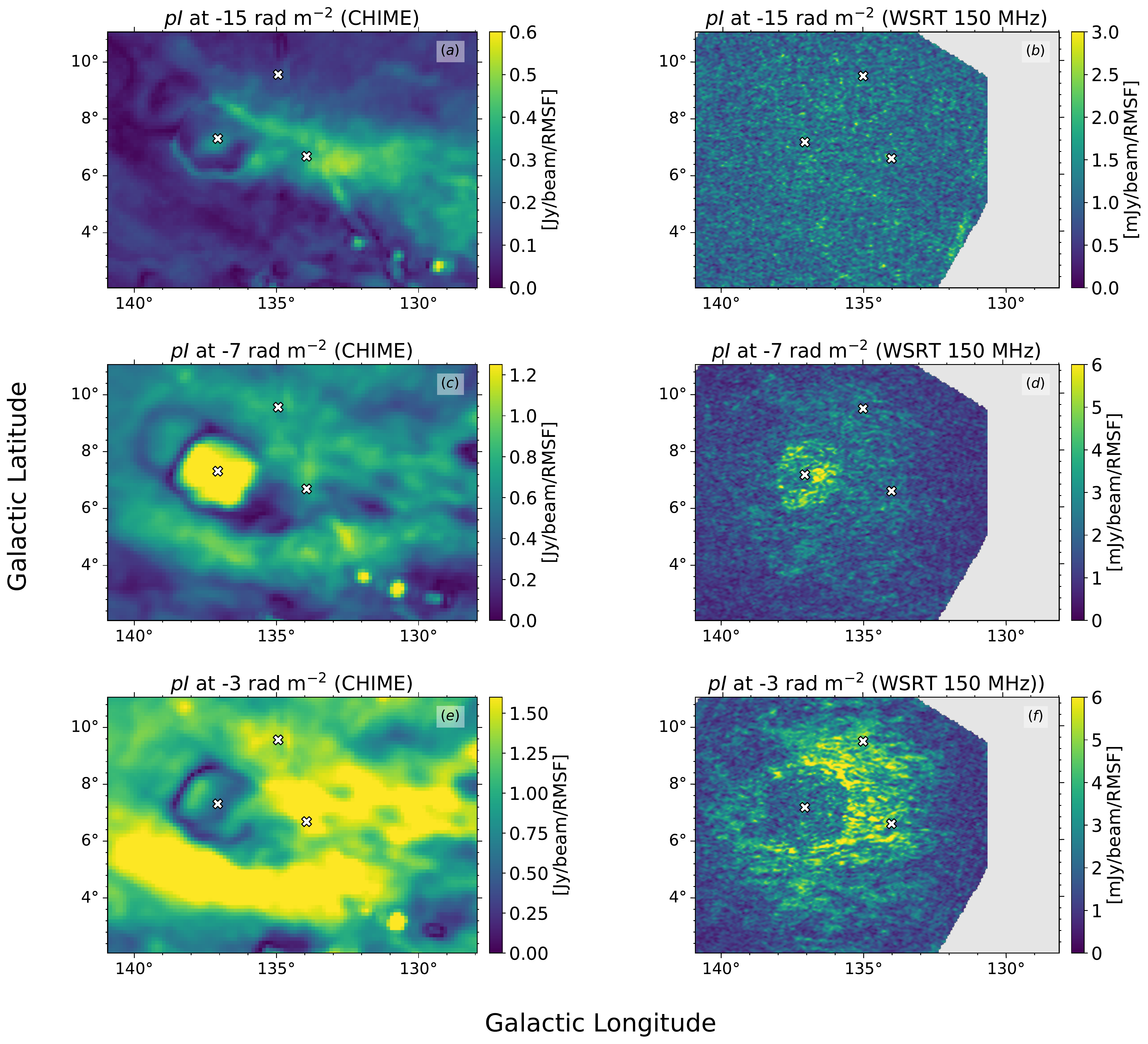}
    \caption{Images of polarized intensity at $\phi = -15~\radmsq$ $(a, b)$, $-7~\radmsq$ $(c, d)$, and $-3~\radmsq$ $(e, f)$ from CHIME $400-729 \MHz$ data (left) and WSRT $150 \MHz$ data (right). Note that the fainter loop above and to the left of the head, visible in all three CHIME channels but most clearly at $-7 \radmsq$, is a grating lobe artifact; see Section~\ref{ssec:artifacts}.}
    \label{fig:fig6}
\end{figure*}

While the tadpole is prominent in polarization angle images, it does not stand out from the background in total intensity in single-channel images. Previous studies (\citealt{haverkorn_2003, bernardi_2009, iacobelli_haverkorn_katgert_2012}) found the head to show strong negative Faraday rotation peaking at $\phi \approx -8 \radmsq$. The data sets used in these studies and their observing parameters are listed in Table~\ref{tab:studies}. 

We show $pI$ images at $\phi = -15$, $-7$, and $-3 \radmsq$ using CHIME and WSRT $150 \MHz$ data in Figure~\ref{fig:fig6}. In both data sets, the head stands out as a strong feature in $pI$ at $-7 \radmsq$ with little surrounding emission. Furthermore, in both data sets the head appears to be a depolarization feature in $pI$ at $-3 \radmsq$, but with differing morphology; in WSRT the entire region of the head shows little or no polarized intensity, while there is a ring of polarized intensity outside the head. This suggests {\em pure Faraday rotation}: the polarized intensity is moved from $\phi = -3 \radmsq$ to $-7 \radmsq$. At CHIME frequencies, the $-3 \radmsq$ image shows a narrow depolarized ring immediately outside the head, but it lacks the clearly defined region of bright emission we find in WSRT. At $\phi = -15 \radmsq$ a structure appears in the CHIME data which is absent in the WSRT data. The tail of the tadpole stands out from its surroundings at this Faraday depth, with the bright emission extending partway around the head, tracing the depolarized ring that appears at the other two Faraday depths shown. The tail also appears as reduced $pI$ at $\phi = -7 \radmsq$ in CHIME data.

\begin{figure*}[tb]
    \plotone{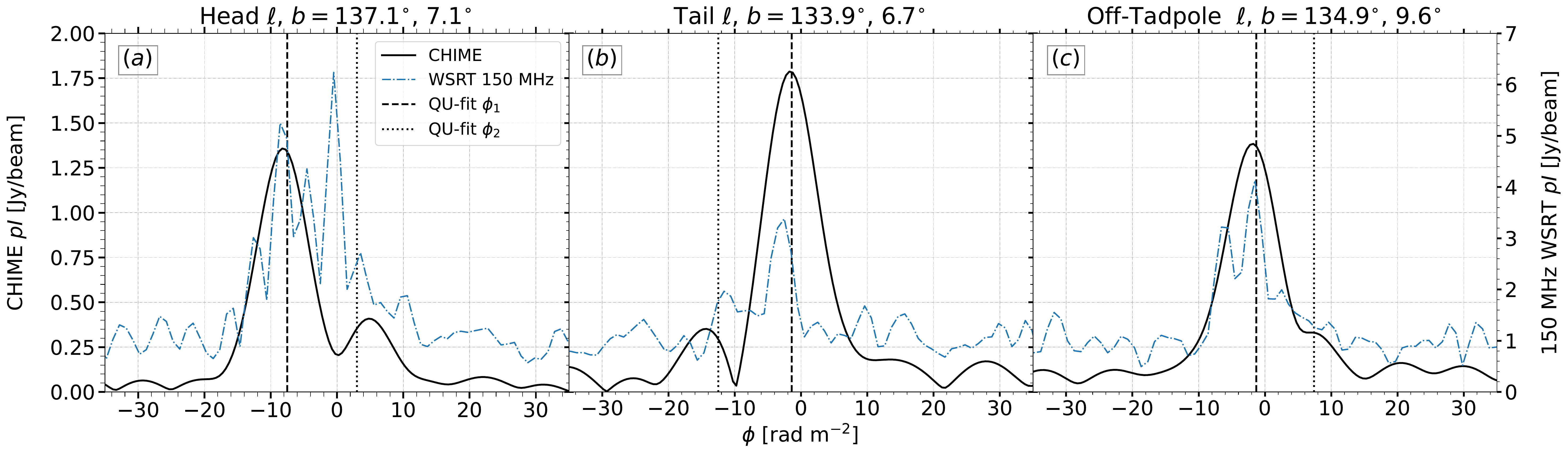}
    \caption{Faraday spectra (magnitudes) from CHIME $400-729 \MHz$ (black solid lines) and WSRT $150 \MHz$ (blue dot-dashed lines) for lines of sight on the tadpole head, tail, and in the surrounding region. These lines of sight correspond to the markers shown in Figure \ref{fig:fig2} and elsewhere. Dashed and dotted vertical lines show the peaks $\phi_1$ and $\phi_2$ from QU fitting (see Section~\ref{ssec:qufitting}). The intensity scale on the left applies to CHIME data; the intensity scale on the right applies to WSRT data.}
    \label{fig:fig7}
\end{figure*}

We show Faraday spectra at three positions---one in the head, one in the tail, and a background position---in Figure~\ref{fig:fig7} for CHIME and $150 \MHz$ WSRT data. In the head, we find multiple peaks in both data sets. In the $150 \MHz$ data, there are peaks at $\approx -8$, $-4$, and $0 \radmsq$, with polarized intensities of $\sim 5$, $4$, and $6 \Jybeam$ respectively. In CHIME data there is also a Faraday depth peak at $\phi \approx -8 \radmsq$ with $pI \approx 1.3 \Jybeam$, while a secondary peak is at $+5 \radmsq$ with $pI \approx 0.4 \Jybeam$. There is no evidence of a peak near $-4 \radmsq$. However, with $\delta \phi = 9.7 \radmsq$ in the CHIME data, we would not expect to resolve two peaks separated by $\approx 4 \radmsq$. In the tail, we see two peaks in both the CHIME and $150 \MHz$ data at $\phi \approx -2$ and $-14 \radmsq$ ($pI \approx 1.75 $ and $ 0.35 \Jybeam)$, and $\phi \approx -4 $ and $ -12 \radmsq$ ($pI \approx 3.4$ and $ 1.8 \Jybeam)$ respectively. The CHIME peaks in both the head and tail are separated by close to $\delta \phi$, suggesting that the separation of the peaks may be an artifact of the RMSF and may not be physically meaningful, as we discuss in Section \ref{sec:complexity} below. Off-tadpole, we find only one peak in the CHIME data at $\phi \approx -2 \radmsq$ with $pI \approx 1.35 \Jybeam$. There is a bump in the spectrum seen at $\sim 10 \radmsq$, however, this bump does not coincide with any notable peak in the 150 $\MHz$ data. Rather, we find two peaks at $\phi \approx -2$ and $-8 \radmsq$ with polarized intensities $\approx 4$ and $ 3 \Jybeam$, respectively.

\begin{figure*}[tb]
\plotone{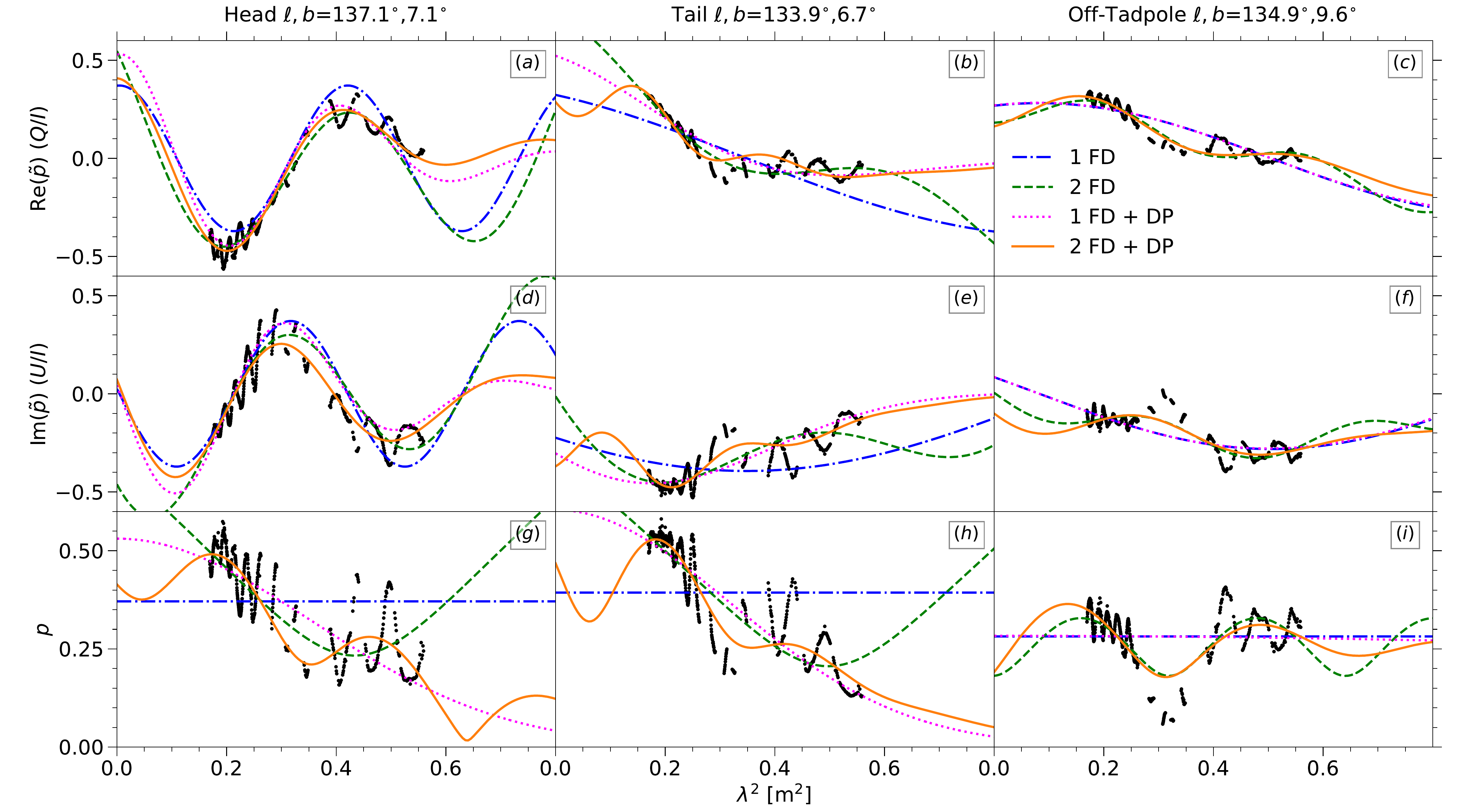}
\caption{Best-fit models from QU fitting for the lines of sight shown in Figure~\ref{fig:fig7}. The panels show $Q/I$ ($a-c$), $U/I$ ($d-f$) and the fractional polarized intensity, $p$ ($g-i$). Black points represent the data. The blue dot-dashed line is the 1-component model (1 FD), the green dashed line is the two-component model (2 FD), the magenta dotted line is the one-component model with beam depolarization (1 FD+DP), and the orange solid line is the two-component model with beam depolarization (2 FD+DP). The fast ripples in the data (an instrumental effect) are not fitted by the models.} \label{fig:fig8}
\end{figure*}

\begin{figure}
\plotone{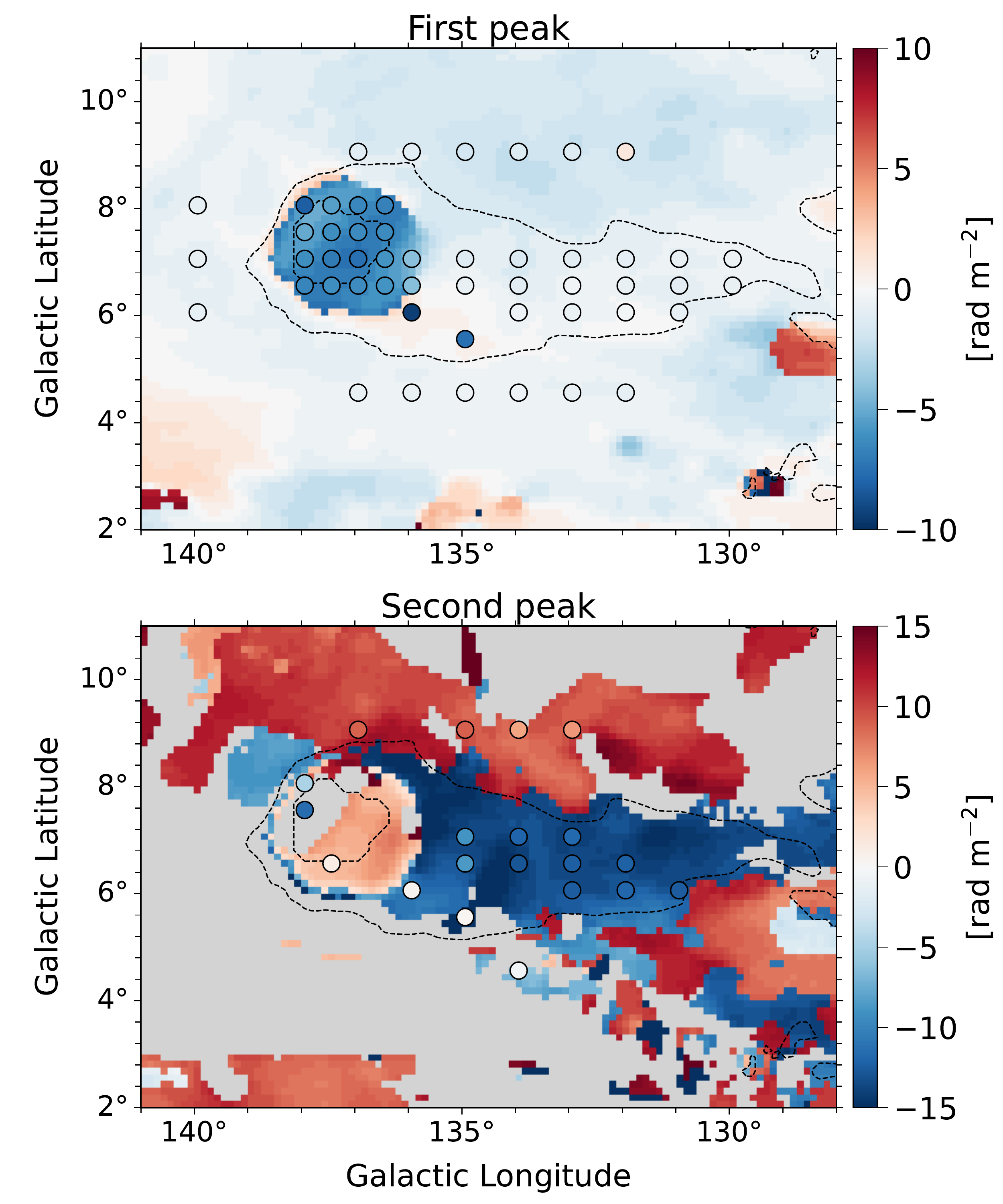}
\caption{Results of QU fitting along representative lines of sight in the tadpole region, compared to Faraday synthesis results. The background images show the Faraday depth of the brightest peak in the Faraday spectrum (top panel) and the second-brightest peak (bottom panel) at each pixel. The circles show the Faraday depths derived from QU fitting ($\phi_1$ in the top panel, $\phi_2$ in the bottom panel) in 51 directions in and around the tadpole. A mask is applied in the bottom panel to exlude peaks in the spectra below the CLEAN threshold ($0.2 \Jybeam$; greyed out regions) and QU-fitted components below $p=0.1$ (missing points).}
    \label{fig:fig9}
\end{figure}

\begin{table*}[tb]
    \centering
    \caption{Results of QU fitting for representative lines of sight}
    \label{tab:qufit}
    \begin{tabular}{llcccccccc}
        \hline
        Location & $\ell$, $b$ & $p_1$ & $\phi_1$ & $\chi_{0,1}$ & $\sigma_1$ & $p_2$ & $\phi_2$ & $\chi_{0,2}$ & $\sigma_2$ \\
        & & & (rad m$^{-2}$) & &(rad m$^{-2}$)& &(rad m$^{-2}$) & & (rad m$^{-2}$)\\
        \hline
        Head & 137.1\arcdeg, 7.1\arcdeg & 0.46 & \bf{-7.5} & 1\arcdeg & 1.4 & 0.08 & \bf{3.0} & 65\arcdeg & 0.02\\
        Tail & 133.9\arcdeg, 6.7\arcdeg & 0.47 & \bf{-1.4} & 163\arcdeg & 1.3 & 0.15 & \bf{-12.5} & 112\arcdeg & 2.0\\
        Off  & 134.9\arcdeg, 9.6\arcdeg & 0.26 & \bf{-1.3} & 176\arcdeg & 0.03 & 0.11 & \bf{7.4} & 106\arcdeg & 1.3\\
        \hline
    \end{tabular}
    \tablecomments{Parameters refer to variables in Equation~\ref{eq:qumodels}.}
\end{table*}

Plotting the fitted peak Faraday depth values (as described in Section~\ref{ssec:RMsynths}) in Galactic coordinates gives the map shown in Figure \ref{fig:fig3}$b$. This image lets us view the Faraday depth morphology of the region more effectively. In this image the head of the tadpole is clearly visible, showing significant negative Faraday depths (around $-7$ to $-8 \radmsq$) compared to the surrounding region (around $-1 \radmsq$). In contrast with the single-frequency images (Stokes $Q$ and $U$ in Figure~\ref{fig:fig2} and polarization angle in Figure~\ref{fig:fig5}) the tail of the tadpole does not stand out in peak Faraday depth.

\subsection{Faraday complexity} \label{sec:complexity}
Using the peak Faraday depths in Figure~\ref{fig:fig3}$b$, we derotated the observed polarization angle to the nominal intrinsic angle by rearranging Equation~\ref{eq:chi}. The result, shown in Figure~\ref{fig:fig3}$c$, reveals the tail as a distinct feature, separate from its background, and spatially uniform in polarization angle. If the tadpole is solely a Faraday rotation phenomenon, with a single Faraday-simple feature representing each LOS, we would not expect it to be visible in a map of derotated $\chi$. The fact that it \textit{does} appear means that either the tadpole contributes significant polarized emission distinct from its surroundings, or there is Faraday complexity along the lines of sight passing through it. The latter possibility is strongly suggested by the sample Faraday depth spectra in Figure~\ref{fig:fig7} and the image slices shown in Figure~\ref{fig:fig6}. 

The separation of the two peaks in the tail in the CHIME data, $\approx 12 \radmsq$, is close to the separation between the main lobe and first sidelobe of the RMSF, which suggests the influence of the $\lambda^2$ coverage. In this context, the Faraday complexity is a true feature but unresolved or marginally resolved. We address this issue in Appendix~\ref{sec:A1}, concluding that while the {\em presence} of multiple peaks in the Faraday spectra is real, the {\em positions} of two peaks separated by $\lesssim \delta \phi$ are modified by the Faraday synthesis process, as was first demonstrated by \citet{Sun:Rudnick:2015}. QU fitting is less subject to this issue; in most but not all cases, QU fitting recovers the true Faraday depths of multiple screens even when separated by $\lesssim \delta \phi$. We investigate the validity of the secondary peaks further in the following section.

\subsection{QU fitting}\label{ssec:qufitting}
In the CHIME Faraday depth spectra, a secondary peak at $\phi \approx -14 \radmsq$ appears as a coherent structure along the tail. Given the proximity of this feature to the first sidelobe of the RMSF (Figure~\ref{fig:fig4}), it is necessary to investigate the validity of these secondary peaks. To this end, we employed QU fitting, using the \texttt{qufit} package in \texttt{RM-Tools}, with the \texttt{pymultinest} nested sampler. We examined 51 representative lines of sight throughout the head and tail of the tadpole and in the surrounding region. We tested four models: screens at one and two Faraday depths (1 FD and 2 FD), with and without beam depolarization (DP). The fractional polarization for this set of models is given by
\begin{equation} \label{eq:qumodels}
\tilde{p}(\lambda^2) = \frac{\tilde{P}(\lambda^2)}{I} = \sum_k p_{0,k} e^{2i(\chi_{0,k} + \phi_k \lambda^2)}e^{-2\sigma_k^2\lambda^4},
\end{equation}
where $\sigma_k$ is the standard deviation of Faraday depths within the beam, the subscript $0$ refers to quantities at the source of emission, and the subscript $k$ refers to the individual Faraday depth components (primary and secondary in our models). The $e^{-2 \sigma_k^2 \lambda^4}$ factor represents beam depolarization caused by any unresolved variation of $\phi$ within the beam. Beam depolarization can be caused by unresolved turbulent cells or a gradient of $\phi$ across the beam. Setting $\sigma_k$ to zero produces the corresponding Faraday screen(s) with no beam depolarization, while setting $p_{0,k}$ to zero for $k>1$ produces the one-component model. 

For the nested sampling, we used uniform distributions of priors for all parameters, with $\phi_k\in[-50, 50]\radmsq$, $p_{0,k}\in[0, 1]$, $\chi_{0,k}\in[0\arcdeg, 180\arcdeg]$ and $\sigma_k\in[0, 100]\radmsq$. We also constrained the Faraday depths and fractional polarizations such that $|\phi_1-\phi_2| \leq100\radmsq$ and $\sum_k p_{0,k}\leq1$. Following \cite{thomson:2021} we use the Bayesian evidence to determine the best fit models, as described in Appendix~\ref{sec:A2}.

For the three representative lines of sight shown in Figure~\ref{fig:fig7}, we found a two-component model with independent beam depolarization factors for each component to be the best fit (the full form of Equation~\ref{eq:qumodels}), although the primary and secondary components of the `off-tadpole' and `head' points respectively have low depolarization. By contrast, both components for the LOS through the tail exhibit significant depolarization. The results of this model and best-fit parameters are summarized in Table~\ref{tab:qufit} and Figure~\ref{fig:fig8} (orange lines) for these three sample lines of sight, and the $\phi$ values are marked by vertical lines on Figure~\ref{fig:fig7}. Figure~\ref{fig:fig8} also shows the three other models we tested for those lines of sight. A comparison of the models is presented in detail in Appendix~\ref{sec:A2}. Note that the ripples seen in the data in Figure~\ref{fig:fig8} that are not fitted by the models are the well-known 30~MHz CHIME ripple caused by reflections between the cylinders and the focal line \citep{chime_overview_2022,chime-21-emission}. The Faraday depths for all 51 lines of sight tested are shown in Figure~\ref{fig:fig9}, with the background images indicating the first (top panel) and second (bottom panel) peaks from the Faraday depth cube, and the colors of the 51 points indicating the corresponding QU-fitted Faraday depths from the two-component model with depolarization.

In the head of the tadpole, most of the 18 tested lines of sight have a primary component near $-7 \radmsq$ (Figure~\ref{fig:fig9}, top panel), with some lines of sight having a weak secondary component (mostly less than $p=0.1$; Figure~\ref{fig:fig9}, bottom panel). Since this secondary component is relatively weak, the head is mostly well-fit with a one-component model with depolarization. In the tail of the tadpole a two-component model (with depolarization) is consistently the best-fit model across the 18 lines of sight tested, in contrast to the surroundings for which several of the 15 tested lines of sight are adequately described by a one- or two-component screen (without depolarization). Although the tail does not stand out from its surroundings in terms of the primary component (which is mostly between $-2$ and $0 \radmsq$ in both the tail and the `off-tadpole' region), it \textit{does} have a relatively coherent secondary component between $-14$ and $-11 \radmsq$, which the surroundings lack. This is the secondary component that also appears in the Faraday depth spectra, although in some cases it is shifted slightly in Faraday depth due to the interaction of the components with the RMSF \citep[see][and Appendix~\ref{sec:A1}]{Sun:Rudnick:2015}. 

We note that the purpose of the models we chose to test was to confirm the Faraday depth values of the primary and secondary peaks in the spectra derived using Faraday synthesis. As we can see from large values of the Bayes odds ratios listed for the models in Appendix~\ref{sec:A2}, the `true' description of the tadpole lines of sight is likely more complicated than this set of models.

\subsection{Artifacts}\label{ssec:artifacts}

The CHIME maps are sensitive to structures on a wide range of angular scales. Some artifacts are described in detail in \citet{chime_overview_2022}. One is evident in the single diagonal stripe in the top left corner of Figures~\ref{fig:fig2} and \ref{fig:fig3}, which is a line at the right ascension of Tau~A. Curved striations, seen in the Stokes $Q$ and $U$ images (Figure~\ref{fig:fig2}) and the Faraday synthesis images (Figure~\ref{fig:fig3}), correspond to fixed zenith angles (or, equivalently, declinations). Point sources appear as a single point with bright copies at the same declination on either side of the source due to grating lobes, resulting in an apparent triple source. The sources themselves appear in Stokes $Q$ in equatorial coordinates due to leakage, with symmetric sidelobes, while in Stokes $U$ only the asymmetric sidelobes, with opposite signs, appear. In the Galactic coordinates shown in this paper, leakage sources appear in both $Q$ and $U$, along with their sidelobes.

Grating lobes also appear for larger-scale structures, having a slight effect on the appearance of the images of the tadpole region. In the CHIME 410~MHz $pI$ map in Figure~\ref{fig:fig5}, a copy of the head of the tadpole can be seen as an outline that stands out in polarized intensity, centered on $\ell\approx 138.5\arcdeg$, $b\approx8.5\arcdeg$. The location and the separation between this and the center of the head agrees with the position of the grating lobes in relation to the main lobes of the point sources. This `ghost' copy of the tadpole also appears in the 410~MHz polarization angle image in Figure~\ref{fig:fig5}, is generally more apparent at the lower frequencies, and is quite evident in the Faraday depth slices shown in Figure~\ref{fig:fig6}.

There is also declination-dependent striping, which appears as curved stripes in Galactic coordinates in Figure~\ref{fig:fig2} and other images, although it is much less pronounced in angle images and Faraday synthesis products. This arises from cross talk between adjacent feeds \citep[Figure 18]{chime_overview_2022}. We could remove the striping using image processing techniques, but this cosmetic improvement to the images is unnecessary for our science. Ultimately, the inclusion of the DRAO~15~m survey (Ordog et al in prep) will allow us to exclude baselines $\lesssim 5 \m$ from the final CHIME-GMIMS data product; we expect this to considerably reduce this striping.

\section{The origin of the tadpole}\label{sec:origins}

\subsection{Neutral Hydrogen Structure} \label{ssec:H_I}
 
\begin{figure}
    \plotone{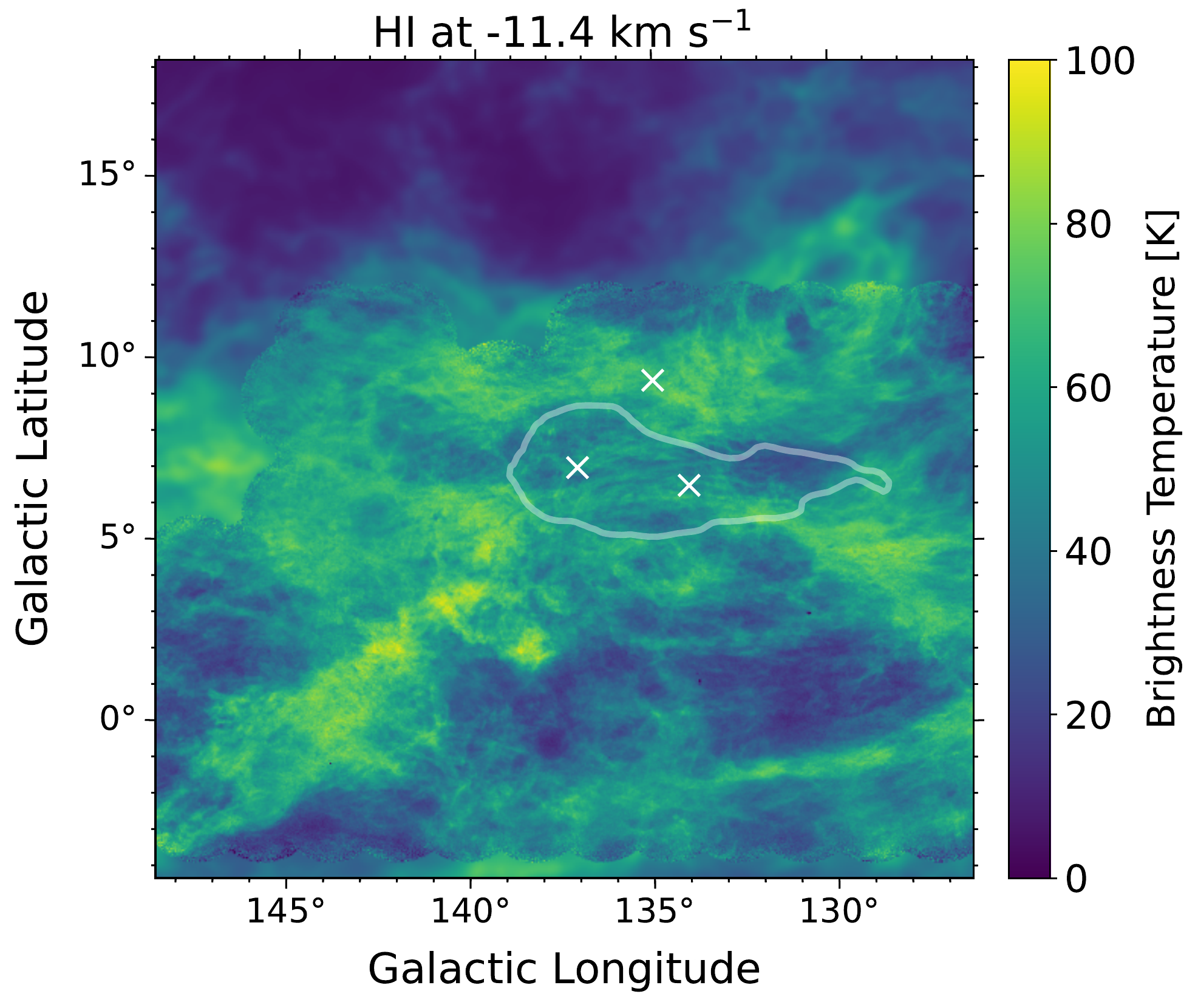}
    \caption{$\HI$ image from ST+HI4PI at local arm velocity $-11.4 \kms$. Superimposed is a contour from CHIME polarization angle representing the tadpole region. `$\times$' marker are as in Figure \ref{fig:fig2}. In parts of the image where ST data are not available (mostly at $b > +12\arcdeg$), we use HI4PI data instead.}
\label{fig:fig10}
\end{figure}

 We turn our investigation to an analysis of the general structure of neutral hydrogen in the Galaxy's ISM in the Fan region. We use DRAO ST+HI4PI \ion{H}{1} observations described in Section~\ref{ssec:st} and shown in Figure~\ref{fig:fig10} for this purpose. We searched the ST+HI4PI data cube at velocities corresponding to the Local, Perseus, and Outer spiral arms \citep{2014ApJ...783..130R} for emission features with a morphological resemblance to the tadpole but were unable to find any. The tadpole most likely lies within the local arm, corresponding to $|V_{\mathrm{LSR}}| \lesssim 30 \kms$ for \HI\ emission. Under this assumption, its physical scale is reasonable; at outer-arm distances, the head would be 170 parsecs in diameter. As discussed by \citet{haverkorn_2003}, this is implausibly large for a single-star ionized region or any other Faraday-rotating feature. Moreover, if the tadpole is local, there will be polarized emission arising at larger distances, which can be Faraday rotated.
 
In an attempt to identify the structure of G137+7 in neutral hydrogen observations in regions of excess, or lack of, hydrogen, \citet{verschuur_1969} made `relative intensity' profiles of \HI\ at each velocity interval \citep[using data from][]{Heiles_1967}. These relative intensity profiles came from drawing a baseline through the mean levels at the two highest and two lowest declinations from the scans. Then, the baselines were subtracted from the \HI\ data to create profiles to show abundances or deficiencies in~\HI. From this, \citet{verschuur_1969} identified \HI\ deficiencies at velocities between $-16$ and $-4$ km s$^{-1}$ that coincide with the structure G137+7 in polarization maps. 

We attempted to replicate the findings of \citet{verschuur_1969}, now using the ST+HI4PI data. The ST+HI4PI data has an angular resolution of ~$1'$, an improvement on the $12'$ resolution of \citet{Heiles_1967}, and both surveys have a spectral resolution of roughly $1 \kms$. Searching the entire cube after subtraction, we note a weak deficit of \HI\ at $-11.4 \kms$ roughly coincident with the tail, showing a drop in \HI\ of $\sim$ 50 percent relative to the immediate surroundings. However, this deficit does not match the morphology of the tail, and dips in \HI\ intensity at this scale are very common in the HI4PI data for this area, as seen in Figure~\ref{fig:fig10}. We do not regard this as a significant feature.

\subsection{Ionized Hydrogen Structure} \label{ssec:H_alpa}

H$\alpha$ profiles offer a means to probe the distribution of ionized gas \citep{wham_nss}. The WHAM survey provides us with a kinematically resolved map of the H$\alpha$ emission in the Galaxy, within approximately $100 \kms$ of our local standard of rest. Areas rich in ionized gas signify an enhanced population of free electrons, which play a crucial role in the Faraday rotation mechanism.

\begin{figure}
\plotone{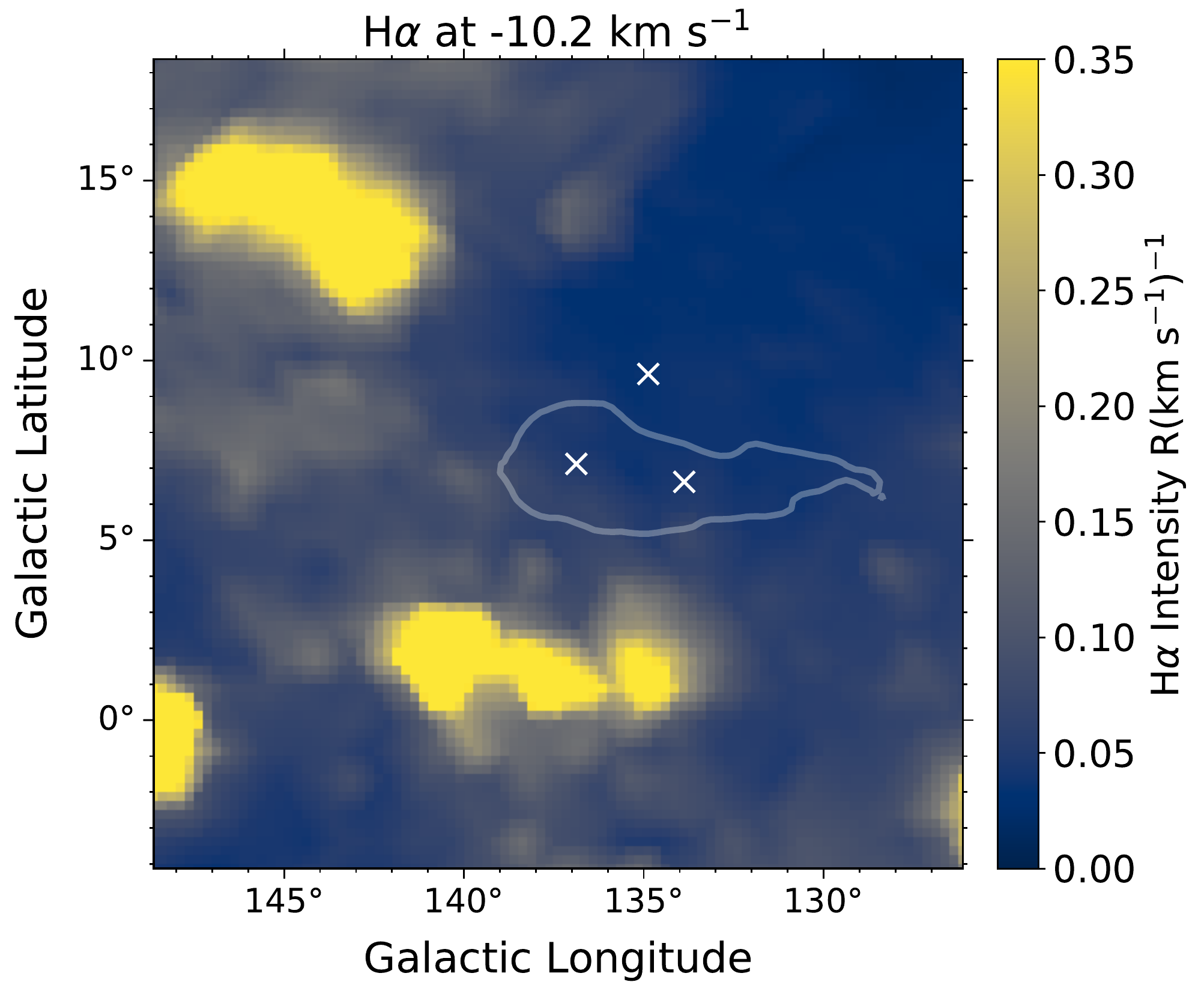}
    \caption{H$\alpha$ image from the Wisconsin H-alpha Mapper survey at local arm velocity $-10.2 \kms$. Contour is as in Figure~\ref{fig:fig10}; `$\times$' marker is as in Figure \ref{fig:fig2}.}
\label{fig:fig11}
\end{figure}

Local arm H$\alpha$ data from the WHAM survey is shown in Figure~\ref{fig:fig11}, where the tadpole feature does not coincide with any region of bright H$\alpha$ emission. We searched the rest of the channels in the data cube and found no correspondences. The regions bright in H$\alpha$ emission that are seen in Figure~\ref{fig:fig11} are from documented features, namely the three \ion{H}{2} regions W3, W4, and W5 near $134\arcdeg \lesssim \ell \lesssim 142\arcdeg$ and $b \approx +1\arcdeg$.

We consider the possibility that the tail is primarily due to residual ionization from a past ionization source such as a passing hot star. If the WIM is $8000 \K$, the recombination coefficient to energy levels $n \geq 2$ is $\alpha^{(2)} \approx 3.04 \times 10^{-13} \text{ cm}^3 \ \text{s}^{-1}$ \citep{wham_nss, Draine_2011}. The recombination time is $t_r \approx (\alpha^{(2)} n_e)^{-1}$. If we assume an electron density $n_e = 0.1 \cucm$ \citep{Hill:Benjamin:2008}, this gives a recombination time of $\sim 1 \Myr$. Under the assumption that the ionizing source has travelled along the path laid out tail-to-head, then $\sim 1 \Myr$ ago the star was in the middle of the current location of the tail ($\ell \approx 132\arcdeg$). The angular separation of the center of the head and halfway along the tail is about 5\arcdeg, meaning the ionizing source would have a proper motion of $\sim 20 \mas \yr^{-1}$ in this model. The thinning of the tail with increasing distance from the head is consistent with this recombination scenario.

\subsection{Proper Motions of Candidate Stars} \label{ssec:proper_motion}

Previous studies of feature G137+7 considered the circular region to be a result of a Str\"omgren sphere of the B2(e) star HD~20336, or alternatively,  a relic Str\"omgren sphere from the white dwarf WD~0314+64, both of which lay within the head of the tadpole \citep{verschuur_1968,iacobelli_haverkorn_katgert_2012}.

We hypothesize that the tail of the tadpole is a trail of ionized gas behind a suitable star, and should indicate motion related to the feature, potentially in the fashion suggested by \citet{haverkorn_2003} with an elongated Str\"{o}mgren sphere. The tail may be similar to tails associated with planetary nebulae  \citep{Ransom2010,Ransom2015}, yielding estimates of the timescales for the interactions of planetary nebulae and the ISM. For a star to be considered a strong candidate for the tadpole, the characteristics we observe require it to be a hot star (Type O or B) with a proper motion comparable to our calculations based on the recombination time (see section \ref{ssec:H_alpa}) and direction aligned with the orientation of the tail.

We check to see whether the motion of HD~20336 meets these criteria. We overlay the inferred positions over the past $3 \Myr$ of HD 20336 given the sky-projected space velocity calculated from Gaia proper motion values \citep{gaia_2022}, corrected for the Local Standard of Rest (LSR) \citep[e.g.][]{soderblom,Ransom2015} given by \citet{Huang2015} atop Figure~\ref{fig:fig2}. The orientation of the tadpole is just outside the error cone of the trajectory we determined for HD~20336, suggesting that an association is possible but not definitive. The error in the path of HD 20336 is largely from the uncertainty in the LSR. The orientation of the tail parallel to the Galactic plane also suggests motion in the plane, as might be expected for a star. Similar corrected space velocity values were calculated for WD 0314+64, finding a proper motion path that agreed less well with the tail. The angular velocities of the B2(e) star and white dwarf are $9.98 \pm 0.02 \mas \yr^{-1}$ and $159.4 \pm 0.1 \mas \yr^{-1}$ respectively. The high angular velocity of the white dwarf rules it out based on the timescale estimates we find for the tail in Section~\ref{ssec:H_alpa} (unless $n_e \sim 1 \cucm$, which is much higher than typical in the WIM), however, the B2(e) star's angular velocity is of the same order of magnitude as our estimates---a result consistent with the star causing the feature. Using the Gaia database, we searched\footnote{We used the capabilities of a natural language model to construct ADQL search queries of the Gaia archive \citep{OpenAI}.} for any other candidate stars that could cause the Faraday rotation feature \citep{anderson2023reconciling}. We identified no new candidates from this query. 

\subsection{Faraday depth and electron column}\label{ssec:FDandElectron}

We can make sense of the absence of the tadpole in both the ST+HI4PI $\HI$ and WHAM surveys by considering that the strongest observed Faraday rotation in the head is $\phi \approx -8 \radmsq$. This requires free electrons that would have been ionized from neutral hydrogen in the ISM. Here we assume that a change in the electron density, not a change in the magnetic field, is the dominant factor in the change in $\phi$. This would leave a deficit in the neutral hydrogen column density, $N(\HI)$, but as demonstrated above, we do not observe this. The magnitude of change in the \HI\ density needed to produce the observed Faraday depth can be estimated using (from Equation~\ref{eq:RM}) $ \phi = 0.81\cdot n_eB_{\parallel}L$. Using the Faraday depth from the head of the tadpole and the canonical total magnetic field strength in the Solar neighborhood of $6 \uG$ \citep{haverkorn:2015}, the change needed in the column density of free electrons is $n_e L\sim 1.5 \times 10^{18} B_6^{-1}  \cm^{-2} \sim 1.4 B_6^{-1} \pc \cucm$, or an emission measure of $\EM \equiv \int n_e^2 \, ds = (n_e L)^2 / L \sim 2 (L/1\pc)^{-1} B_6^{-2} \pc \cm^{-6}$. Here we define $B_6 \equiv B_{||} / 6 \uG$; $B_6 = 1/\sqrt{3}$ if a $6 \uG$ field is randomly distributed amongst three orthogonal components. 

If the free electrons arose from ionization of existing neutral hydrogen, this would result in a decrease in $N($\ion{H}{1}$)$ of $\sim 1.5 \times 10^{18} B_6^{-1} \cm^{-2}$. This is small compared to the typical $N($\ion{H}{1}$) \gtrsim 10^{20} \cm^{-2}$ near the plane (less than $3\%$ in the case of $B_6 = 1/\sqrt{3}$)  and thus would be impossible to see as a deficit in \ion{H}{1} maps. An emission measure of $\approx 2 \pc \cm^{-6}$ would be detectable in H$\alpha$ emission \citep{wham_nss}; if the path length of ionized gas were $L \gtrsim 3 \pc$, the emission measure would be less than the $3 \sigma$ sensitivity of the WHAM survey ($\EM \sim 0.3 \pc \cm^{-6}$). Therefore, Faraday rotation would be sensitive to a change in electron column density that could not be detected in other tracers even if there were no change in the magnetic field. Because we would not expect to detect a change in $n_e L$ in other tracers, we cannot readily distinguish between a change in electron column and a change in $B_{||}$. The motion of ionized gas may distort the magnetic field lines, changing their orientation with respect to the LOS, which can also lead to a gradient in Faraday depth, a so-called ``magnetic wake'' \citep{Ransom2010}.

\section{Summary and future prospects} \label{sec:summary}

In this paper, we have presented the first polarization maps from CHIME in what will be a component of the GMIMS low band north all-sky survey. The wide bandwidth at relatively low frequencies ($400-729 \MHz$) gives us a Faraday depth resolution which is roughly half the largest scale we are sensitive to, enabling investigation of Faraday complexity. We analyzed a relatively small region of the sky where the emission is bright and instrumental properties are best understood; in future work, we will expand the use of CHIME data to analyze the large-scale structure of the Galactic magnetic field.

The polarized structure that we refer to as the `tadpole' is composed of a circular `head' centered on $\ell = 137\arcdeg$, $b = +7\arcdeg$ roughly $2\arcdeg$ in diameter, with a `tail' extending about $10\arcdeg$ to $\ell = 127\arcdeg$. The entire feature has an observed polarization angle rotated significantly compared to the surrounding sky due to Faraday rotation. The largest Faraday depth $|\phi|$ is in the head of the tadpole, with values as large as $-8\radmsq$. The tail appears as a second Faraday depth component at $\approx -12 \radmsq$. The tail has been hinted at previously but is very clear both in single-channel polarization angle images and in Faraday depth images with CHIME. Similarly, what we identify as the head has been seen primarily as a depolarization feature in the past. With the angular and Faraday depth resolution of CHIME, we clearly identify it as a Faraday rotation feature, recovering most of the power that is depolarized with poorer angular or Faraday depth resolution.

The presence of a tail suggests motion along its direction through the interstellar medium. The proper motion of HD 20336 is marginally consistent with the orientation of the tail. The tadpole is not seen in maps of neutral or ionized hydrogen column density, although our estimates suggest that the amount of gas necessary to undergo ionization, resulting in the observed Faraday rotation magnitude, falls below the sensitivity threshold of H$\alpha$ and \ion{H}{1} surveys, especially given the intensity of surrounding emission. We find the recombination time for the ionized electrons to be $\sim$ 1 Myr, and with this estimate, we derive a proper motion of the (unknown) ionizing source of $\sim 20 \mas \yr^{-1}$, which is on the order of magnitude of the B2(e) star's velocity. Although this is suggestive evidence, we cannot be certain that the B2(e) star is linked to the tadpole. 

The CHIME Stokes $Q$ and $U$ cubes covering the $400 - 729 \MHz$ range comprise the primary data set used in this study. This frequency coverage yields a Faraday depth resolution of $9.7 \radmsq$, resolving the maximum $\phi$ scale of $\sim 19 \radmsq$. Given the asymmetries present in the main lobes of some of the spectra we observe on or near the tadpole feature, exploring this region with improved $\phi$ resolution combined with sensitivity to extended features will be valuable.

In the longer term, we will be able to combine $\phi$ resolution and improved spatial resolution with the advent of the upgraded DRAO ST, which will cover $400 - 1800 \MHz$ at $1'$ resolution (compared to the single frequency channel available from the current ST). The single-channel $1'$-resolution $\chi$ map from the DRAO ST+EMLS in Figure~\ref{fig:fig5} already reveals an abundance of structures on scales much smaller than the size of the tadpole, and investigating these in $\phi$ space with high resolution may provide further insights into the nature of the overall structure. Other future studies may also include exploring potential correlations between the tadpole Faraday structure and thermal dust emission features in three-dimensional dust maps.

There are three factors in this study that CHIME makes possible. First, for the purpose of this study, CHIME is effectively a large single antenna, with sensitivity to large structures, coupled with better angular resolution than the single antennas that have been used for polarimetry in this frequency range. Second, the wide bandwidth and many frequency channels enable Faraday synthesis with $\phi_{\mathrm{max-scale}} \approx 2 \delta \phi$, comfortably resolving the maximum scale. Third, the polarized sky is Nyquist sampled. This powerful combination has revealed extended and Faraday complex structure in G137+7, a region that has been a curiosity for six decades.

{\em Data availability:} FITS files containing the CHIME data products, Stokes $I$, $Q$, $U$, and $V$ as a function of frequency and Faraday depth for the regions shown in this paper, are available through the Canadian Astronomical Data Centre \citep{data}.

\begin{acknowledgements}

This paper relies on observations obtained using telescopes located at the Dominion Radio Astrophysical Observatory, which is located on the traditional, ancestral, and unceded territory of the syilx people. We benefit enormously from the stewardship of the land by the syilx Okanagan Nation and the radio frequency interference environment protection work by the syilx Okanagan Nation and DRAO. We acknowledge DRAO staff, especially K.\ Phillips and B.\ Robert, for their work on the site and the telescopes used in this work. DRAO is a national facility operated by the National Research Council Canada. CHIME is funded by grants from the Canada Foundation for Innovation (CFI) 2012 Leading Edge Fund (Project 31170), the CFI 2015 Innovation Fund (Project 33213), and by contributions from the provinces of British Columbia, Québec, and Ontario. Additional support was provided by the University of British Columbia, McGill University, and the University of Toronto. CHIME also benefits from several NSERC Discovery Grants, including RGPIN-2020-05035 and 569654. This research was enabled in part by support provided by the Digital Research Alliance of Canada.

A.S.H. and A.O. acknowledge Interstellar Institute’s program ``II6'' and the Paris-Saclay University’s Institut Pascal for hosting discussions that nourished the development of the ideas behind this work, especially with R.~A.~Benjamin. We thank R.~Mckinven for useful discussions about instrumental polarization in CHIME/FRB data. We also acknowledge helpful discussions with W.~Raja. We thank the anonymous referee for a careful and constructive report which led to an improved paper.

N.M. was supported by an Undergraduate Research Award from UBC Okanagan and an NSERC Undergraduate Student Research Award. A.O. is partly supported by the Dunlap Institute at the University of Toronto. A.B.  acknowledges financial support from the INAF initiative ``IAF Astronomy Fellowships in Italy'', grant name MEGASKAT. M. Haverkorn acknowledges funding from the European Research Council (ERC) under the European Union's Horizon 2020 research and innovation programme (grant agreement No 772663). We acknowledge the support of NSERC, funding reference number 569654. K.W.M. holds the Adam J. Burgasser Chair in Astrophysics and is supported by NSF grants (2008031, 2018490). M.T. is supported by the Banting Fellowship (Natural Sciences and Engineering Research Council Canada) hosted at Stanford University. 

\end{acknowledgements}

\facility{CHIME,DRAO:Synthesis Telescope,Effelsberg}

\software{astropy \citep{2022ApJ...935..167A}; ch\_pipeline; matplotlib \citep{Hunter:2007}; numpy; RM-tools \citep{purcell:2020}.}

\appendix
\restartappendixnumbering

\section{Resolved and unresolved Faraday components in Faraday synthesis}\label{sec:A1}

\begin{figure}
\plottwo{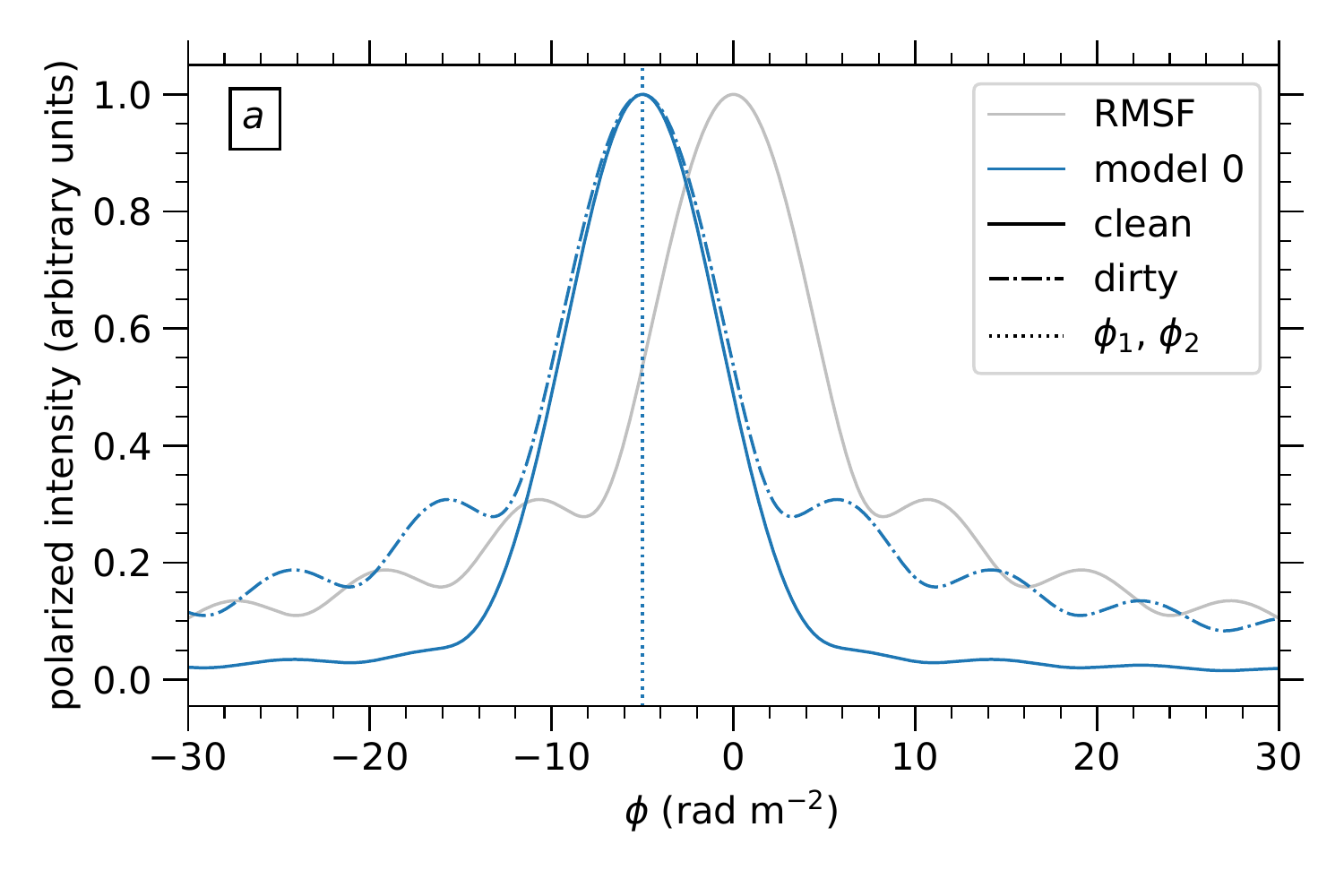}{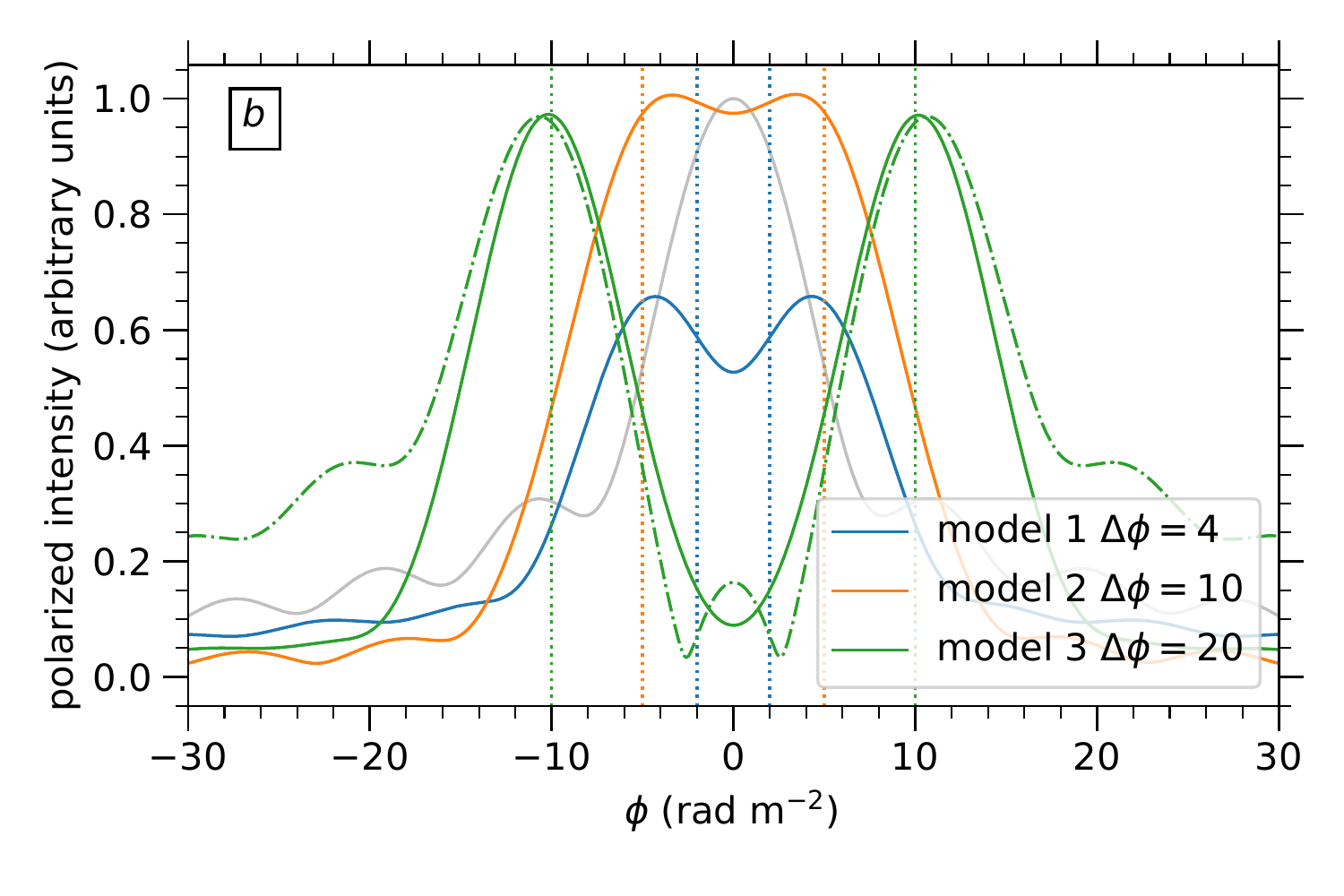}
\plottwo{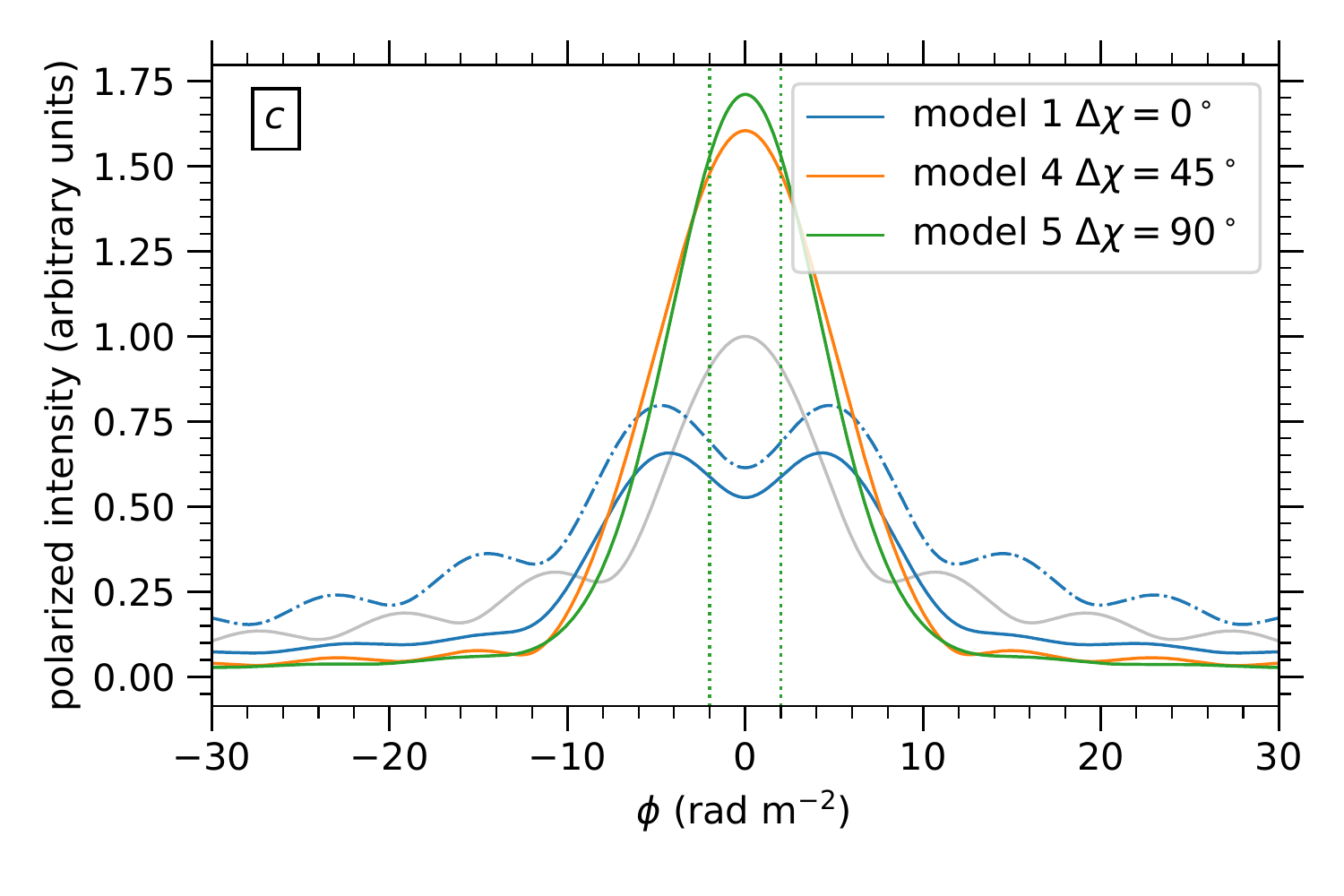}{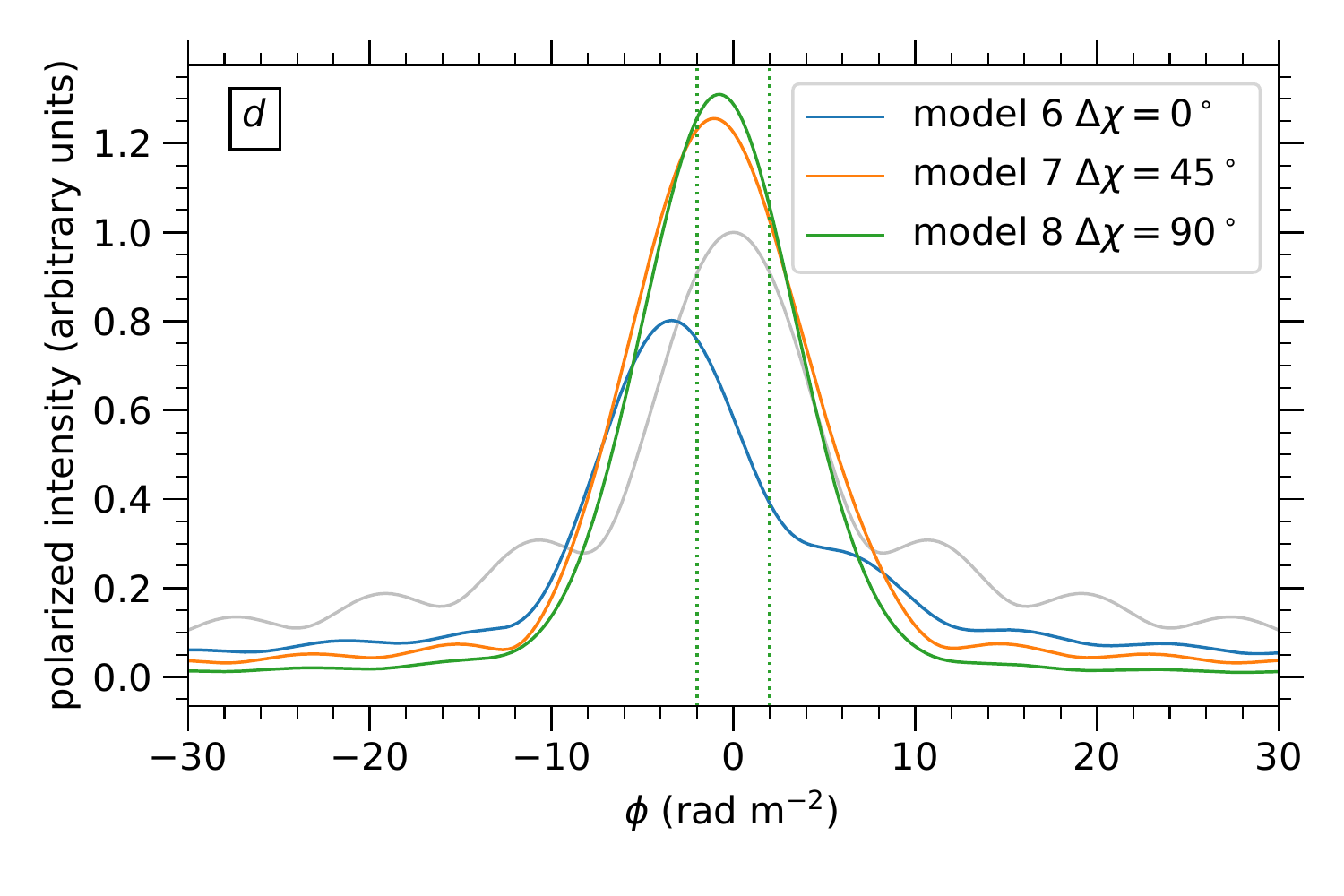}
\caption{Output of models from Table~\ref{tab:models}. Vertical dotted lines show the input Faraday depths $\phi_1$ and $\phi_2$. We show dirty (dot-dashed lines) and clean (solid lines) spectra for a few cases but clean spectra only in most cases to reduce clutter. {\em a}: Our control model with one component at $-5 \radmsq$. {\em b}: Models~1--3 with two components separated by $4$, $10$, and $20 \radmsq$. {\em c:} Models~1, 4, and 5, with two components separated by $4 \radmsq$, which is not resolved by the RMSF. We show the same model but with $\chi_2$ ranging from $0$ (model~1, as in panel $b$) to $90\arcdeg$ (model~5). {\em d:} Models~6--8, with two components separated by $4 \radmsq$, which is not resolved by the RMSF. The second component has reduced $pI$. As in models 1, 3, and 4, we show the same model but with $\chi_2$ ranging from $0$ (model~5) to $90\arcdeg$ (model~7).} \label{fig:figA1}
\end{figure}

\begin{deluxetable}{lcccccccc}
\tablecaption{Model parameters shown in Figure~\ref{fig:figA1}.} \label{tab:models}
\tablehead{ \colhead{Model} & \colhead{$p_{0,1}I$} & \colhead{$\phi_1$} & \colhead{$\sigma_{0,1}$} & \colhead{$\chi_{0,1}$} & \colhead{$p_{0,2}I$} & \colhead{$\phi_2$} & \colhead{$\sigma_{0,2}$} & \colhead{$\chi_{0,2}$}
}
\startdata
0 & $1.0$ & $-5 \radmsq$ & $0.0 \radmsq$ & $0\arcdeg$ & $0.0$ & \nodata & \nodata & \nodata \\
\hline
1 & $1.0$ & $-2 \radmsq$ & $0.0 \radmsq$ & $0\arcdeg$ & $1.0$ & $2 \radmsq$ & $0.0 \radmsq$ & $0\arcdeg$ \\
2 & $1.0$ & $-5 \radmsq$ & $0.0 \radmsq$ & $0\arcdeg$ & $1.0$ & $5 \radmsq$ & $0.0 \radmsq$ & $0\arcdeg$ \\
3 & $1.0$ & $-10 \radmsq$ & $0.0 \radmsq$ & $0\arcdeg$ & $1.0$ & $10 \radmsq$ & $0.0 \radmsq$ & $0\arcdeg$ \\
\hline
4 & $1.0$ & $-2 \radmsq$ & $0.0 \radmsq$ & $0\arcdeg$ & $1.0$ & $2 \radmsq$ & $0.0 \radmsq$ & $45\arcdeg$ \\
5 & $1.0$ & $-2 \radmsq$ & $0.0 \radmsq$ & $0\arcdeg$ & $1.0$ & $2 \radmsq$ & $0.0 \radmsq$ & $90\arcdeg$ \\
\hline
6 & $1.0$ & $-2 \radmsq$ & $0.0 \radmsq$ & $0\arcdeg$ & $0.5$ & $2 \radmsq$ & $0.0 \radmsq$ & $0\arcdeg$ \\
7 & $1.0$ & $-2 \radmsq$ & $0.0 \radmsq$ & $0\arcdeg$ & $0.5$ & $2 \radmsq$ & $0.0 \radmsq$ & $45\arcdeg$ \\
8 & $1.0$ & $-2 \radmsq$ & $0.0 \radmsq$ & $0\arcdeg$ & $0.5$ & $2 \radmsq$ & $0.0 \radmsq$ & $90\arcdeg$ \\
\enddata
\tablecomments{Input parameters used with Equation~\ref{eq:qumodels} to create Figure~\ref{fig:figA1}.}
\end{deluxetable}

In this appendix, we model the impact of resolved and unresolved Faraday depth components on Faraday synthesis spectra, similar to the analysis of \cite{Sun:Rudnick:2015}, but covering the much broader $\lambda^2$ range corresponding to the CHIME frequency coverage. We create a model with Faraday depth components $\phi_k$ with polarization fractions $p_{0,k}$ emitted at polarization angles $\chi_{0,k}$. The observed complex polarization is then given by Equation~\ref{eq:qumodels}. We performed a series of simulations with the inputs listed in Table~\ref{tab:models}. We set $\sigma_k = 0 \radmsq$ in all cases; making $\sigma_k$ non-zero produces results that are similar to reducing $p_{0,k}$ for each component. We then performed Faraday synthesis on the resulting $\tilde{P}(\lambda^2)$ spectra and cleaned the spectra to a threshold of $pI = 0.2$ in the arbitrary units used in these figures, producing the Faraday depth spectra shown in Figure~\ref{fig:figA1}. We used complete frequency coverage from $400-729 \MHz$ in running these simulations, an idealized version of the portion of the CHIME data used in this paper, with the FWHM of $R(\phi)$ being $\delta \phi = 9.7 \radmsq$. Accounting for the gaps in our frequency coverage produces additional sidelobes in the dirty spectrum but does not change the qualitative picture.

First, in Figure~\ref{fig:figA1}$a$, we show a model with a single component at $\phi = -5 \radmsq$. The dirty spectrum is an exact replica of the RMSF, shifted to the input Faraday depth. Cleaning removes the sidelobes effectively.

Next, in Figure~\ref{fig:figA1}$b$, we show cases in which there are two components which are separated by $4 \radmsq$, which is unresolved (model 1); separated by $10 \radmsq$, which is marginally resolved (model 2), and separated by $20 \radmsq$, more than twice $\delta \phi$ (model 3). In the resolved case (model~3), two peaks appear at their input Faraday depths. Sidelobes are present but removed precisely by cleaning, leaving two peaks that closely resemble Gaussians. We performed the same experiment with a wide range of input angles $\chi_{0,2}$; there is no appreciable change in the resulting Faraday spectrum. For the $\Delta \phi = 10 \radmsq$ case (model~2), there is a flattened appearance to the blended peaks in the Faraday spectrum. There are two peaks, but they are separated by $\approx 8 \radmsq$, less than the separation of the input Faraday depths. Lastly, in the unresolved case ($\Delta \phi = 4 \radmsq$; model~1), there are two distinct peaks but they are separated by $\approx 9 \radmsq$.

In Figure~\ref{fig:figA1}$c$, we again show two components with equal polarized intensity separated by $4 \radmsq$, which is not resolved by the RMSF. Model~1, with the same $\chi_0$ in both components, is as in Figure~\ref{fig:figA1}$b$. When the two components are emitted at different angles ($45\arcdeg$ different in model 3 and $90\arcdeg$ different in model 4), we see only a single component centered between the two input components. These two peaks are true features in that they correspond to two distinct but unresolved input components, but their observed Faraday depths are not accurate. Instead, the Faraday depths of the observed peaks are separated by $\approx \delta \phi$. Because these peaks represent true features, they are not removed by RM clean, but their position is also not changed (or perhaps changed very slightly). In contrast, the sidelobes at $\pm 15 \radmsq$ and $\pm 23 \radmsq$ are removed by RM clean.

In Figure~\ref{fig:figA1}$d$, we show a similar experiment but the intensity of one of the input components is reduced by $50\%$. In the in-phase model (model 5), the two peaks remain present and remain pushed out to be separated by $\approx \delta \phi$. In the out-of-phase models (6 and 7), only one peak is evident, centered at the intensity-weighted mean of the two input components.

\begin{figure}
\centering
\includegraphics[width=0.5\textwidth]{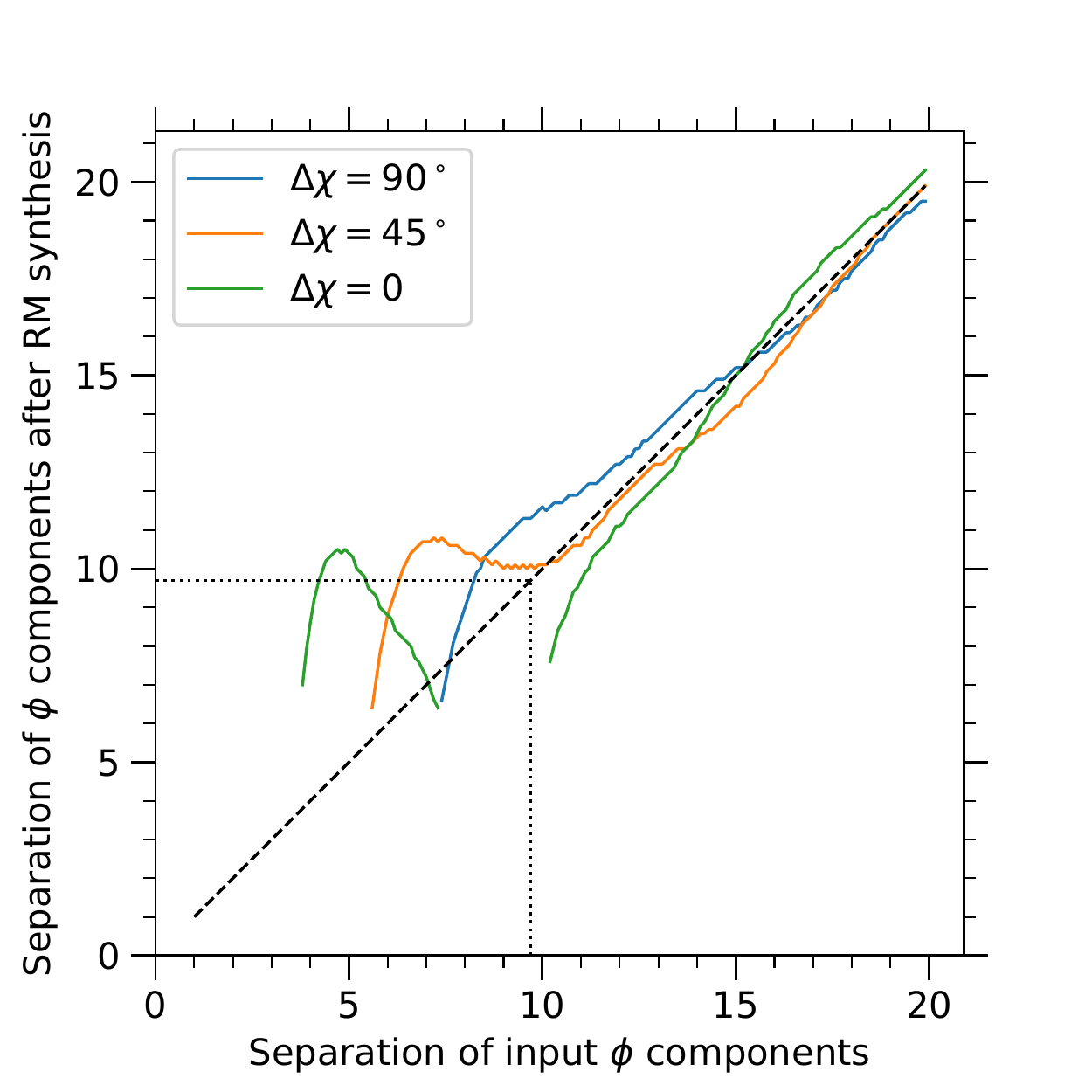}
\caption{Output separation of Faraday depth components (calculated using Faraday synthesis) as a function of the input separation of Faraday depth components in the simulation. The dashed line shows a 1:1 line, and the dotted box shows the FWHM of $R(\phi)$ ($9.7 \radmsq$). We ran this experiment for $1 < \phi < 20 \radmsq$ in $0.1 \radmsq$ increments; outputs for which there are not separate peaks in the output Faraday spectrum are blank.} \label{fig:figA2}
\end{figure}

We ran models of types 1, 4 and 5 (equal input polarized intensities with three different angular separations of the components), but now for a range of input Faraday depth separations between the components and then determined the separation of the peaks in Faraday spectra. We show the results in Figure~\ref{fig:figA2} (similar to Figure 4 of \cite{Sun:Rudnick:2015}, but with a denser sampling of $\Delta\phi$). At $\Delta \phi \gtrsim 12 \radmsq$, the components are resolved by Faraday synthesis and the output peaks correspond well with the input peaks. We do not observe peaks in the Faraday spectrum which are appreciably closer than the FWHM of $R(\phi)$, $9.7 \radmsq$.

What causes the observed peaks in the Farady spectrum to be at separations wider than the input values? This is a depolarization effect. Absent depolarization, we would expect two unresolved components to appear in a single, broadened Gaussian-like form. When the Faraday-rotated emission from the two components is out of phase, depolarization occurs. With two Faraday screens of equal $pI$, the Faraday spectrum $\tilde{F}(\phi)$ is the sum of two copies of the RMSF with centroids separated by $\phi_2 - \phi_1$. The imaginary part of $\tilde{R}(\phi)$ is antisymmetric, so for certain separations of $\phi_1$ and $\phi_2$, the imaginary parts of the copies of $\tilde{R}(\phi)$ interfere and depolarize. From these investigations, we conclude that Faraday synthesis results should be interpreted with caution when the spacing of peaks is $\lesssim \delta \phi$ (Equation~\ref{eq:res}), in agreement with the similar results at higher frequencies presented by \cite{Sun:Rudnick:2015}. In this case, QU fitting, which is less sensitive to this type of interaction between Faraday depth components, should be used to refine the results.

\section{QU fitting results}
\label{sec:A2}
Here we present a comparison between the Faraday synthesis and QU fitting results for the CHIME data in terms of the Faraday depth spectra, along with a comparison of the four models tested. For the three lines of sight shown in Figures~\ref{fig:fig6}~and~\ref{fig:fig7} we determine the complex polarized fraction, $\tilde{p}(\lambda^2)$, from the best-fit parameters resulting from QU fitting for the four models we tested: one- and two-component Faraday screens (1 FD and 2 FD), with and without beam depolarization (DP). These are shown in Figure~\ref{fig:fig8} for comparison in the $\lambda^2$ domain. In Figure~\ref{fig:fig9} we show the Faraday depths derived from the two-component model that includes beam depolarization, 2 FD + DP.

\begin{figure}
\plotone{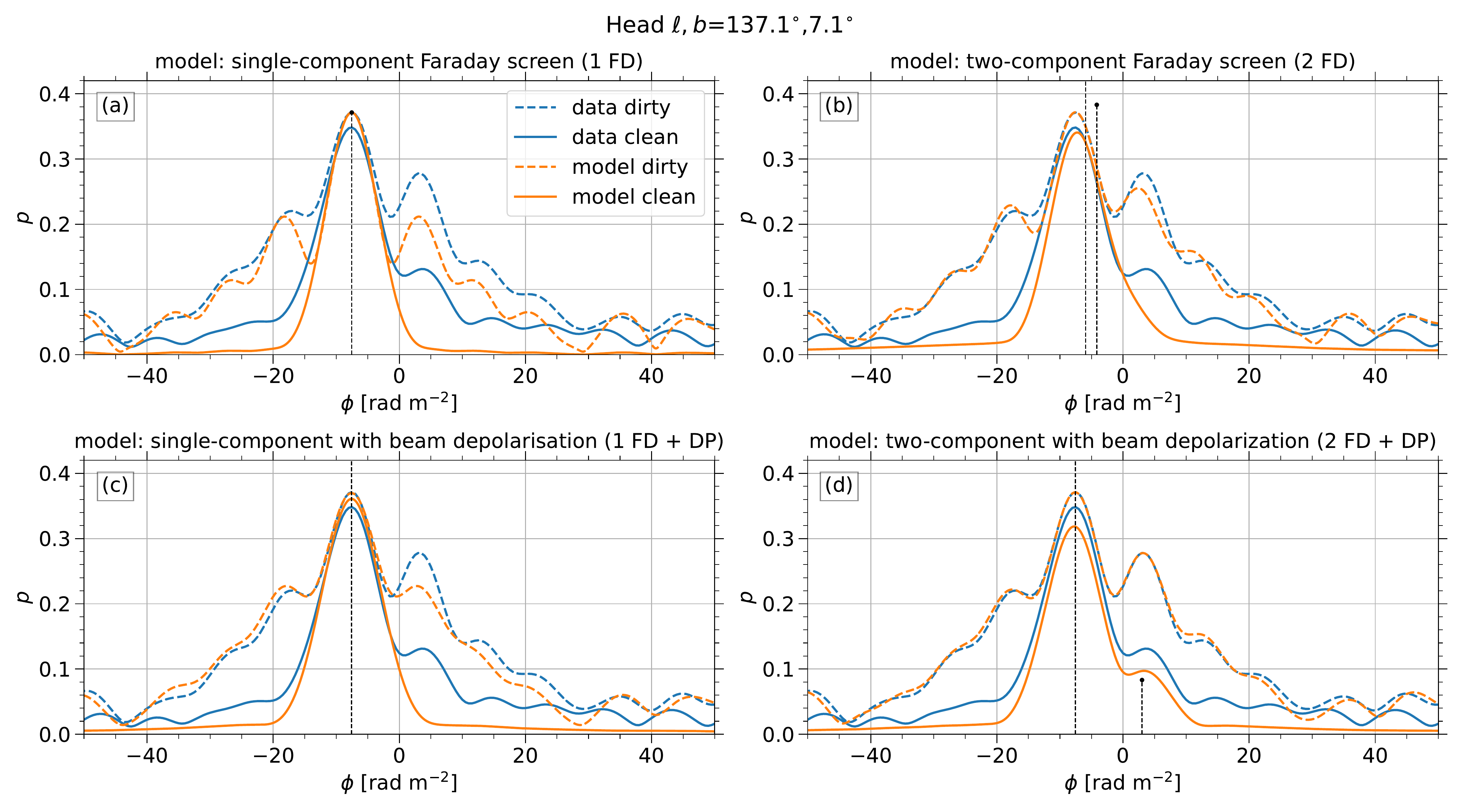}
\caption{Results of QU fitting used as input to Faraday synthesis for a LOS on the head of the tadpole for the four models described in Section~\ref{ssec:qufitting}. Blue lines represent the magnitudes of the spectra derived from applying Faraday synthesis to the CHIME $\tilde{p}(\lambda^2)$ data. Orange lines are the magnitudes of the spectra derived from applying Faraday synthesis to the model $\tilde{p}(\lambda^2)$ determined from QU fitting. Solid (dashed) lines represent clean (dirty) spectra. Vertical, black dotted lines indicate the locations of the QU-fitted Faraday depth components, and the black dot indicates the polarized fraction corresponding to that component.} \label{fig:figB3}
\end{figure}

\begin{figure}
\plotone{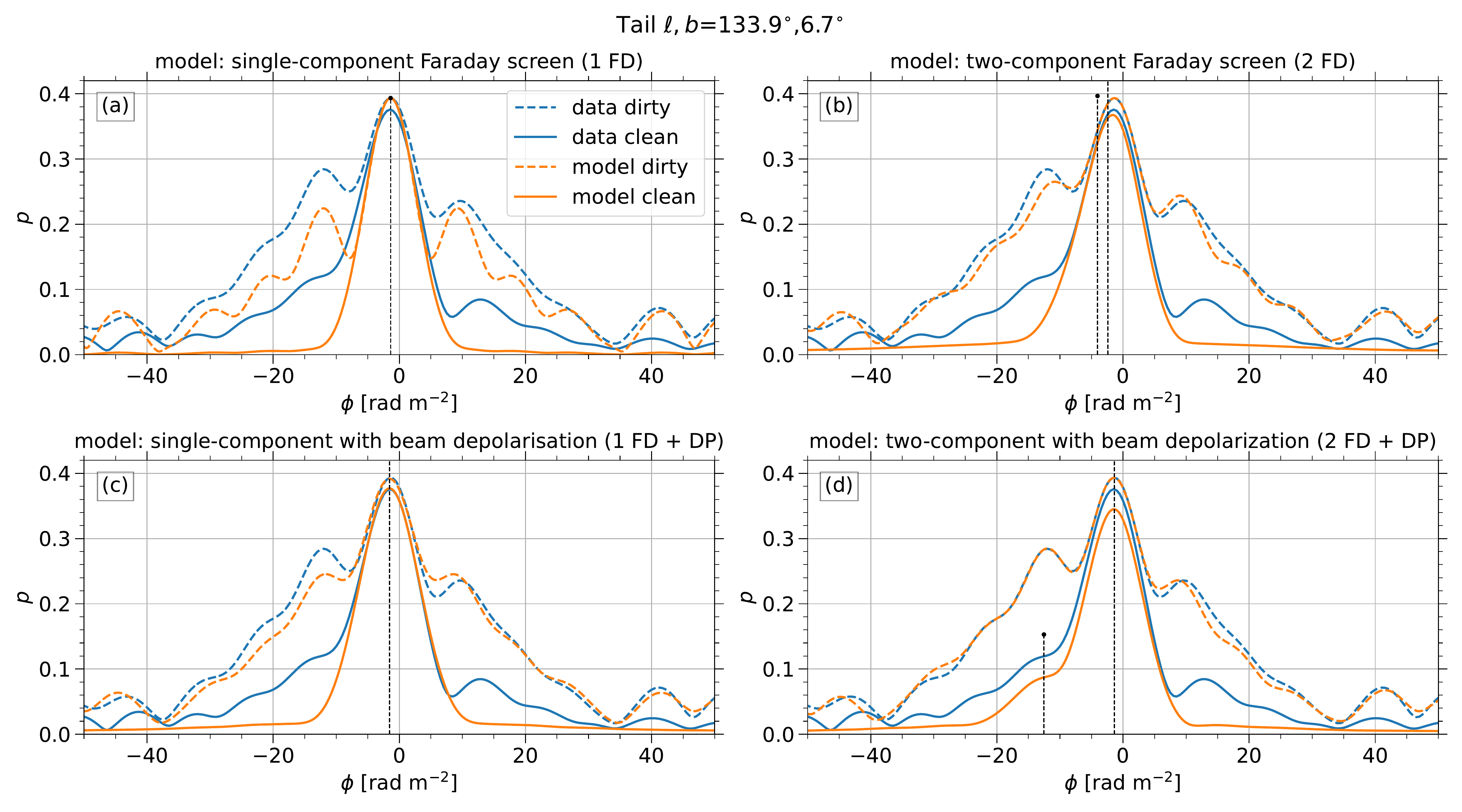}
\caption{Results of QU fitting used as input to Faraday synthesis for a LOS on the tail of the tadpole. See Figure~\ref{fig:figB3} for full description.} \label{fig:figB4}
\end{figure}

\begin{figure}
\plotone{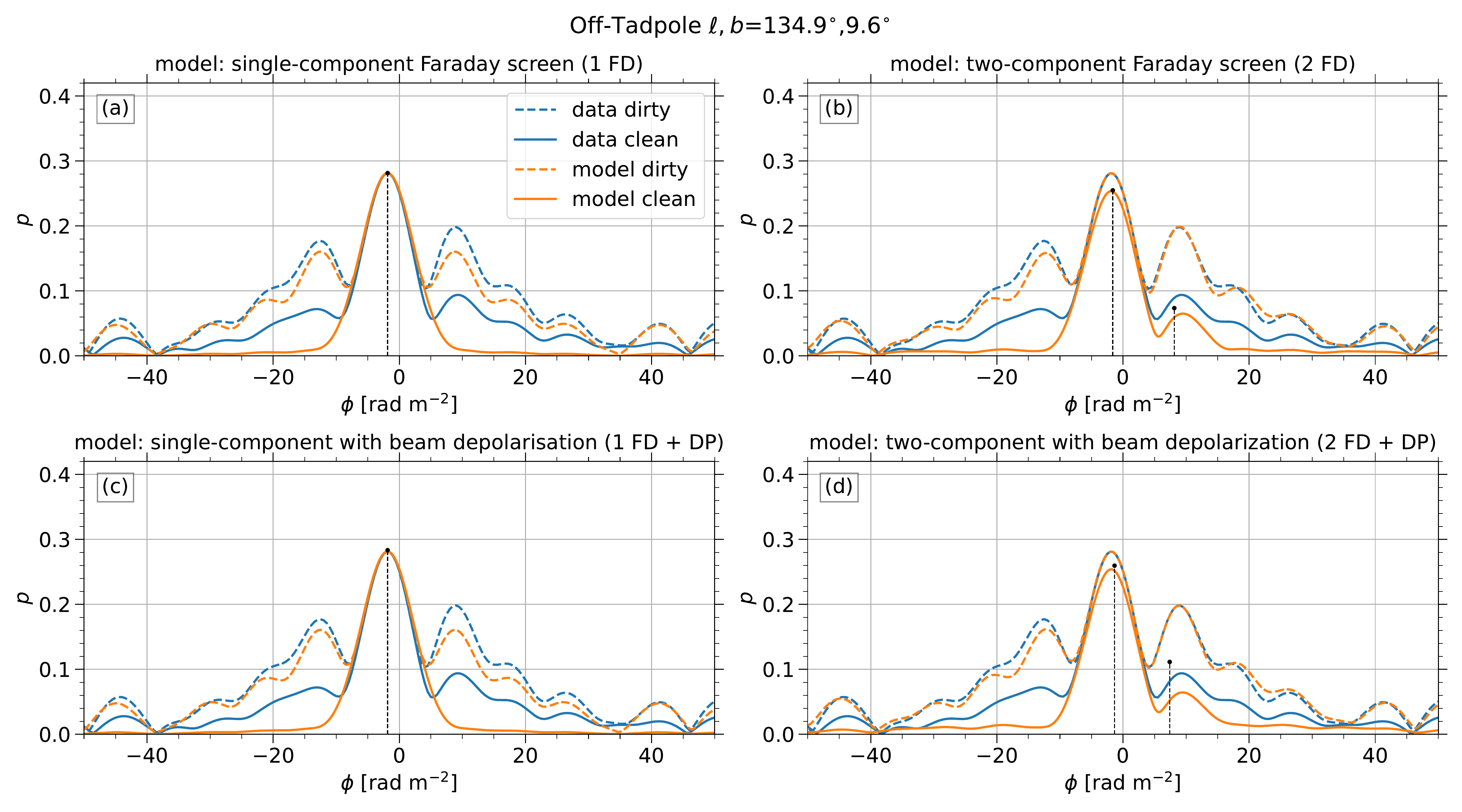}
\caption{Results of QU fitting used as input to Faraday synthesis for a LOS in the off-tadpole region. See Figure~\ref{fig:figB3} for full description.} \label{fig:figB5}
\end{figure}

We apply Faraday synthesis to each model $\tilde{p}(\lambda^2)$ determined from QU fitting. We show the resulting Faraday depth spectra, compared to the spectra derived from the data $\tilde{p}(\lambda^2)$ in Figures~\ref{fig:figB3}--\ref{fig:figB5}. For all three lines of sight, the two-component, beam depolarized model (2 FD + DP) spectrum agrees well with the spectrum derived from the data $\tilde{p}(\lambda^2)$. Note that small discrepancies between the data spectra shown in Figure~\ref{fig:fig7} and the spectra shown here arise because the latter were calculated from $Q/I$ and $U/I$ (no spectral index) rather than $Q$ and $U$ (including spectral index) as were used for the main analysis. The off-tadpole LOS shown if Figure~\ref{fig:figB5}$d$ is an example of the type of scenario described in Appendix~\ref{sec:A1} and depicted in Figure~\ref{fig:figA1}$d$ (model~8), in which two Faraday depth components (denoted by the vertical black lines in both of the two-component models) become slightly shifted with respect to each other through the Faraday synthesis process.

In Table~\ref{tab:qufitall} we summarize the best-fit parameters of all models for these three lines of sight, along with the Bayesian evidence, $Z$, and the Bayes odds ratio, $\Delta \ln(Z)$, to quantify the comparison between the models. We compare all models, m, to 2 FD + DP, which has the highest value of $ln(Z)$, such that 
\begin{equation}
    \Delta \ln(Z_{\text{m}}) = \ln(Z_{\text{2 FD + DP}}) - \ln(Z_{\text{m}}).
\end{equation}
For the selected points in the head and tail of the tadpole, a two-component beam depolarized model (2 FD + DP) is clearly the best fit, while for the selected off-tadpole point a 2-component model without depolarization (2 FD) may be sufficient within the set of models explored here. As noted in Section~\ref{ssec:qufitting} a full description of the lines of sight toward the tadpole likely includes more complexity than this basic set of models. The primary Faraday depth components, $\phi_1$, are reasonably consistent between these models, while in the head and tail of the tadpole the secondary component, $\phi_2$, is strongly dependent on whether or not beam depolarization is included in the model.

\begin{table*}[tb]
    \centering
    \caption{QU fitting results for representative lines of sight}
    \label{tab:qufitall}
    \begin{tabular}{lllcccccccccc}
        \hline
        Location & $\ell$, $b$ & Model & $\ln(Z)$ & $\Delta \ln(Z)$ & $p_1$ & $\phi_1$ & $\chi_{0,1}$ & $\sigma_1$ & $p_2$ & $\phi_2$ & $\chi_{0,2}$ & $\sigma_2$ \\
        & & & ($\times 10^3$) &($\times 10^3$) & & (rad m$^{-2}$) & &(rad m$^{-2}$)& &(rad m$^{-2}$) & & (rad m$^{-2}$)\\
        \hline
        Head & 137\arcdeg, 7\arcdeg & 1 FD & -53.4 & 37.5 & 0.4 & -7.5 & 2.0\arcdeg \\
        & & 2 FD & -27.0 & 11.1 & 0.6 & -5.9 & 143.8\arcdeg & & 0.4 & -4.2 & 9.6\arcdeg\\
        & & 1 FD + DP & -27.7 & 11.8 & 0.5 & -7.6 & 2.4\arcdeg & 1.4\\
        & & 2 FD + DP & -15.9 & 0.0 & 0.5 & -7.5 & 0.6\arcdeg & 1.4 & 0.1 & 3.0 & 65.3\arcdeg & 0.02 \\
        \hline
        Tail & 134\arcdeg, 6.5\arcdeg & 1 FD & -57.3 & 49.2 & 0.4 & -1.4 & 162.7\arcdeg \\
        & & 2 FD & -13.4 & 5.3 & 0.6 & -2.4 & 15.7\arcdeg & & 0.4 & -4.0 & 152.2\arcdeg \\
        & & 1 FD + DP & -16.7 & 8.6 & 0.6 & -1.6 & 164.9\arcdeg & 1.6\\
        & & 2 FD + DP & -8.1 & 0.0 & 0.5 & -1.4 & 163.2\arcdeg & 1.3 & 0.2 & -12.5 & 112.5\arcdeg & 2.0 \\
        \hline
        Off & 135\arcdeg, 9\arcdeg & 1 FD & -14.9 & 9.8 & 0.3 & -1.9 & 8.8\arcdeg \\
        & & 2 FD & -5.7 & 0.6 & 0.3 & -1.6 & 1.5\arcdeg & & 0.1 & 8.1 & 93.2\arcdeg \\
        & & 1 FD + DP & -14.9 & 9.8 & 0.3 & -1.9 & 8.8\arcdeg & 0.18\\
        & & 2 FD + DP & -5.1 & 0.0 & 0.3 & -1.3 & 175.5\arcdeg & 0.03 & 0.1 & 7.4 & 106.3\arcdeg & 1.3 \\
        \hline
    \end{tabular}
\end{table*}

\bibliography{main}{}

\begin{thebibliography}{}
\expandafter\ifx\csname natexlab\endcsname\relax\def\natexlab#1{#1}\fi
\providecommand{\url}[1]{\href{#1}{#1}}
\providecommand{\dodoi}[1]{doi:~\href{http://doi.org/#1}{\nolinkurl{#1}}}
\providecommand{\doeprint}[1]{\href{http://ascl.net/#1}{\nolinkurl{http://ascl.net/#1}}}
\providecommand{\doarXiv}[1]{\href{https://arxiv.org/abs/#1}{\nolinkurl{https://arxiv.org/abs/#1}}}

\bibitem[{{Anderson} {et~al.}(2023){Anderson}, {Koblischke}, \&
  {Eyer}}]{anderson2023reconciling}
{Anderson}, R.~I., {Koblischke}, N.~W., \& {Eyer}, L. 2023, {Reconciling
  astronomical distance scales with variable red giant stars},
  \dodoi{10.48550/arXiv.2303.04790}

\bibitem[{{Astropy Collaboration} {et~al.}(2022){Astropy Collaboration},
  {Price-Whelan}, {Lim}, {Earl}, {Starkman}, {Bradley}, {Shupe}, {Patil},
  {Corrales}, {Brasseur}, {N{\"o}the}, {Donath}, {Tollerud}, {Morris},
  {Ginsburg}, {Vaher}, {Weaver}, {Tocknell}, {Jamieson}, {van Kerkwijk},
  {Robitaille}, {Merry}, {Bachetti}, {G{\"u}nther}, {Aldcroft},
  {Alvarado-Montes}, {Archibald}, {B{\'o}di}, {Bapat}, {Barentsen},
  {Baz{\'a}n}, {Biswas}, {Boquien}, {Burke}, {Cara}, {Cara}, {Conroy},
  {Conseil}, {Craig}, {Cross}, {Cruz}, {D'Eugenio}, {Dencheva}, {Devillepoix},
  {Dietrich}, {Eigenbrot}, {Erben}, {Ferreira}, {Foreman-Mackey}, {Fox},
  {Freij}, {Garg}, {Geda}, {Glattly}, {Gondhalekar}, {Gordon}, {Grant},
  {Greenfield}, {Groener}, {Guest}, {Gurovich}, {Handberg}, {Hart},
  {Hatfield-Dodds}, {Homeier}, {Hosseinzadeh}, {Jenness}, {Jones}, {Joseph},
  {Kalmbach}, {Karamehmetoglu}, {Ka{\l}uszy{\'n}ski}, {Kelley}, {Kern},
  {Kerzendorf}, {Koch}, {Kulumani}, {Lee}, {Ly}, {Ma}, {MacBride}, {Maljaars},
  {Muna}, {Murphy}, {Norman}, {O'Steen}, {Oman}, {Pacifici}, {Pascual},
  {Pascual-Granado}, {Patil}, {Perren}, {Pickering}, {Rastogi}, {Roulston},
  {Ryan}, {Rykoff}, {Sabater}, {Sakurikar}, {Salgado}, {Sanghi}, {Saunders},
  {Savchenko}, {Schwardt}, {Seifert-Eckert}, {Shih}, {Jain}, {Shukla}, {Sick},
  {Simpson}, {Singanamalla}, {Singer}, {Singhal}, {Sinha}, {Sip{\H{o}}cz},
  {Spitler}, {Stansby}, {Streicher}, {{\v{S}}umak}, {Swinbank}, {Taranu},
  {Tewary}, {Tremblay}, {de Val-Borro}, {Van Kooten}, {Vasovi{\'c}}, {Verma},
  {de Miranda Cardoso}, {Williams}, {Wilson}, {Winkel}, {Wood-Vasey}, {Xue},
  {Yoachim}, {Zhang}, {Zonca}, \& {Astropy Project
  Contributors}}]{2022ApJ...935..167A}
{Astropy Collaboration}, {Price-Whelan}, A.~M., {Lim}, P.~L., {et~al.} 2022,
  \apj, 935, 167, \dodoi{10.3847/1538-4357/ac7c74}

\bibitem[{{Bell} \& {En{\ss}lin}(2012)}]{2012A&A...540A..80B}
{Bell}, M.~R., \& {En{\ss}lin}, T.~A. 2012, \aap, 540, A80,
  \dodoi{10.1051/0004-6361/201118672}

\bibitem[{{Berkhuijsen} {et~al.}(1964){Berkhuijsen}, {Brouw}, {Muller}, \&
  {Tinbergen}}]{Berkhuijsen_Brouw_etal_1964}
{Berkhuijsen}, E.~M., {Brouw}, W.~N., {Muller}, C.~A., \& {Tinbergen}, J. 1964,
  \bain, 17, 465

\bibitem[{{Bernardi} {et~al.}(2009){Bernardi}, {de Bruyn}, {Brentjens},
  {Ciardi}, {Harker}, {Jeli{\'c}}, {Koopmans}, {Labropoulos}, {Offringa},
  {Pandey}, {Schaye}, {Thomas}, {Yatawatta}, \& {Zaroubi}}]{bernardi_2009}
{Bernardi}, G., {de Bruyn}, A.~G., {Brentjens}, M.~A., {et~al.} 2009, \aap,
  500, 965, \dodoi{10.1051/0004-6361/200911627}

\bibitem[{{Bracco} {et~al.}(2022){Bracco}, {Ntormousi}, {Jeli{\'c}},
  {Padovani}, {{\v{S}}iljeg}, {Erceg}, {Turi{\'c}}, {Ceraj}, \&
  {{\v{S}}nidari{\'c}}}]{bracco:ntormousi:2022}
{Bracco}, A., {Ntormousi}, E., {Jeli{\'c}}, V., {et~al.} 2022, \aap, 663, A37,
  \dodoi{10.1051/0004-6361/202142453}

\bibitem[{{Brentjens} \& {de Bruyn}(2005)}]{brentjens_de_bruyn_2005}
{Brentjens}, M.~A., \& {de Bruyn}, A.~G. 2005, \aap, 441, 1217,
  \dodoi{10.1051/0004-6361:20052990}

\bibitem[{{Brouw} \& {Spoelstra}(1976)}]{brouw_spoelstra_1976}
{Brouw}, W.~N., \& {Spoelstra}, T.~A.~T. 1976, \aaps, 26, 129

\bibitem[{{Brown} {et~al.}(2003){Brown}, {Taylor}, \&
  {Jackel}}]{Brown:Taylor:2003}
{Brown}, J.~C., {Taylor}, A.~R., \& {Jackel}, B.~J. 2003, \apjs, 145, 213,
  \dodoi{10.1086/346082}

\bibitem[{{Burn}(1966)}]{burn_1966}
{Burn}, B.~J. 1966, \mnras, 133, 67, \dodoi{10.1093/mnras/133.1.67}

\bibitem[{{Carretti} {et~al.}(2019){Carretti}, {Haverkorn}, {Staveley-Smith},
  {Bernardi}, {Gaensler}, {Kesteven}, {Poppi}, {Brown}, {Crocker}, {Purcell},
  {Schnitzeler}, \& {Sun}}]{Carretti:2019}
{Carretti}, E., {Haverkorn}, M., {Staveley-Smith}, L., {et~al.} 2019, \mnras,
  489, 2330, \dodoi{10.1093/mnras/stz806}

\bibitem[{{CHIME \& GMIMS collaborations}(2024)}]{data}
{CHIME \& GMIMS collaborations}. 2024, {Faraday tomography with CHIME: the
  `tadpole' feature G137+7},  Canadian Astronomical Data Centre,
  \dodoi{10.11570/24.0001}

\bibitem[{{CHIME Collaboration}(2022)}]{chime_overview_2022}
{CHIME Collaboration}. 2022, \apjs, 261, 29, \dodoi{10.3847/1538-4365/ac6fd9}

\bibitem[{{CHIME Collaboration}(2023)}]{chime-21-emission}
---. 2023, \apj, 947, 16, \dodoi{10.3847/1538-4357/acb13f}

\bibitem[{{CHIME/FRB Collaboration}(2021)}]{2021ApJS..257...59C}
{CHIME/FRB Collaboration}. 2021, \apjs, 257, 59,
  \dodoi{10.3847/1538-4365/ac33ab}

\bibitem[{{Condon} {et~al.}(1998){Condon}, {Cotton}, {Greisen}, {Yin},
  {Perley}, {Taylor}, \& {Broderick}}]{nvss}
{Condon}, J.~J., {Cotton}, W.~D., {Greisen}, E.~W., {et~al.} 1998, \aj, 115,
  1693, \dodoi{10.1086/300337}

\bibitem[{{Dickey} {et~al.}(2019){Dickey}, {Landecker}, {Thomson}, {Wolleben},
  {Sun}, {Carretti}, {Douglas}, {Fletcher}, {Gaensler}, {Gray}, {Haverkorn},
  {Hill}, {Mao}, \& {McClure-Griffiths}}]{dickey:2019}
{Dickey}, J.~M., {Landecker}, T.~L., {Thomson}, A. J.~M., {et~al.} 2019, \apj,
  871, 106, \dodoi{10.3847/1538-4357/aaf85f}

\bibitem[{{Dickey} {et~al.}(2022){Dickey}, {West}, {Thomson}, {Landecker},
  {Bracco}, {Carretti}, {Han}, {Hill}, {Ma}, {Mao}, {Ordog}, {Brown},
  {Douglas}, {Erceg}, {Jeli{\'c}}, {Kothes}, \& {Wolleben}}]{dickey:2022}
{Dickey}, J.~M., {West}, J., {Thomson}, A. J.~M., {et~al.} 2022, \apj, 940, 75,
  \dodoi{10.3847/1538-4357/ac94ce}

\bibitem[{Draine(2011)}]{Draine_2011}
Draine, B. 2011, Physics of the interstellar and Intergalactic Medium
  (Princeton University Press)

\bibitem[{{Erceg} {et~al.}(2022){Erceg}, {Jeli{\'c}}, {Haverkorn}, {Bracco},
  {Shimwell}, {Tasse}, {Dickey}, {Ceraj}, {Drabent}, {Hardcastle}, \&
  {Turi{\'c}}}]{erceg:jelic:2022}
{Erceg}, A., {Jeli{\'c}}, V., {Haverkorn}, M., {et~al.} 2022, \aap, 663, A7,
  \dodoi{10.1051/0004-6361/202142244}

\bibitem[{{Farnsworth} {et~al.}(2011){Farnsworth}, {Rudnick}, \&
  {Brown}}]{Farnsworth:Rudnick:2011}
{Farnsworth}, D., {Rudnick}, L., \& {Brown}, S. 2011, \aj, 141, 191,
  \dodoi{10.1088/0004-6256/141/6/191}

\bibitem[{{Ferri{\`e}re} {et~al.}(2021){Ferri{\`e}re}, {West}, \&
  {Jaffe}}]{ferriere:west:2021}
{Ferri{\`e}re}, K., {West}, J.~L., \& {Jaffe}, T.~R. 2021, \mnras, 507, 4968,
  \dodoi{10.1093/mnras/stab1641}

\bibitem[{{Ferri{\`e}re}(2001)}]{Ferriere:2001}
{Ferri{\`e}re}, K.~M. 2001, Reviews of Modern Physics, 73, 1031,
  \dodoi{10.1103/RevModPhys.73.1031}

\bibitem[{{Foster} {et~al.}(2013){Foster}, {Kothes}, \&
  {Brown}}]{Foster:Kothes:2013}
{Foster}, T., {Kothes}, R., \& {Brown}, J.~C. 2013, \apjl, 773, L11,
  \dodoi{10.1088/2041-8205/773/1/L11}

\bibitem[{{Gaensler} {et~al.}(2001){Gaensler}, {Dickey}, {McClure-Griffiths},
  {Green}, {Wieringa}, \& {Haynes}}]{Gaensler:Dickey:2001}
{Gaensler}, B.~M., {Dickey}, J.~M., {McClure-Griffiths}, N.~M., {et~al.} 2001,
  \apj, 549, 959, \dodoi{10.1086/319468}

\bibitem[{{Gaia Collaboration} {et~al.}(2023){Gaia Collaboration}, {Vallenari},
  {Brown}, {Prusti}, {de Bruijne}, {Arenou}, {Babusiaux}, {Biermann},
  {Creevey}, {Ducourant}, {Evans}, {Eyer}, {Guerra}, {Hutton}, {Jordi},
  {Klioner}, {Lammers}, {Lindegren}, {Luri}, {Mignard}, {Panem}, {Pourbaix},
  {Randich}, {Sartoretti}, {Soubiran}, {Tanga}, {Walton}, {Bailer-Jones},
  {Bastian}, {Drimmel}, {Jansen}, {Katz}, {Lattanzi}, {van Leeuwen}, {Bakker},
  {Cacciari}, {Casta{\~n}eda}, {De Angeli}, {Fabricius}, {Fouesneau},
  {Fr{\'e}mat}, {Galluccio}, {Guerrier}, {Heiter}, {Masana}, {Messineo},
  {Mowlavi}, {Nicolas}, {Nienartowicz}, {Pailler}, {Panuzzo}, {Riclet}, {Roux},
  {Seabroke}, {Sordo}, {Th{\'e}venin}, {Gracia-Abril}, {Portell}, {Teyssier},
  {Altmann}, {Andrae}, {Audard}, {Bellas-Velidis}, {Benson}, {Berthier},
  {Blomme}, {Burgess}, {Busonero}, {Busso}, {C{\'a}novas}, {Carry}, {Cellino},
  {Cheek}, {Clementini}, {Damerdji}, {Davidson}, {de Teodoro}, {Nu{\~n}ez
  Campos}, {Delchambre}, {Dell'Oro}, {Esquej}, {Fern{\'a}ndez-Hern{\'a}ndez},
  {Fraile}, {Garabato}, {Garc{\'\i}a-Lario}, {Gosset}, {Haigron}, {Halbwachs},
  {Hambly}, {Harrison}, {Hern{\'a}ndez}, {Hestroffer}, {Hodgkin}, {Holl},
  {Jan{\ss}en}, {Jevardat de Fombelle}, {Jordan}, {Krone-Martins}, {Lanzafame},
  {L{\"o}ffler}, {Marchal}, {Marrese}, {Moitinho}, {Muinonen}, {Osborne},
  {Pancino}, {Pauwels}, {Recio-Blanco}, {Reyl{\'e}}, {Riello}, {Rimoldini},
  {Roegiers}, {Rybizki}, {Sarro}, {Siopis}, {Smith}, {Sozzetti}, {Utrilla},
  {van Leeuwen}, {Abbas}, {{\'A}brah{\'a}m}, {Abreu Aramburu}, {Aerts},
  {Aguado}, {Ajaj}, {Aldea-Montero}, {Altavilla}, {{\'A}lvarez}, {Alves},
  {Anders}, {Anderson}, {Anglada Varela}, {Antoja}, {Baines}, {Baker},
  {Balaguer-N{\'u}{\~n}ez}, {Balbinot}, {Balog}, {Barache}, {Barbato},
  {Barros}, {Barstow}, {Bartolom{\'e}}, {Bassilana}, {Bauchet}, {Becciani},
  {Bellazzini}, {Berihuete}, {Bernet}, {Bertone}, {Bianchi}, {Binnenfeld},
  {Blanco-Cuaresma}, {Blazere}, {Boch}, {Bombrun}, {Bossini}, {Bouquillon},
  {Bragaglia}, {Bramante}, {Breedt}, {Bressan}, {Brouillet}, {Brugaletta},
  {Bucciarelli}, {Burlacu}, {Butkevich}, {Buzzi}, {Caffau}, {Cancelliere},
  {Cantat-Gaudin}, {Carballo}, {Carlucci}, {Carnerero}, {Carrasco},
  {Casamiquela}, {Castellani}, {Castro-Ginard}, {Chaoul}, {Charlot}, {Chemin},
  {Chiaramida}, {Chiavassa}, {Chornay}, {Comoretto}, {Contursi}, {Cooper},
  {Cornez}, {Cowell}, {Crifo}, {Cropper}, {Crosta}, {Crowley}, {Dafonte},
  {Dapergolas}, {David}, {David}, {de Laverny}, {De Luise}, {De March}, {De
  Ridder}, {de Souza}, {de Torres}, {del Peloso}, {del Pozo}, {Delbo},
  {Delgado}, {Delisle}, {Demouchy}, {Dharmawardena}, {Di Matteo}, {Diakite},
  {Diener}, {Distefano}, {Dolding}, {Edvardsson}, {Enke}, {Fabre}, {Fabrizio},
  {Faigler}, {Fedorets}, {Fernique}, {Fienga}, {Figueras}, {Fournier},
  {Fouron}, {Fragkoudi}, {Gai}, {Garcia-Gutierrez}, {Garcia-Reinaldos},
  {Garc{\'\i}a-Torres}, {Garofalo}, {Gavel}, {Gavras}, {Gerlach}, {Geyer},
  {Giacobbe}, {Gilmore}, {Girona}, {Giuffrida}, {Gomel}, {Gomez},
  {Gonz{\'a}lez-N{\'u}{\~n}ez}, {Gonz{\'a}lez-Santamar{\'\i}a},
  {Gonz{\'a}lez-Vidal}, {Granvik}, {Guillout}, {Guiraud},
  {Guti{\'e}rrez-S{\'a}nchez}, {Guy}, {Hatzidimitriou}, {Hauser}, {Haywood},
  {Helmer}, {Helmi}, {Sarmiento}, {Hidalgo}, {Hilger}, {H{\l}adczuk}, {Hobbs},
  {Holland}, {Huckle}, {Jardine}, {Jasniewicz}, {Jean-Antoine Piccolo},
  {Jim{\'e}nez-Arranz}, {Jorissen}, {Juaristi Campillo}, {Julbe}, {Karbevska},
  {Kervella}, {Khanna}, {Kontizas}, {Kordopatis}, {Korn}, {K{\'o}sp{\'a}l},
  {Kostrzewa-Rutkowska}, {Kruszy{\'n}ska}, {Kun}, {Laizeau}, {Lambert},
  {Lanza}, {Lasne}, {Le Campion}, {Lebreton}, {Lebzelter}, {Leccia}, {Leclerc},
  {Lecoeur-Taibi}, {Liao}, {Licata}, {Lindstr{\o}m}, {Lister}, {Livanou},
  {Lobel}, {Lorca}, {Loup}, {Madrero Pardo}, {Magdaleno Romeo}, {Managau},
  {Mann}, {Manteiga}, {Marchant}, {Marconi}, {Marcos}, {Marcos Santos},
  {Mar{\'\i}n Pina}, {Marinoni}, {Marocco}, {Marshall}, {Martin Polo},
  {Mart{\'\i}n-Fleitas}, {Marton}, {Mary}, {Masip}, {Massari},
  {Mastrobuono-Battisti}, {Mazeh}, {McMillan}, {Messina}, {Michalik}, {Millar},
  {Mints}, {Molina}, {Molinaro}, {Moln{\'a}r}, {Monari}, {Mongui{\'o}},
  {Montegriffo}, {Montero}, {Mor}, {Mora}, {Morbidelli}, {Morel}, {Morris},
  {Muraveva}, {Murphy}, {Musella}, {Nagy}, {Noval}, {Oca{\~n}a}, {Ogden},
  {Ordenovic}, {Osinde}, {Pagani}, {Pagano}, {Palaversa}, {Palicio},
  {Pallas-Quintela}, {Panahi}, {Payne-Wardenaar}, {Pe{\~n}alosa Esteller},
  {Penttil{\"a}}, {Pichon}, {Piersimoni}, {Pineau}, {Plachy}, {Plum}, {Poggio},
  {Pr{\v{s}}a}, {Pulone}, {Racero}, {Ragaini}, {Rainer}, {Raiteri}, {Rambaux},
  {Ramos}, {Ramos-Lerate}, {Re Fiorentin}, {Regibo}, {Richards}, {Rios Diaz},
  {Ripepi}, {Riva}, {Rix}, {Rixon}, {Robichon}, {Robin}, {Robin}, {Roelens},
  {Rogues}, {Rohrbasser}, {Romero-G{\'o}mez}, {Rowell}, {Royer}, {Ruz Mieres},
  {Rybicki}, {Sadowski}, {S{\'a}ez N{\'u}{\~n}ez}, {Sagrist{\`a} Sell{\'e}s},
  {Sahlmann}, {Salguero}, {Samaras}, {Sanchez Gimenez}, {Sanna},
  {Santove{\~n}a}, {Sarasso}, {Schultheis}, {Sciacca}, {Segol}, {Segovia},
  {S{\'e}gransan}, {Semeux}, {Shahaf}, {Siddiqui}, {Siebert}, {Siltala},
  {Silvelo}, {Slezak}, {Slezak}, {Smart}, {Snaith}, {Solano}, {Solitro},
  {Souami}, {Souchay}, {Spagna}, {Spina}, {Spoto}, {Steele},
  {Steidelm{\"u}ller}, {Stephenson}, {S{\"u}veges}, {Surdej}, {Szabados},
  {Szegedi-Elek}, {Taris}, {Taylor}, {Teixeira}, {Tolomei}, {Tonello}, {Torra},
  {Torra}, {Torralba Elipe}, {Trabucchi}, {Tsounis}, {Turon}, {Ulla}, {Unger},
  {Vaillant}, {van Dillen}, {van Reeven}, {Vanel}, {Vecchiato}, {Viala},
  {Vicente}, {Voutsinas}, {Weiler}, {Wevers}, {Wyrzykowski}, {Yoldas}, {Yvard},
  {Zhao}, {Zorec}, {Zucker}, \& {Zwitter}}]{gaia_2022}
{Gaia Collaboration}, {Vallenari}, A., {Brown}, A.~G.~A., {et~al.} 2023, \aap,
  674, A1, \dodoi{10.1051/0004-6361/202243940}

\bibitem[{{Haffner} {et~al.}(2003){Haffner}, {Reynolds}, {Tufte}, {Madsen},
  {Jaehnig}, \& {Percival}}]{wham_nss}
{Haffner}, L.~M., {Reynolds}, R.~J., {Tufte}, S.~L., {et~al.} 2003, \apjs, 149,
  405, \dodoi{10.1086/378850}

\bibitem[{{Haffner} {et~al.}(2009){Haffner}, {Dettmar}, {Beckman}, {Wood},
  {Slavin}, {Giammanco}, {Madsen}, {Zurita}, \& {Reynolds}}]{Haffner:2009}
{Haffner}, L.~M., {Dettmar}, R.~J., {Beckman}, J.~E., {et~al.} 2009, Reviews of
  Modern Physics, 81, 969, \dodoi{10.1103/RevModPhys.81.969}

\bibitem[{{Haffner} {et~al.}(2010){Haffner}, {Reynolds}, {Madsen}, {Hill},
  {Barger}, {Jaehnig}, {Mierkiewicz}, {Percival}, \& {Chopra}}]{Haffner2010}
{Haffner}, L.~M., {Reynolds}, R.~J., {Madsen}, G.~J., {et~al.} 2010, in
  Astronomical Society of the Pacific Conference Series, Vol. 438, The Dynamic
  Interstellar Medium: A Celebration of the Canadian Galactic Plane Survey, ed.
  R.~{Kothes}, T.~L. {Landecker}, \& A.~G. {Willis}, 388,
  \dodoi{10.48550/arXiv.1008.0612}

\bibitem[{{Hausen} {et~al.}(2002){Hausen}, {Reynolds}, \&
  {Haffner}}]{Hausen:Reynolds:2002}
{Hausen}, N.~R., {Reynolds}, R.~J., \& {Haffner}, L.~M. 2002, \aj, 124, 3336,
  \dodoi{10.1086/344603}

\bibitem[{{Haverkorn}(2015)}]{haverkorn:2015}
{Haverkorn}, M. 2015, in Astrophysics and Space Science Library, Vol. 407,
  Magnetic Fields in Diffuse Media, ed. A.~{Lazarian}, E.~M. {de Gouveia Dal
  Pino}, \& C.~{Melioli}, 483, \dodoi{10.1007/978-3-662-44625-6_17}

\bibitem[{{Haverkorn} {et~al.}(2003){Haverkorn}, {Katgert}, \& {de
  Bruyn}}]{haverkorn_2003}
{Haverkorn}, M., {Katgert}, P., \& {de Bruyn}, A.~G. 2003, \aap, 404, 233,
  \dodoi{10.1051/0004-6361:20030530}

\bibitem[{{Heald}(2009)}]{Heald:2009}
{Heald}, G. 2009, in Cosmic Magnetic Fields: From Planets, to Stars and
  Galaxies, ed. K.~G. {Strassmeier}, A.~G. {Kosovichev}, \& J.~E. {Beckman},
  Vol. 259, 591--602, \dodoi{10.1017/S1743921309031421}

\bibitem[{{Heiles}(1967)}]{Heiles_1967}
{Heiles}, C. 1967, \apjs, 15, 97, \dodoi{10.1086/190164}

\bibitem[{{Heiles}(2002)}]{heiles_2001}
{Heiles}, C. 2002, in Astronomical Society of the Pacific Conference Series,
  Vol. 278, Single-Dish Radio Astronomy: Techniques and Applications, ed.
  S.~{Stanimirovic}, D.~{Altschuler}, P.~{Goldsmith}, \& C.~{Salter}, 131--152,
  \dodoi{10.48550/arXiv.astro-ph/0107327}

\bibitem[{{Heiles} \& {Haverkorn}(2012)}]{heiles:haverkorn:2012}
{Heiles}, C., \& {Haverkorn}, M. 2012, \ssr, 166, 293,
  \dodoi{10.1007/s11214-012-9866-4}

\bibitem[{{HI4PI Collaboration} {et~al.}(2016){HI4PI Collaboration}, {Ben
  Bekhti}, {Fl{\"o}er}, {Keller}, {Kerp}, {Lenz}, {Winkel}, {Bailin},
  {Calabretta}, {Dedes}, {Ford}, {Gibson}, {Haud}, {Janowiecki}, {Kalberla},
  {Lockman}, {McClure-Griffiths}, {Murphy}, {Nakanishi}, {Pisano}, \&
  {Staveley-Smith}}]{ben_bekhti_et_al._2016}
{HI4PI Collaboration}, {Ben Bekhti}, N., {Fl{\"o}er}, L., {et~al.} 2016, \aap,
  594, A116, \dodoi{10.1051/0004-6361/201629178}

\bibitem[{{Hill} {et~al.}(2008){Hill}, {Benjamin}, {Kowal}, {Reynolds},
  {Haffner}, \& {Lazarian}}]{Hill:Benjamin:2008}
{Hill}, A.~S., {Benjamin}, R.~A., {Kowal}, G., {et~al.} 2008, \apj, 686, 363,
  \dodoi{10.1086/590543}

\bibitem[{{Huang}(2015)}]{Huang2015}
{Huang}, Y. 2015, in IAU General Assembly, Vol.~29, 2251816

\bibitem[{{Hunter}(2007)}]{Hunter:2007}
{Hunter}, J.~D. 2007, Computing in Science and Engineering, 9, 90,
  \dodoi{10.1109/MCSE.2007.55}

\bibitem[{{Hutschenreuter} {et~al.}(2022){Hutschenreuter}, {Anderson}, {Betti},
  {Bower}, {Brown}, {Br{\"u}ggen}, {Carretti}, {Clarke}, {Clegg}, {Costa},
  {Croft}, {Van Eck}, {Gaensler}, {de Gasperin}, {Haverkorn}, {Heald}, {Hull},
  {Inoue}, {Johnston-Hollitt}, {Kaczmarek}, {Law}, {Ma}, {MacMahon}, {Mao},
  {Riseley}, {Roy}, {Shanahan}, {Shimwell}, {Stil}, {Sobey}, {O'Sullivan},
  {Tasse}, {Vacca}, {Vernstrom}, {Williams}, {Wright}, \&
  {En{\ss}lin}}]{hutschenreuter:2022}
{Hutschenreuter}, S., {Anderson}, C.~S., {Betti}, S., {et~al.} 2022, \aap, 657,
  A43, \dodoi{10.1051/0004-6361/202140486}

\bibitem[{{Iacobelli} {et~al.}(2013){Iacobelli}, {Haverkorn}, \&
  {Katgert}}]{iacobelli_haverkorn_katgert_2012}
{Iacobelli}, M., {Haverkorn}, M., \& {Katgert}, P. 2013, \aap, 549, A56,
  \dodoi{10.1051/0004-6361/201220175}

\bibitem[{IAU(1973)}]{iau:1973}
IAU. 1973, Transactions of the International Astronomical Union, 15, 165–167,
  \dodoi{10.1017/S0251107X00031606}

\bibitem[{{Ideguchi} {et~al.}(2014){Ideguchi}, {Takahashi}, {Akahori},
  {Kumazaki}, \& {Ryu}}]{Ideguchi:Takahashi:2014}
{Ideguchi}, S., {Takahashi}, K., {Akahori}, T., {Kumazaki}, K., \& {Ryu}, D.
  2014, \pasj, 66, 5, \dodoi{10.1093/pasj/pst007}

\bibitem[{{Jenkins}(2013)}]{Jenkins:2013}
{Jenkins}, E.~B. 2013, \apj, 764, 25, \dodoi{10.1088/0004-637X/764/1/25}

\bibitem[{{Kothes} {et~al.}(2014){Kothes}, {Sun}, {Reich}, \&
  {Foster}}]{2014ApJ...784L..26K}
{Kothes}, R., {Sun}, X.~H., {Reich}, W., \& {Foster}, T.~J. 2014, \apjl, 784,
  L26, \dodoi{10.1088/2041-8205/784/2/L26}

\bibitem[{{Landecker} {et~al.}(2000){Landecker}, {Dewdney}, {Burgess}, {Gray},
  {Higgs}, {Hoffmann}, {Hovey}, {Karpa}, {Lacey}, {Prowse}, {Purton}, {Roger},
  {Willis}, {Wyslouzil}, {Routledge}, \& {Vaneldik}}]{Landecker:2000}
{Landecker}, T.~L., {Dewdney}, P.~E., {Burgess}, T.~A., {et~al.} 2000, \aaps,
  145, 509, \dodoi{10.1051/aas:2000257}

\bibitem[{{Landecker} {et~al.}(2010){Landecker}, {Reich}, {Reid}, {Reich},
  {Wolleben}, {Kothes}, {Uyan{\i}ker}, {Gray}, {Del Rizzo}, {F{\"u}rst},
  {Taylor}, \& {Wielebinski}}]{Landecker:2010}
{Landecker}, T.~L., {Reich}, W., {Reid}, R.~I., {et~al.} 2010, \aap, 520, A80,
  \dodoi{10.1051/0004-6361/200913921}

\bibitem[{{Mao} {et~al.}(2012{\natexlab{a}}){Mao}, {McClure-Griffiths},
  {Gaensler}, {Brown}, {van Eck}, {Haverkorn}, {Kronberg}, {Stil}, {Shukurov},
  \& {Taylor}}]{2012ApJ...755...21M}
{Mao}, S.~A., {McClure-Griffiths}, N.~M., {Gaensler}, B.~M., {et~al.}
  2012{\natexlab{a}}, \apj, 755, 21, \dodoi{10.1088/0004-637X/755/1/21}

\bibitem[{{Mao} {et~al.}(2012{\natexlab{b}}){Mao}, {McClure-Griffiths},
  {Gaensler}, {Haverkorn}, {Beck}, {McConnell}, {Wolleben}, {Stanimirovi{\'c}},
  {Dickey}, \& {Staveley-Smith}}]{2012ApJ...759...25M}
---. 2012{\natexlab{b}}, \apj, 759, 25, \dodoi{10.1088/0004-637X/759/1/25}

\bibitem[{{Masui} {et~al.}(2019){Masui}, {Shaw}, {Ng}, {Smith}, {Vanderlinde},
  \& {Paradise}}]{masui:shaw:2019}
{Masui}, K.~W., {Shaw}, J.~R., {Ng}, C., {et~al.} 2019, \apj, 879, 16,
  \dodoi{10.3847/1538-4357/ab229e}

\bibitem[{{Mckinven} {et~al.}(2021){Mckinven}, {Michilli}, {Masui}, {Cubranic},
  {Gaensler}, {Ng}, {Bhardwaj}, {Leung}, {Boyle}, {Brar}, {Cassanelli}, {Li},
  {Mena-Parra}, {Rahman}, \& {Stairs}}]{McKinven:2021}
{Mckinven}, R., {Michilli}, D., {Masui}, K., {et~al.} 2021, \apj, 920, 138,
  \dodoi{10.3847/1538-4357/ac126a}

\bibitem[{{Mckinven} {et~al.}(2023){Mckinven}, {Gaensler}, {Michilli}, {Masui},
  {Kaspi}, {Bhardwaj}, {Cassanelli}, {Chawla}, {Dong}, {Fonseca}, {Leung},
  {Li}, {Ng}, {Patel}, {Petroff}, {Pearlman}, {Pleunis}, {Rafiei-Ravandi},
  {Rahman}, {Sand}, {Shin}, {Scholz}, {Stairs}, {Smith}, {Su}, \&
  {Tendulkar}}]{McKinven:2023}
{Mckinven}, R., {Gaensler}, B.~M., {Michilli}, D., {et~al.} 2023, \apj, 950,
  12, \dodoi{10.3847/1538-4357/acc65f}

\bibitem[{OpenAI(2023)}]{OpenAI}
OpenAI. 2023, {ChatGPT}.
\newblock \url{https://beta.openai.com/chatgpt/}

\bibitem[{{Ordog} {et~al.}(2017){Ordog}, {Brown}, {Kothes}, \&
  {Landecker}}]{ordog:2017}
{Ordog}, A., {Brown}, J.~C., {Kothes}, R., \& {Landecker}, T.~L. 2017, \aap,
  603, A15, \dodoi{10.1051/0004-6361/201730740}

\bibitem[{{O'Sullivan} {et~al.}(2012){O'Sullivan}, {Brown}, {Robishaw},
  {Schnitzeler}, {McClure-Griffiths}, {Feain}, {Taylor}, {Gaensler},
  {Landecker}, {Harvey-Smith}, \& {Carretti}}]{O'Sullivan:Brown:2012}
{O'Sullivan}, S.~P., {Brown}, S., {Robishaw}, T., {et~al.} 2012, \mnras, 421,
  3300, \dodoi{10.1111/j.1365-2966.2012.20554.x}

\bibitem[{{Purcell} {et~al.}(2020){Purcell}, {Van Eck}, {West}, {Sun}, \&
  {Gaensler}}]{purcell:2020}
{Purcell}, C.~R., {Van Eck}, C.~L., {West}, J., {Sun}, X.~H., \& {Gaensler},
  B.~M. 2020, {RM-Tools: Rotation measure (RM) synthesis and Stokes
  QU-fitting}, Astrophysics Source Code Library, record ascl:2005.003.
\newblock \doeprint{2005.003}

\bibitem[{{Ransom} {et~al.}(2015){Ransom}, {Kothes}, {Geisbuesch}, {Reich}, \&
  {Landecker}}]{Ransom2015}
{Ransom}, R.~R., {Kothes}, R., {Geisbuesch}, J., {Reich}, W., \& {Landecker},
  T.~L. 2015, \apj, 799, 198, \dodoi{10.1088/0004-637X/799/2/198}

\bibitem[{{Ransom} {et~al.}(2010){Ransom}, {Kothes}, {Wolleben}, \&
  {Landecker}}]{Ransom2010}
{Ransom}, R.~R., {Kothes}, R., {Wolleben}, M., \& {Landecker}, T.~L. 2010,
  \apj, 724, 946, \dodoi{10.1088/0004-637X/724/2/946}

\bibitem[{{Reich} {et~al.}(1990){Reich}, {Fuerst}, {Reich}, \&
  {Reif}}]{1990A&AS...85..633R}
{Reich}, W., {Fuerst}, E., {Reich}, P., \& {Reif}, K. 1990, \aaps, 85, 633

\bibitem[{{Reich} {et~al.}(2004){Reich}, {F{\"u}rst}, {Reich}, {Uyaniker},
  {Wielebinski}, \& {Wolleben}}]{2004mim..proc...45R}
{Reich}, W., {F{\"u}rst}, E., {Reich}, P., {et~al.} 2004, in The Magnetized
  Interstellar Medium, ed. B.~{Uyaniker}, W.~{Reich}, \& R.~{Wielebinski},
  45--50

\bibitem[{{Reid} {et~al.}(2014){Reid}, {Menten}, {Brunthaler}, {Zheng}, {Dame},
  {Xu}, {Wu}, {Zhang}, {Sanna}, {Sato}, {Hachisuka}, {Choi}, {Immer},
  {Moscadelli}, {Rygl}, \& {Bartkiewicz}}]{2014ApJ...783..130R}
{Reid}, M.~J., {Menten}, K.~M., {Brunthaler}, A., {et~al.} 2014, \apj, 783,
  130, \dodoi{10.1088/0004-637X/783/2/130}

\bibitem[{{Rudnick} \& {Cotton}(2023)}]{Rudnick:Cotton:2023}
{Rudnick}, L., \& {Cotton}, W.~D. 2023, \mnras, 522, 1464,
  \dodoi{10.1093/mnras/stad1090}

\bibitem[{{Schnitzeler} {et~al.}(2007){Schnitzeler}, {Katgert}, \& {de
  Bruyn}}]{schnitzeler:2007}
{Schnitzeler}, D.~H.~F.~M., {Katgert}, P., \& {de Bruyn}, A.~G. 2007, \aap,
  471, L21, \dodoi{10.1051/0004-6361:20077635}

\bibitem[{{Schnitzeler} {et~al.}(2009){Schnitzeler}, {Katgert}, \& {de
  Bruyn}}]{Schnitzeler:2009}
---. 2009, \aap, 494, 611, \dodoi{10.1051/0004-6361:20078912}

\bibitem[{{Soderblom} {et~al.}(1989){Soderblom}, {Pendleton}, \&
  {Pallavicini}}]{soderblom}
{Soderblom}, D.~R., {Pendleton}, J., \& {Pallavicini}, R. 1989, \aj, 97, 539,
  \dodoi{10.1086/115003}

\bibitem[{{Sokoloff} {et~al.}(1998){Sokoloff}, {Bykov}, {Shukurov},
  {Berkhuijsen}, {Beck}, \& {Poezd}}]{sokoloff:1998}
{Sokoloff}, D.~D., {Bykov}, A.~A., {Shukurov}, A., {et~al.} 1998, \mnras, 299,
  189, \dodoi{10.1046/j.1365-8711.1998.01782.x}

\bibitem[{{Sun} {et~al.}(2015){Sun}, {Rudnick}, {Akahori}, {Anderson}, {Bell},
  {Bray}, {Farnes}, {Ideguchi}, {Kumazaki}, {O'Brien}, {O'Sullivan}, {Scaife},
  {Stepanov}, {Stil}, {Takahashi}, {van Weeren}, \&
  {Wolleben}}]{Sun:Rudnick:2015}
{Sun}, X.~H., {Rudnick}, L., {Akahori}, T., {et~al.} 2015, \aj, 149, 60,
  \dodoi{10.1088/0004-6256/149/2/60}

\bibitem[{{Tahani} {et~al.}(2018){Tahani}, {Plume}, {Brown}, \&
  {Kainulainen}}]{Tahani:2018}
{Tahani}, M., {Plume}, R., {Brown}, J.~C., \& {Kainulainen}, J. 2018, \aap,
  614, A100, \dodoi{10.1051/0004-6361/201732219}

\bibitem[{{Tahani} {et~al.}(2022{\natexlab{a}}){Tahani}, {Glover}, {Lupypciw},
  {West}, {Kothes}, {Plume}, {Inutsuka}, {Lee}, {Grenier}, {Knee}, {Brown},
  {Doi}, {Robishaw}, \& {Haverkorn}}]{2022A&A...660L...7T}
{Tahani}, M., {Glover}, J., {Lupypciw}, W., {et~al.} 2022{\natexlab{a}}, \aap,
  660, L7, \dodoi{10.1051/0004-6361/202243322}

\bibitem[{{Tahani} {et~al.}(2022{\natexlab{b}}){Tahani}, {Lupypciw}, {Glover},
  {Plume}, {West}, {Kothes}, {Inutsuka}, {Lee}, {Robishaw}, {Knee}, {Brown},
  {Doi}, {Grenier}, \& {Haverkorn}}]{2022A&A...660A..97T}
{Tahani}, M., {Lupypciw}, W., {Glover}, J., {et~al.} 2022{\natexlab{b}}, \aap,
  660, A97, \dodoi{10.1051/0004-6361/202141170}

\bibitem[{{Takahashi}(2023)}]{2023PASJ...75S..50T}
{Takahashi}, K. 2023, \pasj, 75, S50, \dodoi{10.1093/pasj/psac111}

\bibitem[{{Taylor} {et~al.}(2009){Taylor}, {Stil}, \& {Sunstrum}}]{Taylor:2009}
{Taylor}, A.~R., {Stil}, J.~M., \& {Sunstrum}, C. 2009, \apj, 702, 1230,
  \dodoi{10.1088/0004-637X/702/2/1230}

\bibitem[{{Taylor} {et~al.}(2003){Taylor}, {Gibson}, {Peracaula}, {Martin},
  {Landecker}, {Brunt}, {Dewdney}, {Dougherty}, {Gray}, {Higgs}, {Kerton},
  {Knee}, {Kothes}, {Purton}, {Uyaniker}, {Wallace}, {Willis}, \&
  {Durand}}]{Taylor:2003}
{Taylor}, A.~R., {Gibson}, S.~J., {Peracaula}, M., {et~al.} 2003, \aj, 125,
  3145, \dodoi{10.1086/375301}

\bibitem[{{Thomson} {et~al.}(2019){Thomson}, {Landecker}, {Dickey},
  {McClure-Griffiths}, {Wolleben}, {Carretti}, {Fletcher}, {Federrath}, {Hill},
  {Mao}, {Gaensler}, {Haverkorn}, {Clark}, {Van Eck}, \& {West}}]{thomson:2019}
{Thomson}, A. J.~M., {Landecker}, T.~L., {Dickey}, J.~M., {et~al.} 2019,
  \mnras, 487, 4751, \dodoi{10.1093/mnras/stz1438}

\bibitem[{{Thomson} {et~al.}(2021){Thomson}, {Landecker}, {McClure-Griffiths},
  {Dickey}, {Campbell}, {Carretti}, {Clark}, {Federrath}, {Gaensler}, {Han},
  {Haverkorn}, {Hill}, {Mao}, {Ordog}, {Pratley}, {Reich}, {Van Eck}, {West},
  \& {Wolleben}}]{thomson:2021}
{Thomson}, A. J.~M., {Landecker}, T.~L., {McClure-Griffiths}, N.~M., {et~al.}
  2021, \mnras, 507, 3495, \dodoi{10.1093/mnras/stab1805}

\bibitem[{{Thomson} {et~al.}(2023){Thomson}, {McConnell}, {Lenc}, {Galvin},
  {Rudnick}, {Heald}, {Hale}, {Duchesne}, {Anderson}, {Carretti}, {Federrath},
  {Gaensler}, {Harvey-Smith}, {Haverkorn}, {Hotan}, {Ma}, {Murphy},
  {McClure-Griffiths}, {Moss}, {O'Sullivan}, {Raja}, {Seta}, {Van Eck}, {West},
  {Whiting}, \& {Wieringa}}]{thomson:2023}
{Thomson}, A. J.~M., {McConnell}, D., {Lenc}, E., {et~al.} 2023, \pasa, 40,
  e040, \dodoi{10.1017/pasa.2023.38}

\bibitem[{{Uyaniker} {et~al.}(1998){Uyaniker}, {Fuerst}, {Reich}, {Reich}, \&
  {Wielebinski}}]{Uyaniker:1998}
{Uyaniker}, B., {Fuerst}, E., {Reich}, W., {Reich}, P., \& {Wielebinski}, R.
  1998, \aaps, 132, 401, \dodoi{10.1051/aas:1998449}

\bibitem[{{Uyaniker} {et~al.}(1999){Uyaniker}, {F{\"u}rst}, {Reich}, {Reich},
  \& {Wielebinski}}]{Uyaniker:1999}
{Uyaniker}, B., {F{\"u}rst}, E., {Reich}, W., {Reich}, P., \& {Wielebinski}, R.
  1999, \aaps, 138, 31, \dodoi{10.1051/aas:1999494}

\bibitem[{{Uyaniker} {et~al.}(2003){Uyaniker}, {Landecker}, {Gray}, \&
  {Kothes}}]{Uyaniker:Landecker:2003}
{Uyaniker}, B., {Landecker}, T.~L., {Gray}, A.~D., \& {Kothes}, R. 2003, \apj,
  585, 785, \dodoi{10.1086/346234}

\bibitem[{{Van Eck} {et~al.}(2017){Van Eck}, {Haverkorn}, {Alves}, {Beck}, {de
  Bruyn}, {En{\ss}lin}, {Farnes}, {Ferri{\`e}re}, {Heald}, {Horellou},
  {Horneffer}, {Iacobelli}, {Jeli{\'c}}, {Mart{\'\i}-Vidal}, {Mulcahy},
  {Reich}, {R{\"o}ttgering}, {Scaife}, {Schnitzeler}, {Sobey}, \&
  {Sridhar}}]{vaneck:haverkorn:2017}
{Van Eck}, C.~L., {Haverkorn}, M., {Alves}, M.~I.~R., {et~al.} 2017, \aap, 597,
  A98, \dodoi{10.1051/0004-6361/201629707}

\bibitem[{{Van Eck} {et~al.}(2019){Van Eck}, {Haverkorn}, {Alves}, {Beck},
  {Best}, {Carretti}, {Chy{\.z}y}, {En{\ss}lin}, {Farnes}, {Ferri{\`e}re},
  {Heald}, {Iacobelli}, {Jeli{\'c}}, {Reich}, {R{\"o}ttgering}, \&
  {Schnitzeler}}]{vaneck:haverkorn:2019}
---. 2019, \aap, 623, A71, \dodoi{10.1051/0004-6361/201834777}

\bibitem[{{Van Eck} {et~al.}(2021){Van Eck}, {Brown}, {Ordog}, {Kothes},
  {Landecker}, {Cooper}, {Rae}, {Del Rizzo}, {Gray}, {Ransom}, {Reid}, \&
  {Uyaniker}}]{vaneck:2021}
{Van Eck}, C.~L., {Brown}, J.~C., {Ordog}, A., {et~al.} 2021, \apjs, 253, 48,
  \dodoi{10.3847/1538-4365/abe389}

\bibitem[{{Verschuur}(1968)}]{verschuur_1968}
{Verschuur}, G.~L. 1968, The Observatory, 88, 15

\bibitem[{{Verschuur}(1969)}]{verschuur_1969}
---. 1969, \aj, 74, 597, \dodoi{10.1086/110841}

\bibitem[{{West} {et~al.}(2007){West}, {English}, {Normandeau}, \&
  {Landecker}}]{W3_west_landecker}
{West}, J.~L., {English}, J., {Normandeau}, M., \& {Landecker}, T.~L. 2007,
  \apj, 656, 914, \dodoi{10.1086/510609}

\bibitem[{{Westerhout} {et~al.}(1962){Westerhout}, {Brouw}, {Muller}, \&
  {Tinbergen}}]{westerhout_brouw_muller_tinbergen_1962}
{Westerhout}, G., {Brouw}, W.~N., {Muller}, C.~A., \& {Tinbergen}, J. 1962,
  \aj, 67, 590, \dodoi{10.1086/108852}

\bibitem[{{Wielebinski} {et~al.}(1962){Wielebinski}, {Shakeshaft}, \&
  {Pauliny-Toth}}]{wielebinski_shakeshaft_pauliny-toth}
{Wielebinski}, R., {Shakeshaft}, J.~R., \& {Pauliny-Toth}, I.~I.~K. 1962, The
  Observatory, 82, 158

\bibitem[{{Wolfire} {et~al.}(2003){Wolfire}, {McKee}, {Hollenbach}, \&
  {Tielens}}]{Wolfire:2003}
{Wolfire}, M.~G., {McKee}, C.~F., {Hollenbach}, D., \& {Tielens}, A.~G.~G.~M.
  2003, \apj, 587, 278, \dodoi{10.1086/368016}

\bibitem[{{Wolleben} {et~al.}(2006){Wolleben}, {Landecker}, {Reich}, \&
  {Wielebinski}}]{Wolleben:2006}
{Wolleben}, M., {Landecker}, T.~L., {Reich}, W., \& {Wielebinski}, R. 2006,
  \aap, 448, 411, \dodoi{10.1051/0004-6361:20053851}

\bibitem[{{Wolleben} {et~al.}(2019){Wolleben}, {Landecker}, {Carretti},
  {Dickey}, {Fletcher}, {McClure-Griffiths}, {McConnell}, {Thomson}, {Hill},
  {Gaensler}, {Han}, {Haverkorn}, {Leahy}, {Reich}, \&
  {Taylor}}]{wolleben:2019}
{Wolleben}, M., {Landecker}, T.~L., {Carretti}, E., {et~al.} 2019, \aj, 158,
  44, \dodoi{10.3847/1538-3881/ab22b0}

\bibitem[{{Wolleben} {et~al.}(2021){Wolleben}, {Landecker}, {Douglas}, {Gray},
  {Ordog}, {Dickey}, {Hill}, {Carretti}, {Brown}, {Gaensler}, {Han},
  {Haverkorn}, {Kothes}, {Leahy}, {McClure-Griffiths}, {McConnell}, {Reich},
  {Taylor}, {Thomson}, \& {West}}]{wolleben:2021}
{Wolleben}, M., {Landecker}, T.~L., {Douglas}, K.~A., {et~al.} 2021, \aj, 162,
  35, \dodoi{10.3847/1538-3881/abf7c1}

\end{thebibliography}
\bibliographystyle{aasjournal}


\end{document}